\documentclass[useAMS,usenatbib]{mn2e}

\usepackage{times}
\usepackage{graphicx}
\usepackage{subfigure}

\newcommand{\gsim}{\mathrel{\hbox{\rlap{\lower.55ex \hbox {$\sim$}}
                   \kern-.3em \raise.4ex \hbox{$>$}}}}
\newcommand{\lsim}{\mathrel{\hbox{\rlap{\lower.55ex \hbox {$\sim$}}
                   \kern-.3em \raise.4ex \hbox{$<$}}}}

\title[Stellar and multiple star properties from simulations]{Stellar, brown dwarf, and multiple star properties from hydrodynamical simulations of star cluster formation}
\author[M.R. Bate]{Matthew R. Bate\thanks{E-mail:
mbate@astro.ex.ac.uk}\\ School of Physics, University of Exeter, Stocker
Road, Exeter EX4 4QL}

\date{Accepted for publication in MNRAS}
\begin{document}
\maketitle
\begin{abstract}
We report the statistical properties of stars, brown dwarfs and multiple systems obtained from the largest hydrodynamical simulation of star cluster formation to date that resolves masses down to the opacity limit for fragmentation (a few Jupiter masses).  The simulation is essentially identical to that of Bate, Bonnell \& Bromm except that the initial molecular cloud is larger and more massive.  It produces more than 1250 stars and brown dwarfs, providing unprecedented statistical information that can be compared with observational surveys.  The calculation uses sink particles to model the stars and brown dwarfs.  Part of the calculation is re-run with smaller sink particle accretion radii and gravitational softening to investigate the effect of these approximations on the results.

We find that hydrodynamical/sink particle simulations can reproduce many of the observed stellar properties very well.  Binarity as a function of primary mass, the frequency of very-low-mass (VLM) binaries, general trends for the separation and mass ratio distributions of binaries, and the relative orbital orientations of triples systems are all in reasonable agreement with observations.  We also examine the radial variations of binarity, velocity dispersion, and mass function in the resulting stellar cluster and the distributions of disc truncation radii due to dynamical interactions.  For VLM binaries, because their separations are typically close, we find that their frequency is sensitive to the sink particle accretion radii and gravitational softening used in the calcuations.  Using small accretion radii and gravitational softening results in a frequency of VLM binaries similar to that expected from observational surveys ($\approx 20$ percent).  We also find that VLM binaries evolve from wide, unequal-mass systems towards close equal-mass systems as they form. The two main deficiencies of the calculations are that they over produce brown dwarfs relative to stars and that there are too few unequal mass binaries with K and G-dwarf primaries.  The former of these is likely due to the absence of radiative feedback and/or magnetic fields.
\end{abstract}
\begin{keywords}
ISM: clouds, stars: binaries: general, stars: formation, stars: low-mass, brown dwarfs, stars: luminosity function, mass function, stars: kinematics.
\end{keywords}

\section{Introduction}

Understanding the origin of the statistical properties of stellar systems is the fundamental goal of
a complete theory of star formation.  In terms of their impact on galaxy formation and evolution the
most important statistical properties are probably the stellar initial mass function (IMF) and the
star formation rate and efficiency.  However, for understanding the formation and evolution of
stellar clusters, stellar systems themselves, protoplanetary discs and planetary systems many more
statistical properties are important.  Furthermore, there are currently many models that have been
proposed for the origin of the IMF (see the recent review, \citealt*{BonLarZin2007} or the introduction of \citealt{BatBon2005}).  Many of these are able to explain qualitatively the observed 
form of the IMF, but most of these do not predict other statistical properties.  A complete model must
be able to explain the origin of all the statistical properties of stellar systems, and how these depend
on variations in environment and initial conditions.  Along with the IMF and star formation rate and efficiency, these other
statistical properties include the structure of stellar clusters and stellar velocity dispersions, the
properties of multiple stellar systems, jets, protoplanetary discs, and the rotation rates and magnetic fields of stars.  In particular, when considering binary, triple, and higher-order multiple stellar systems there are many statistical properties that require understanding such as their frequencies, their mass ratios, their orbital separations and eccentricities, relations between orbits and mass ratios in hierarchical systems, and relative stellar rotations.

To investigate the origin of a wide range of statistical properties of stars directly through hydrodynamical calculations is difficult because it is necessary to produce a large number of objects (to get statistically significant results) and to use high resolution (to model low-mass objects such as brown dwarfs, multiple systems, and circumstellar discs).  One approach is to perform a large number of high-resolution calculations of the collapse of isolated small molecular cloud cores (e.g. \citealt*{DelClaBat2004}; \citealt{Delgadoetal2004}; \citealt*{GooWhiWar2004c, GooWhiWar2004a, GooWhiWar2004b}; \citealt*{GooWhiWar2006}).  Such calculations have been able to qualitatively match some of the observed statistical properties of stellar systems.  For example, \citet{Delgadoetal2004} found that multiplicity is an increasing function of primary mass (though they obtained a steeper function than is observed).  \citet{GooWhiWar2004b} found that star formation in small cores might be a good explanation for the somewhat unusual stellar mass function in Taurus (namely the relatively high proportion of stars with masses $\approx 1$~M$_\odot$).  However, such calculations are not applicable to denser star-forming regions since they neglect interactions between cores and protostellar systems.  Furthermore, they use an arbitrary population of dense cores for their initial conditions which may or may not be a good representations of real dense cores.

Over the past few years, we have performed large-scale hydrodynamical calculations of the collapse and fragmentation of turbulent molecular clouds to investigate the origins of stellar properties \citep{BatBonBro2002a, BatBonBro2002b, BatBonBro2003, BatBon2005,Bate2005} (hence forth, the latter three of these papers will be referred to as BBB2003, BB2005, and B2005, respectively).  In these large-scale calculations, dense cores are formed self-consistently from hydrodynamical flows on larger scales and interactions between dense cores and protostellar systems occur naturally.   These calculations have differed from most other large-scale hydrodynamical star formation calculations in that they modelled clouds that were large enough to produce dozens of stars and yet {\em simultaneously} they resolved down to and beyond the opacity limit for fragmentation.  Thus, they resolved the entire mass function, capturing the formation of all stars and brown dwarfs.  They also allowed discs with sizes down to $\approx 10$ AU and binaries with separations of a few AU to be resolved.  Earlier similar large-scale hydrodynamical calculations (\citealt*{KleBurBat1998, KleBur2000, KleBur2001, Klessen2001}; \citealt{Bonnelletal2001a, BonBat2002}; \citealt*{BonBatVin2003}) formed large numbers of stars, but were unable to resolve brown dwarfs, most binaries and discs.  All these calculations used smoothed particle hydrodynamics (SPH) with sink particles to model the star-forming clouds.  Most recently, grid-based adaptive mesh refinement (AMR) calculations have also begun to compete, forming up to a few dozen objects and resolving discs and binaries (\citealt{Lietal2004}; \citealt*{OffKleMcK2008}).  However, regardless of whether SPH or AMR has been used, even the largest high-resolution large-scale calculations published to date have only formed a few dozen stars and brown dwarfs making it difficult to compare the results with observations in any detail.

In this paper, we report the results from two large-scale hydrodynamical calculations of the collapse and fragmentation of turbulent molecular clouds.  The calculations follow the evolution of 500 M$_\odot$ clouds (similar to the calculation presented by BBB2003, but an order of magnitude more massive) to form hundreds of stars and brown dwarfs.  Two versions of the same calculation are performed, one with sink particles with radii of 5 AU (as in BBB2003) and a re-run version that has sink particle radii of only 0.5 AU but which is not followed as far.  The large accretion radii calculation forms 1254 stars and brown dwarfs in 1.5 initial cloud free-fall times.  This large number of objects allows us, for the first time, to make a meaningful comparison of the statistical properties of stars and binary and multiple systems with observations.

The paper is structured as follows. In Section \ref{method}, we briefly describe the numerical method and the initial conditions for the simulations.  In Section \ref{results}, we present our results and compare them with the results of observational surveys.  Our conclusions are given in Section \ref{conclusions}.

\section{Computational method}
\label{method}

The calculations presented here were performed using a three-dimensional 
SPH code.  The SPH code is 
based on a version originally developed by Benz 
\citep{Benz1990, Benzetal1990}.
The smoothing lengths of particles are variable in 
time and space, subject to the constraint that the number 
of neighbours for each particle must remain approximately 
constant at $N_{\rm neigh}=50$.  The SPH equations are 
integrated using a second-order leap-frog 
integrator with individual time steps for each particle.
Gravitational forces between particles and a particle's 
nearest neighbours are calculated using a binary tree.  
We use the standard form of artificial viscosity 
\citep{MonGin1983, Monaghan1992} with strength 
parameters $\alpha_{\rm_v}=1$ and $\beta_{\rm v}=2$.
Further details can be found in \citet{BatBonPri1995}.
The code has been parallelised by M.\ Bate using OpenMP.

\subsection{Equation of state}

To model the thermal behaviour of the gas without performing radiative transfer,
we use a barotropic equation of state for the thermal pressure of the
gas $p = K \rho^{\eta}$, where $K$ is a measure of the entropy
of the gas.  The value of the effective polytropic exponent $\eta$, 
varies with density as
\begin{equation}\label{eta}
\eta = \cases{\begin{array}{rl}
1, & \rho \leq  \rho_{\rm crit}, \cr
7/5, & \rho > \rho_{\rm crit}. \cr
\end{array}}
\end{equation}
We take the mean molecular weight of the gas to be $\mu = 2.46$.
The value of $K$ is defined such that when the gas is 
isothermal $K=c_{\rm s}^2$, with the sound speed
$c_{\rm s} = 1.84 \times 10^4$ cm s$^{-1}$ at 10 K,
and the pressure is continuous when the value of $\eta$ changes.

The value of the critical density above which the
gas becomes non-isothermal is set to 
$\rho_{\rm crit}=10^{-13}~ {\rm g~cm}^{-3}$.
This equation of state has been chosen to match closely the 
relationship between temperature and density during the 
spherically-symmetric collapse of molecular 
cloud cores with solar metallicity as calculated with 
frequency-dependent radiative transfer \citep[e.g.,][]{MasInu2000}.
The equation of state is discussed further by BBB2003.

The heating of the molecular gas that begins at the critical 
density inhibits fragmentation at higher densities.  
This effect is known as the opacity limit for fragmentation 
\citep{LowLyn1976, Rees1976, Silk1977a, Silk1977b, BoyWhi2005}.  
It results in the formation of distinct pressure-supported 
fragments within collapsing gas because the temperature 
increases quickly enough with density that the Jeans mass increases,
and the high density region that was collapsing becomes 
Jeans stable.  These regions stop collapsing and can only contract 
as they accrete mass.  The value of the initial 
mass of a fragment presumably also gives the minimum 
mass for a brown dwarf, since any subsequent accretion 
will only increase a fragment's mass.  This minimum mass
depends on the value of the critical density and is approximately
equal to the Jeans mass at that density and temperature.
The lowest mass object produced by the calculations was $\approx
4$ Jupiter masses (M$_{\rm J}$).

\subsection{Sink particles}

\begin{table*}
\begin{tabular}{lccccccccccc}\hline
Calculation & Initial Gas & Initial  & Jeans & Mach & Accretion & Gravity & End & No. Stars & No. Brown  & Mass of Stars \&  & Mean  \\
& Mass  & Radius & Mass & Number & Radii & Softening & Time & Formed & Dwarfs Formed & Brown Dwarfs & Mass  \\
 & M$_\odot$ & pc & M$_\odot$ & & AU & AU & $t_{\rm ff}$ & & & M$_\odot$ & M$_\odot$ \\ \hline
BBB2003 & 50 & 0.188 & 1 & 6.4 & 5 & 4 & 1.40 & $\geq$23 & $\leq$27 & 5.9 & 0.12  \\
Main & 500 & 0.404 & 1 & 13.7 & 5 & 4 & 1.50 & $\geq$459 & $\leq$795 & 191 & 0.15  \\
& & &  & & & & 1.04 & $\geq$102 & $\leq$119 & 32.6 & 0.15  \\  % Median mass is 0.059  % 69 BDs finished accreting
Re-run & 500 & 0.404 & 1 & 13.7 & 0.5 & 0 & 1.04 & $\geq$94 & $\leq$164 & 32.0 & 0.12  \\ \hline  % Median mass is 0.045   % 100 BDs finished accreting
\end{tabular}
\caption{\label{table1} The parameters and overall statistical results for the BBB2003 calculation and the two calculations presented here.  The initial conditions were similar except that the two calculations presented here are of more massive, larger clouds than that presented by BBB2003.  In particular, the initial densities and mean thermal Jeans masses were identical.  In each case, the magnitudes of the initial turbulent velocity fields were scaled so that the kinetic energy equalled the magnitude of the gravitational potential energy.  The calculations were run for different numbers of initial cloud free-fall times.  Brown dwarfs are defined as having final masses less than 0.075 M$_\odot$.  The numbers of stars (brown dwarfs) are lower (upper) limits because some of the brown dwarfs were still accreting when the calculations were stopped.  The only difference between the main and re-run calculations presented here are in the accretion radii and gravitational softening of the sink particles and the fact that evolution of the re-run calculation could not be followed as long due to computational limitations.}
\end{table*}

As the pressure-supported fragments accrete, their
central density increases, and it becomes computationally impractical
to follow their internal evolution because of the short dynamical
time-scales involved.  Therefore, when the central density of 
a pressure-supported fragment exceeds 
$\rho_{\rm s} = 1000 \rho_{\rm crit}$, 
we insert a sink particle into the calculation \citep{BatBonPri1995}.
This value of $\rho_{\rm s}$ is a factor of ten higher than in earlier
calculations (e.g., BBB2003) which allows more
time for an object to merge or be disrupted before being replaced
by a sink particle.

In the main calculation discussed in this paper, a sink particle is formed by 
replacing the SPH gas particles contained within $r_{\rm acc}=5$ AU 
of the densest gas particle in a pressure-supported fragment 
by a point mass with the same mass and momentum.  Any gas that 
later falls within this radius is accreted by the point mass 
if it is bound and its specific angular momentum is less than 
that required to form a circular orbit at radius $r_{\rm acc}$ 
from the sink particle.  Thus, gaseous discs around sink 
particles can only be resolved if they have radii $\gsim 10$ AU.
Sink particles interact with the gas only via gravity and accretion.
The angular momentum accreted by a sink particle is recorded
but plays no further role in the calculation.

Since all sink particles are created from pressure-supported 
fragments, their initial masses are several M$_{\rm J}$, 
as given by the opacity limit for fragmentation.  
Subsequently, they may accrete large amounts of material 
to become higher-mass brown dwarfs ($\lsim 75$ M$_{\rm J}$) or 
stars ($\gsim 75$ M$_{\rm J}$), but {\it all} the stars and brown
dwarfs begin as these low-mass pressure-supported fragments.

In the main calculation, the gravitational acceleration between two 
sink particles is Newtonian for $r\geq 4$ AU, but is softened within this 
radius using spline softening \citep{Benz1990}.  The maximum acceleration 
occurs at a distance of $\approx 1$ AU; therefore, this is the
minimum separation that a binary can have even if, in reality,
the binary's orbit would have been hardened.  

Part of the main calculation was re-run from just before the first
star formed with sink particle accretion radii of $r_{\rm acc}=0.5$ AU
and with {\it no gravitational softening between sink particles}.
This was done to investigate the dependence of the results on
these approximations.  This partial re-run (referred to henceforth
as the re-run calculation) could not be followed
as long as the main calculation due to the smaller timesteps 
required.

Sink particles were permitted to merge in either calculation if they
passed within 0.02 AU of each other (i.e., $\approx 4$~R$_\odot$).
This radius was chosen because recently formed protostars are
thought to have relatively large radii \citep[e.g.,][]{Larson1969}.
Again, this differs from previous similar calculations.  In the main
calculation, 23 mergers occurred.  In the re-run calculation, 20
mergers occurred (in a shorter period of time).

The benefits and potential problems associated with introducing 
sink particles are discussed in more detail in BBB2003 and will
be further examined in this paper.

\subsubsection{Identification of multiple stellar systems}
\label{calcmultiples}

With the calculations presented in this paper producing many hundreds of stars and brown dwarfs it is important to automate the analysis as much as possible.  Much of this is straightforward.  However, in order to analyse binaries and multiple stellar systems we first need to identify them.  This is done as follows.

At the end of each calculation we essentially construct a structure `tree'.  We begin with every star or brown dwarf  (sink particle) being a `node'.  We then loop over all pairs of nodes calculating the closest pair of `nodes' that are gravitationally bound to each other (i.e. the sum of their relative gravitational and kinetic energies is negative).  This pair of `nodes' then becomes a new node and the original nodes are removed.  For example, if the two nodes are single stars these nodes are replaced by a new node containing a binary that is located at the binary's centre of mass and has the binary's mass and centre of mass velocity.  If one node is a binary and the other is a single star, the new node contains a triple system.  This process is then repeated until no new nodes are formed.  The result is a structure tree that contains single objects (e.g. some that might have been ejected), binaries or multiples that are not bound to any other node), and some nodes which may contain clusters of dozens or hundreds of stars and brown dwarfs, many of which may also be binaries or multiples within these clusters.

The observant reader may note later in the paper that there are a few binaries that have separations of several thousand AU.  These have been checked manually.  They are wide binaries in the periphery of the cluster.  They are composed of ejected objects that happen to be gravitationally bound to one another due to their similar ejection velocities.

\subsection{Initial conditions}

\begin{figure*}
\centering \vspace{-0.0cm}
    \includegraphics[width=15.8cm]{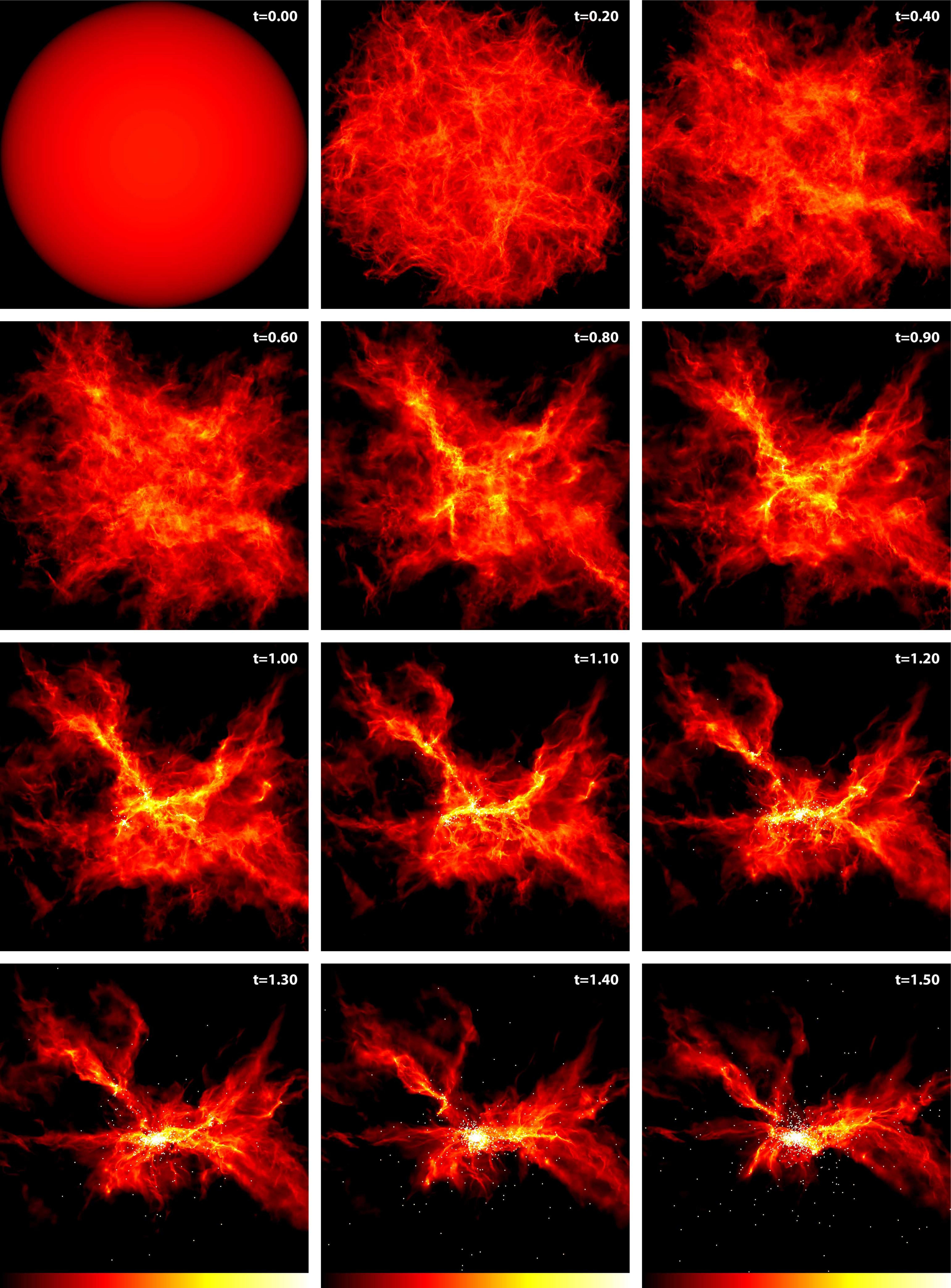}
\caption{The global evolution of the main calculation.  Shocks lead to the dissipation of the turbulent energy that initially supports the cloud, allowing parts of the cloud to collapse.  Star formation begins at $t=0.715t_{\rm ff}$ in a collapsing dense core.  By $t=1.20t_{\rm ff}$ the cloud has produced five main sub-clusters, and by the end of the calculation four out of five of these sub-clusters have merged into a single large cluster.
Each panel is 0.8 pc (165,000 AU) across.  Time is given in units of the initial free-fall time, $t_{\rm ff}=1.90\times 10^5$ yr.  The panels show the logarithm of column density, $N$, through the cloud, with the scale covering $-1.4<\log N<1.0$ with $N$ measured in g~cm$^{-2}$. }
\label{mainevolution}
\end{figure*}

\begin{figure*}
\centering \vspace{-0.0cm}
    \includegraphics[width=15.8cm]{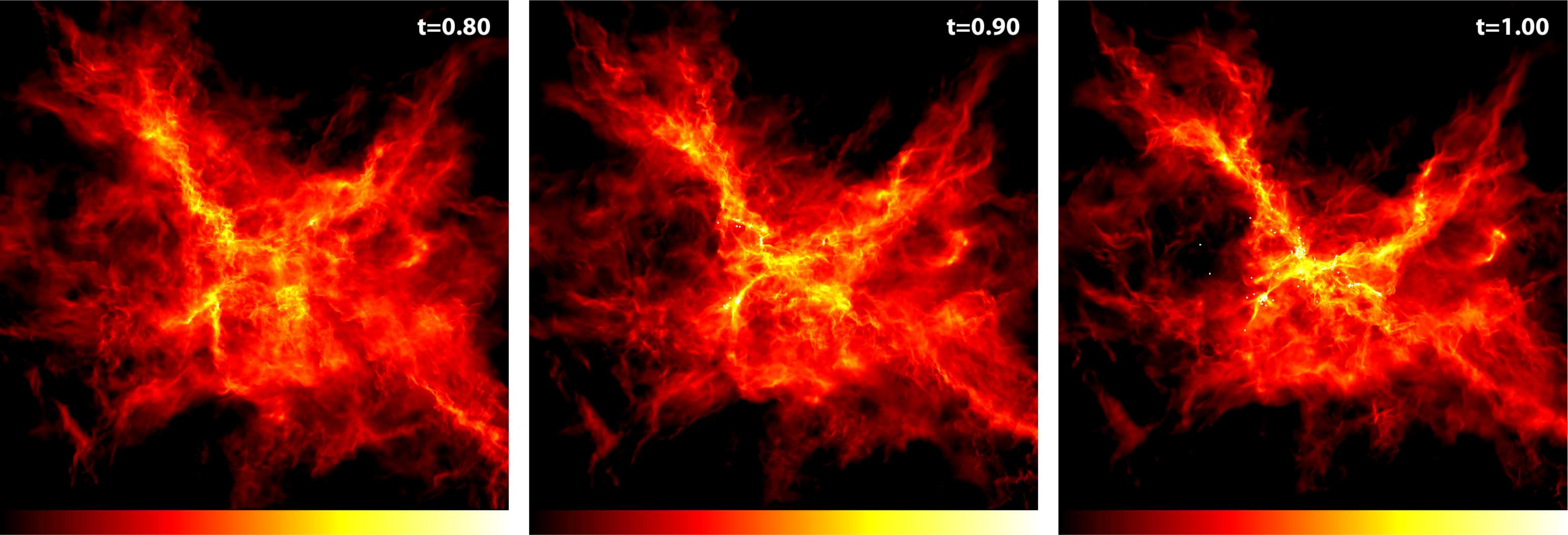}
\caption{The global evolution of the re-run calculation with smaller sink particle accretion radii and no gravitational softening between sink particles.  The global evolution is very similar to the main calculation, but due to the chaotic nature of the dynamics on small-scales the detailed structure of the multiple systems and the ejections differ.  The calculation is only followed to just over one free-fall time because it is much more computationally expensive.
Each panel is 0.8 pc (165,000 AU) across.  Time is given in units of the initial free-fall time, $t_{\rm ff}=1.90\times 10^5$ yr.  The panels show the logarithm of column density, $N$, through the cloud, with the scale covering $-1.4<\log N<1.0$ with $N$ measured in g~cm$^{-2}$. }
\label{rerunevolution}
\end{figure*}

The initial conditions are essentially identical to the calculation of
\citet{BatBonBro2002a, BatBonBro2002b} and BBB2003, except
that the cloud has ten times the mass and a larger radius so as to
give the same initial density, and a larger Mach number so as to
balance the turbulent and gravitational energies initially.  A 500-M$_\odot$ molecular 
cloud was set up as a uniform-density sphere.  The cloud's
radius was set to 0.404 pc (83300 AU).  At the initial temperature of 
10 K, the mean thermal Jeans 
mass is 1 M$_\odot$ (i.e., the cloud contains 500 thermal Jeans masses).  

Although the cloud was uniform in density, we imposed an initial 
supersonic `turbulent' velocity field in the same manner
as \citet*{OstStoGam2001} and BBB2003.  
We generated a divergence-free random Gaussian velocity field with 
a power spectrum $P(k) \propto k^{-4}$, where $k$ is the wavenumber.  
In three dimensions, this results in a
velocity dispersion that varies with distance, $\lambda$, 
as $\sigma(\lambda) \propto \lambda^{1/2}$ in agreement with the 
observed Larson scaling relations for molecular clouds 
\citep{Larson1981}.
The velocity field was generated on a $128^3$ uniform grid and the
velocities of the particles were interpolated from the grid.  As in
BBB2003, the velocity field is normalised so that the kinetic energy 
of the turbulence equals the magnitude of the gravitational potential 
energy of the cloud.
Thus, the initial root-mean-square (rms) Mach number of the turbulence 
was ${\cal M}=13.7$. This is higher than that in BBB2003 (which was ${\cal M}=6.4$).

The initial free-fall times of the cloud was $t_{\rm ff}=6.0\times 10^{12}$~s or 
$1.90\times 10^5$ years (the same as in BBB2003).

\subsection{Resolution}

The local Jeans mass must be resolved throughout the calculations to model fragmentation correctly 
(\citealt{BatBur1997, Trueloveetal1997, Whitworth1998, Bossetal2000}; \citealt*{HubGooWhi2006}).  This requires $\gsim 1.5 N_{\rm neigh}$ SPH particles per Jeans mass; $N_{\rm neigh}$ is insufficient (BBB2003).  The minimum Jeans mass occurs at the maximum density during the isothermal phase of the collapse, $\rho_{\rm crit} = 10^{-13}$ g~cm$^{-3}$, and is $\approx 0.0011$ M$_\odot$ (1.1 M$_{\rm J}$).  Thus, we used $3.5 \times 10^7$ particles to model the 500-M$_\odot$ clouds.

The main calculation required approximately 100,000 CPU hours on a 1.65GHz IBM p570 compute node of the United Kingdom Astrophysical Fluids Facility (UKAFF) while the re-run calculation took approximately half as long.

\section{Results}
\label{results}

The main calculation is the largest simulation of star cluster formation to date in which collapsing gas is resolved down to the opacity limit for fragmentation.  The simulation is similar to that presented by BBB2003, but is of a more massive cloud.  The main purpose of performing the simulation was simply to provide much more accurate statistical information.  BBB2003 only formed 50 stars and brown dwarfs, whereas the main calculation here forms 1254 stars and brown dwarfs in $1.50 t_{\rm ff}$ (285,350 years) and even the re-run calculation that uses smaller accretion radii and no gravitational softening produces 258 objects in $1.038 t_{\rm ff}$ (197,460 years).  See Table \ref{table1} for a summary of the statistics, including the numbers of stars and brown dwarfs produced by the end of the two calculations, the total mass that has been converted to stars and brown dwarfs, and the mean stellar mass.

In BBB2003, although binaries and higher-order multiple systems were produced by the simulation, with such small numbers of objects little could be said about their statistical properties.  Even adding together the results of the three simulations presented by BBB2003, BB2005 and B2005 (which had different initial conditions or thermal physics) only provides 22 binary systems, 15 of which are components of triple and/or quadruple systems.  By contrast the new calculations presented here provide a wealth of binary and high-order multiple systems. The main calculation produced 90 binary, 23 triple, and 25 quadruple systems, including 38 very-low-mass (VLM) multiples in which all components are VLM  (masses less than 0.1 M$_\odot$).  Note that throughout the rest of this paper we will usually refer to VLM objects rather than brown dwarfs in order to allow better comparison to be made with observational surveys that often combine studies of very-low-mass stars and high-mass brown dwarfs in order to increase the sample sizes.  At times we will also make a distinction between VLM objects and low-mass brown dwarfs.  The latter are the subset of VLM objects whose masses are less than 0.03 M$_\odot$ (30 Jupiter masses).  The re-run calculation produced 17 binary, 6 triple, and 17 quadruple systems including 13 VLM multiples.  Thus, we have the ability not just to examine the frequencies of binary stars and VLM objects, but binarity as a function of primary mass, and the separation and mass ratio distributions.

The star formation process itself is similar to that seen in BBB2003, BB2005, B2005.   Figure \ref{mainevolution} shows snapshots of column density from the main calculation illustrating the global evolution.  The initial turbulent velocity field generates structures with those that are strongly self-gravitating collapsing to form stellar groups and clusters.  The main difference from the earlier calculations is that with such a large cloud at least 5 sub-clusters containing dozens to hundreds of objects form ($t\approx 1.10-1.20t_{\rm ff}$), and then merge together to form a single dense stellar cluster by the end of the calculation.  Such hierarchical build-up of a stellar cluster was previously highlighted in the lower-resolution simulation of a 1000 M$_\odot$ cloud performed by \citet{BonBatVin2003}.  The evolution of the cloud and the formation and merger of the sub-clusters is best viewed in a animation.  Animations of the main calculation can be downloaded from http://www.astro.ex.ac.uk/people/mbate/Cluster/ in both the colour scheme of Figure \ref{mainevolution} and as a 3-D red-cyan movie.  Unfortunately, the resolved circumstellar discs and binary systems are not visible on the scale of Figure \ref{mainevolution}, however, with well over 100 multiple systems it is impossible to display these in a paper.  In Figure \ref{rerunevolution} we display the global evolution of the re-run calculation.  There are no substantial differences on large-scales between the two calculations, with the exception of the different pattern of ejected objects visible at $t=1.00t_{\rm ff}$ (c.f.\ the two panels in Figures \ref{mainevolution} and \ref{rerunevolution}).  Since the dynamics of individual stellar systems is chaotic, even changing the sink particle parameters on very small scales affects the outcomes of dynamical interactions.  In the following subsections of the paper, we examine the statistical properties of the stellar systems.

\subsection{The initial mass function}

\begin{figure}
%\vspace{-1.5cm}
\centering
%\hspace{-0.5cm}
    \includegraphics[width=8.4cm]{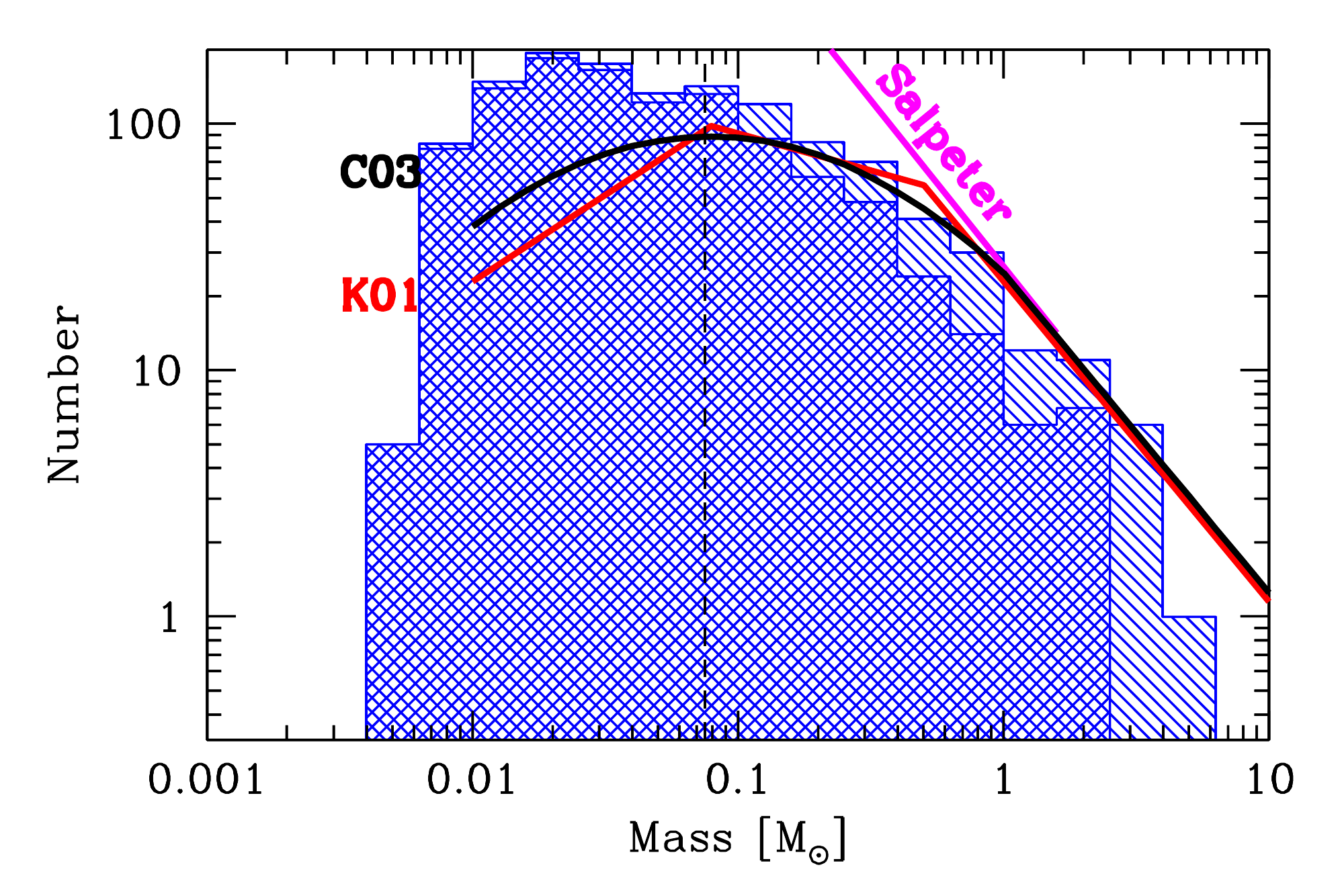}
%\vspace{-5.0cm}
\caption{Histograms giving the initial mass function of the 1254 stars and brown dwarfs that had been produced by the end of the main calculation.  The single hashed region gives all objects while the double hashed region gives those objects that have stopped accreting.  Parameterisations of the observed IMF by \citet{Salpeter1955}, \citet{Kroupa2001} and \citet{Chabrier2003} are given by the magenta line, red broken power law, and black curve, respectively.  The numerical IMF broadly follows the form of the observed IMF, with a Salpeter-like slope above $\sim 0.5$ M$_\odot$ and a turnover at low masses.  However, it clearly over produces brown dwarfs by a factor of $\approx 4$.}
\label{imf}
\end{figure}

\begin{figure*}
\centering
    \includegraphics[width=8.4cm]{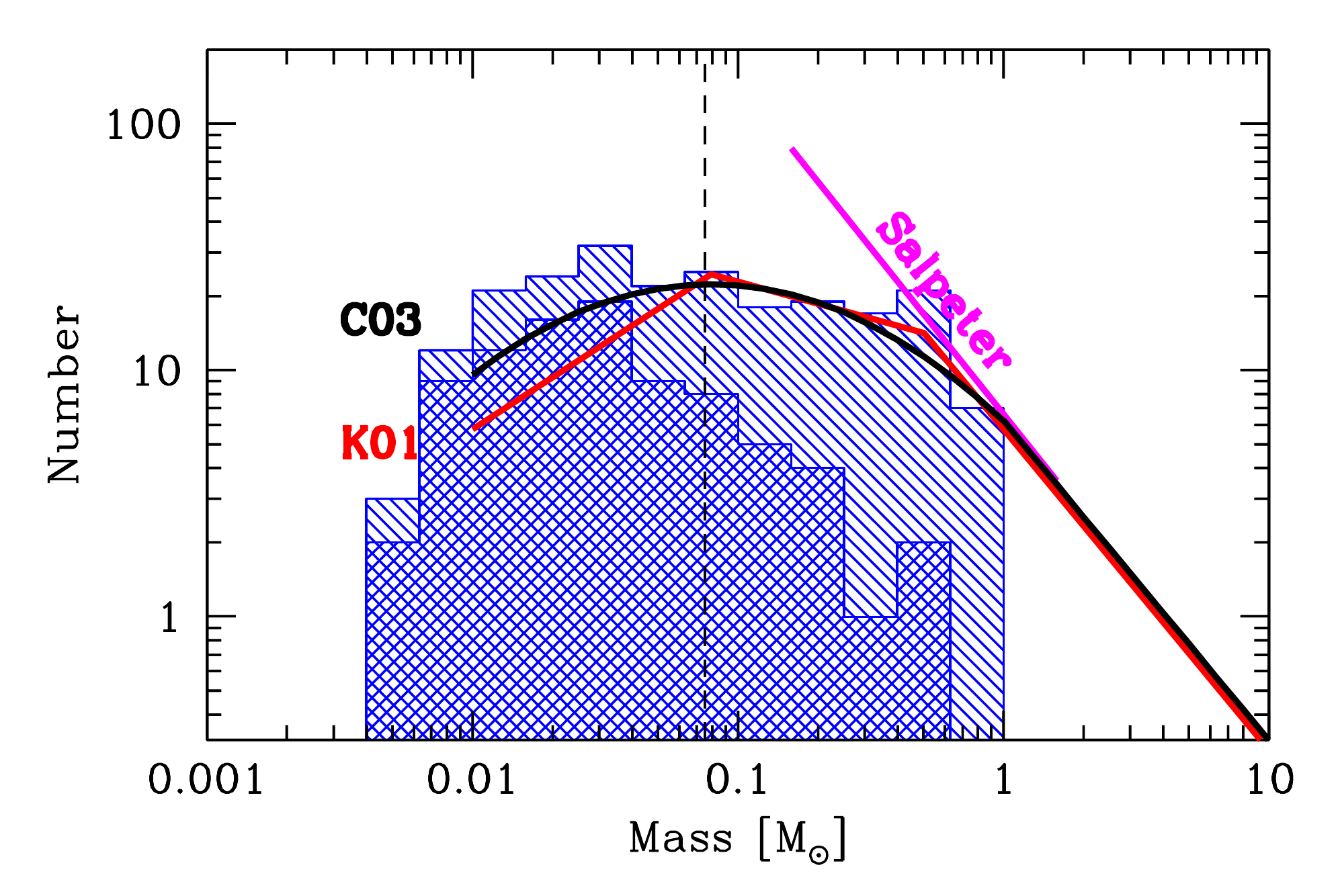}
    \includegraphics[width=8.4cm]{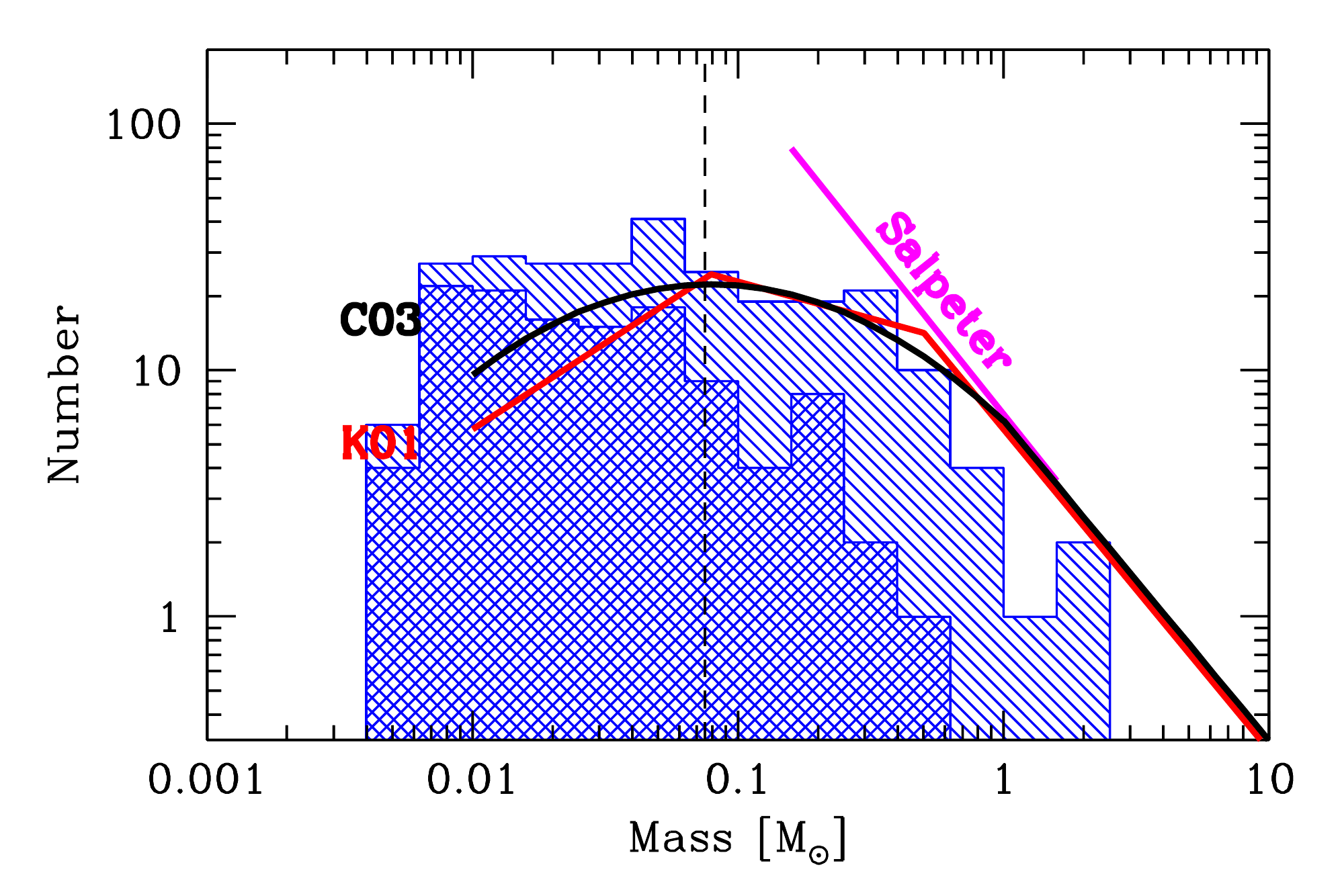}
\caption{Histograms giving the initial mass function of the 221 stars and brown dwarfs at $t=1.038t_{\rm ff}$ in the main calculation (left), and the 258 objects formed at the same time in the re-run calculation with smaller sink particle accretion radii and no gravitational softening between sink particles (right).  The re-run calculation appears to produce a few more very low-mass brown dwarfs (masses less than 10 Jupiter-masses), but even this difference is not statistically significant (see Figure \ref{imfcum}) so we conclude that changing the sink particle parameters does not adversely affect the resulting IMF.  Comparing the left panel with the IMF in Figure \ref{imf} at the end of the main calculation, we find that much of the over-production of brown dwarfs occurs late in the calculation (see also Figure \ref{massform}).}
\label{imfcomp}
\end{figure*}

The initial mass function produced by the end of the main calculation is shown in Figure \ref{imf} and is compared with the parameterisations of the observed IMF given by \citet{Chabrier2003}, \citet{Kroupa2001}, and \citet{Salpeter1955}.  The IMFs obtained from BBB2003 and B2005 were, within the statistical uncertainties, consistent with the observed IMF.  However, the IMF from the main calculation reported on here is much more accurately determined and is clearly not consistent with the observed IMF.  The computed IMF has a similar overall form to the observed IMF, with a reasonable Salpeter-type slope at the high-mass end, a flattening below a solar-mass, and an eventual turn over.  However, it significantly over produces brown dwarfs.  The calculation produces 459 stars and 795 brown dwarfs (masses $<0.075 $ M$_\odot$).  Even taking into account that 46 of the brown dwarfs are still accreting when the calculation is stopped and may eventually reach stellar masses, the ratio of brown dwarfs to stars is at least 3:2 whereas recent observations suggest that the IMF produces more stars than brown dwarfs \citep{Greissletal2007,Andersenetal2008}.  \citet{Andersenetal2008} find that the ratio of stars with masses $0.08-1.0$ M$_\odot$ to brown dwarfs with masses $0.03-0.08$ M$_\odot$ is $N(0.08-1.0)/N(0.03-0.08)\approx 5\pm 2$. For the main calculation, this ratio is $408/326 = 1.25$.  Although the IMF below 0.03 M$_\odot$ is not yet well constrained observationally the number of objects seems to be decreasing for lower masses.  Thus, it is unlikely that the true ratio of brown dwarfs to stars exceeds 1:3.  The main calculation, therefore, over produces brown dwarfs relative to stars by a factor of $\approx 4$ compared with the observed IMF.

\subsubsection{The dependence of the IMF on numerical approximations and missing physics}

There are several potential causes of brown dwarf over production that may be divided into two categories: numerical effects or neglected physical processes.  Arguably, the main numerical approximation made in the calculations is that of sink particles.  High-density gas is replaced by a sink particle whenever the maximum density exceeds $10^{-10}$ g~cm$^{-3}$ and the gas within a radius of 5 AU is accreted onto the sink particle producing a gravitating point mass containing a few Jupiter masses of material.  These sink particles then interact with each other ballistically, which, for example, might plausibly artificially enhance ejections and the production of low-mass objects.

In order to investigate the effect of the sink particle approximation on the results, we re-ran part of the main calculation with smaller sink particles (accretion radii of 0.5 AU) and without gravitational softening between sink particles (they were allowed to merge if them came within 4 $R_\odot$ of each other).  This calculation was only followed to 1.038 $t_{\rm ff}$ due to its much more time consuming nature.  The small accretion radius calculation produced 258 stars and brown dwarfs in the same time period that the main calculation produced 221 objects.  Because the calculations are chaotic identical results should not be expected.  The main question to answer is whether or not the results are statistically different.  

\begin{figure}
\centering
    \includegraphics[width=8.4cm]{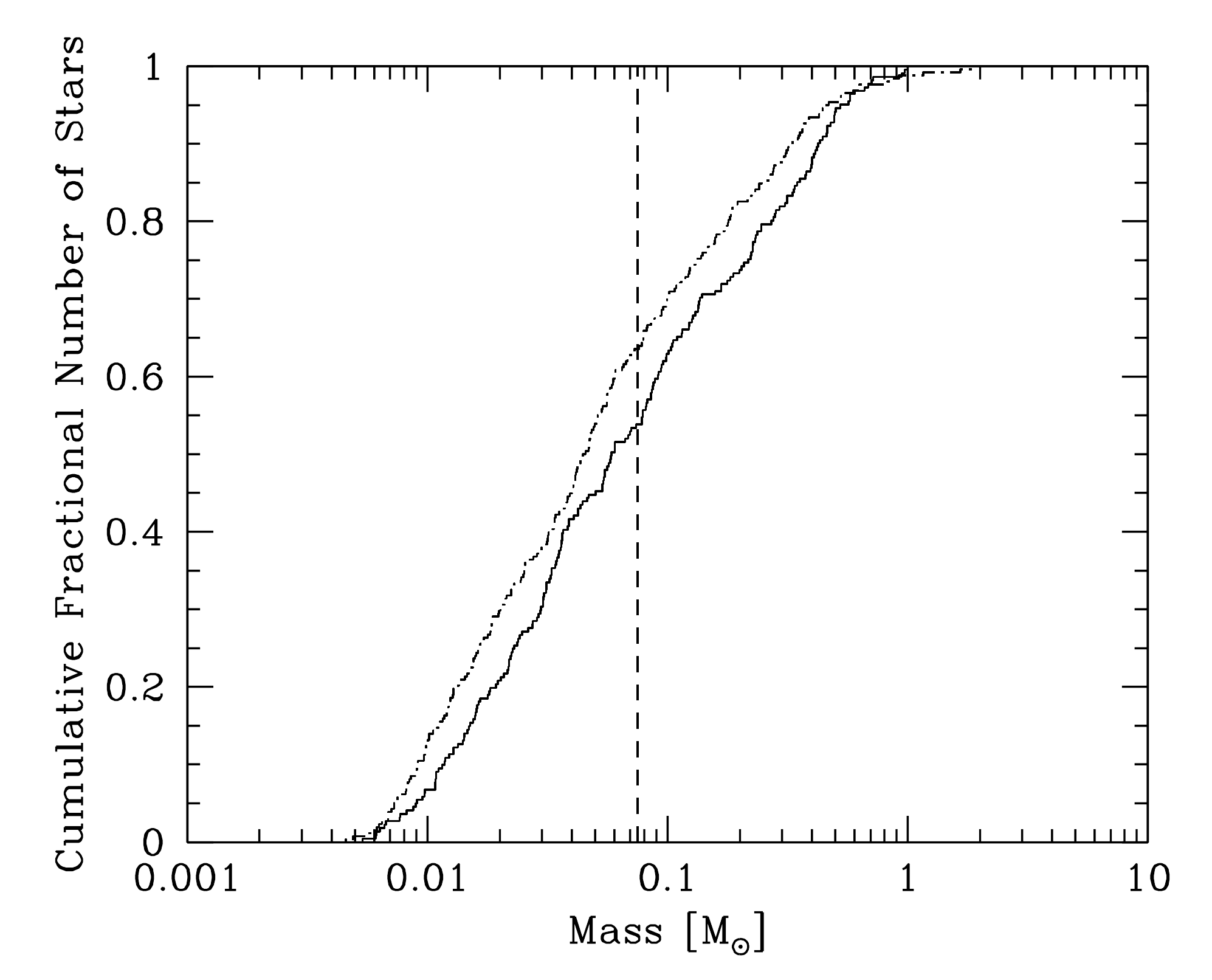}
\caption{The cumulative IMFs from the main calculation (solid line) and the re-run calculation with small accretion radii (dot-dashed line) both at 1.038 $t_{\rm ff}$ (see Figure \ref{imfcomp} for differential graphs of the IMFs).  The calculation with the smaller accretion radii seems to produce more very low-mass brown dwarfs with masses less than 10 Jupiter masses.  However, even with this apparent difference, a Kolmogorov-Smirnov (K-S) test on the two distributions gives a 13\% probability that the two IMFs were drawn from the same underlying distribution (i.e. they are statistically indistinguishable). Thus, the results do not seem to be adversely affected by the sink particle approximation.}
\label{imfcum}
\end{figure}

In Figures \ref{imfcomp} and \ref{imfcum} we compare the IMFs produced by the main calculation and the smaller sink particle calculation at the same time.  The smaller sink particle calculation produces twice as many objects with masses less than 10 Jupiter masses than the main calculation, but overall the two IMFs are very similar.  A K-S test run on the two distributions shows that they have a 13\% probability of being drawn from the same underlying IMF (i.e. they are statistically indistinguishable).  Removing objects with less than 10 Jupiter-masses from the K-S test results in a 38\% probability of the two distributions being drawn from the same underlying IMF.  We conclude that variations in the sink particle accretion radii and gravitational softening may have an effect on the production of extremely low-mass objects.  However, changes to the sink particle parameters do not significantly alter the overall results and, thus, the use of sink particles is probably not responsible for the significant over production of brown dwarfs.

\begin{figure}
\centering
    \includegraphics[width=8.4cm]{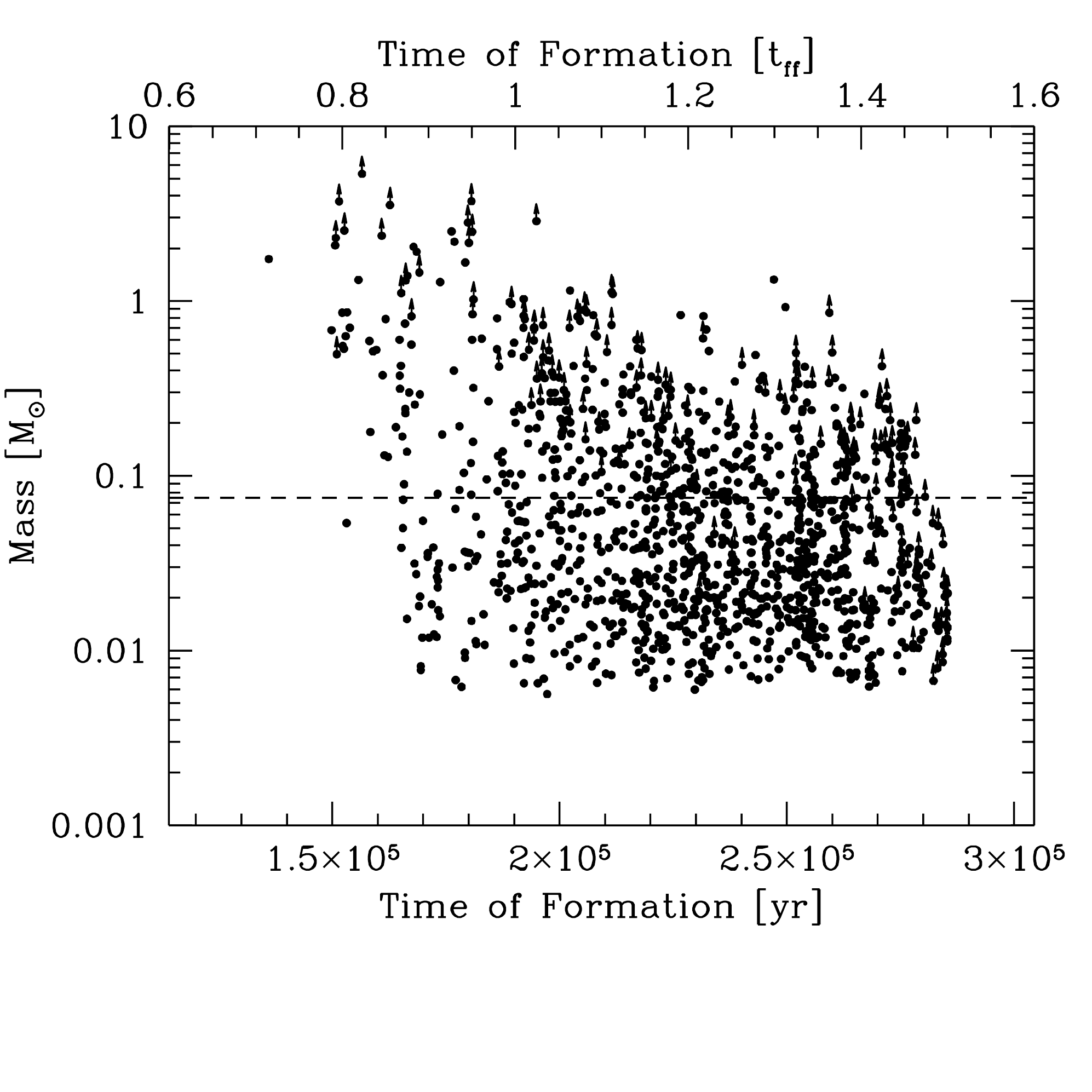}\vspace{-1cm}
\caption{Time of formation and mass of each star and brown dwarf at the end of the main calculation.  It is clear that the objects that are the most massive at the end of the calculation are actually some of the first to collapse and form sink particles.  Furthermore, the longer the calculation proceeds, the higher the ratio brown dwarfs to stars becomes.  Objects that are still accreting significantly at the end of the calculation are represented with vertical arrows.  The horizontal dashed line marks the star/brown dwarf boundary.  Time is measured from the beginning of the calculation in terms of the free-fall time of the initial cloud (top) or years (bottom).}
\label{massform}
\end{figure}

It seems most likely that the over production of brown dwarfs is related to physical processes that are not included in the calculations.  \citet{WhiBat2006} showed that replacing the barotropic equation of state by radiative transfer can lead to temperatures up to an order of magnitude higher near young low-mass protostars and, thus, potentially strongly inhibits fragmentation.  \citet{Krumholz2006} made a similar argument analytically.  Furthermore, in purely hydrodynamical/sink particles star cluster formation calculations, many of the brown dwarfs formed originate via disk fragmentation (e.g. \citealt{BatBonBro2002a} found that 3/4 of the brown dwarfs originated from disc fragmentation).  \citet{Rafikov2005}, \citet{MatLev2005}, \citet{KraMat2006}, and \citet{WhiSta2006} have all pointed out that accurate treatments of radiative transfer are likely to significantly decrease disc fragmentation.  Along with the likely effect of radiative feedback on fragmentation, we note that as the main calculation progresses the ratio of low-mass to high-mass objects increases.  This can be seen in Figure \ref{massform} which plots the final mass of an object versus its time of formation, as well as by comparing Figure \ref{imf} with the left panel of Figure \ref{imfcomp} which show the IMF from the main calculations at $t=1.50t_{\rm ff}$ and $t=1.038t_{\rm ff}$, respectively.  Radiative feedback is likely to heat the entire central cluster region later in the calculation, potentially curtailing off the formation of many of the late low-mass objects.

Another possibility is the effect of magnetic fields.  Recently, \citet{PriBat2007} showed that stronger magnetic fields generally inhibit disc formation and binary formation \citep[see also][]{HenFro2008,HenTey2008}.  \citet{PriBat2008} ran star cluster formation simulations similar to BBB2003, but with magnetic fields.  They found that the extra pressure support provided by magnetic fields generally decreased the rate of star formation and the importance of dynamical interactions between objects.  Stronger magnetic fields resulted in a decrease in the ratio of brown dwarfs to stars (though the total numbers of objects formed in the calculations were small, ranging from 15 to 69).

In summary, we have shown for the first time that purely hydrodynamical simulations of star cluster formation over produce brown dwarfs.  This result is statistically robust.  This disagreement with observations is most likely due to the neglect of the physical processes of radiative feedback and/or magnetic fields.

\subsubsection{The origin of the initial mass function}
\label{origin}

\begin{figure}
\centering\vspace{-0.2cm}
    \includegraphics[width=8.4cm]{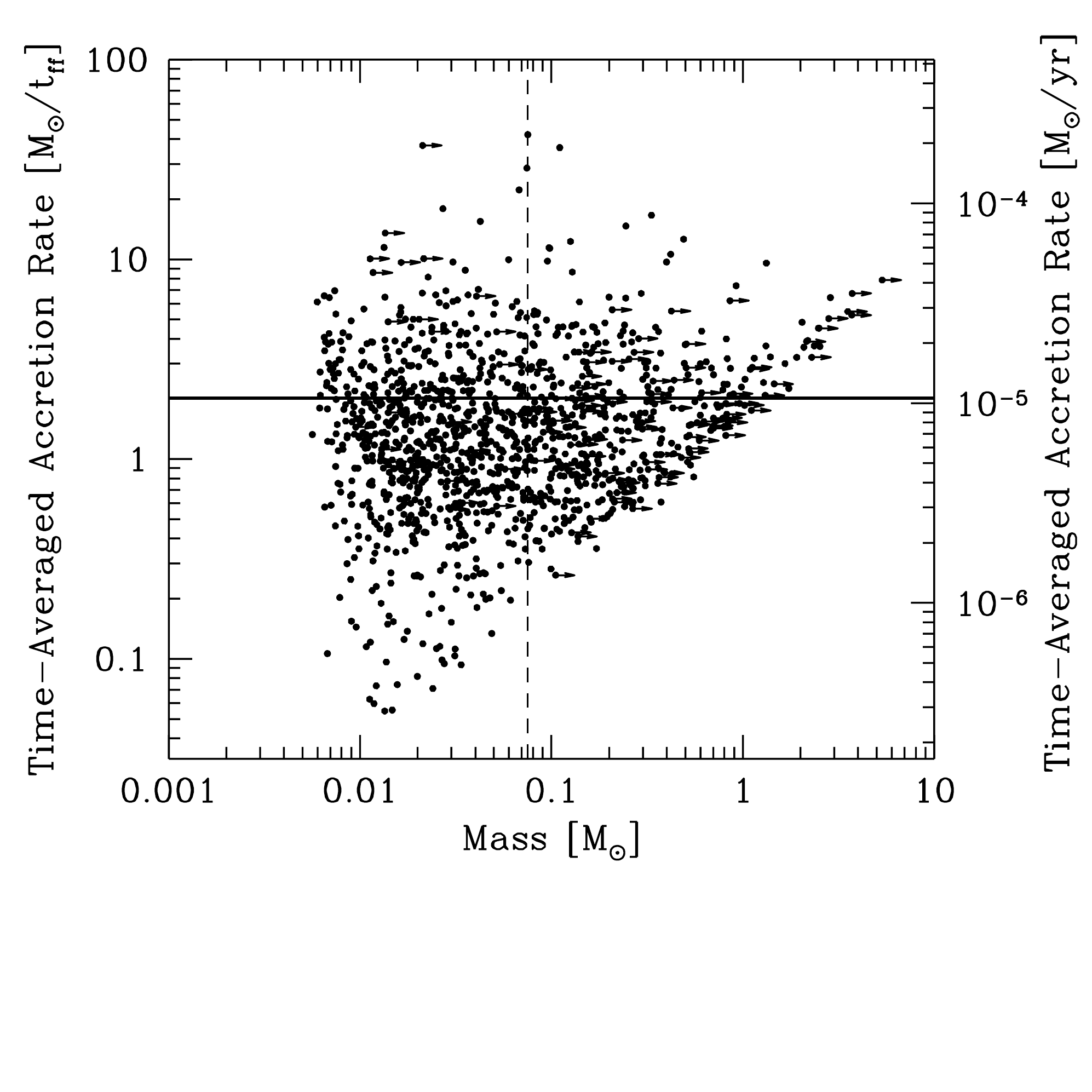}\vspace{-1.5cm}
\caption{The time-averaged accretion rates of the objects formed in the main calculation versus their final masses.  The accretion rates are calculated as the final mass of an object divided by the time between its formation and the termination of its accretion or the end of the calculation.  Objects that are still accreting significantly at the end of the calculation are represented with horizontal arrows.  There is no dependence of mean accretion rate on final mass for objects with less than $\sim 0.5$ M$_\odot$ (and a large dispersion).  However, there is a low-accretion rate region of exclusion for the most massive objects since only objects with mean accretion rates greater than their mass divided by their age can reach these high masses during the calculation.  The horizontal solid line gives the mean of the accretion rates: $1.02\times 10^{-5}$ M$_\odot$~yr$^{-1}$. The accretion rates are given in M$_\odot/t_{\rm ff}$ on the left-hand axes and M$_\odot$~yr$^{-1}$ on the right-hand axes. The vertical dashed line marks the star/brown dwarf boundary.
}
\label{accrate}
\end{figure}

\begin{figure}
\centering\vspace{-0.2cm}
    \includegraphics[width=8.4cm]{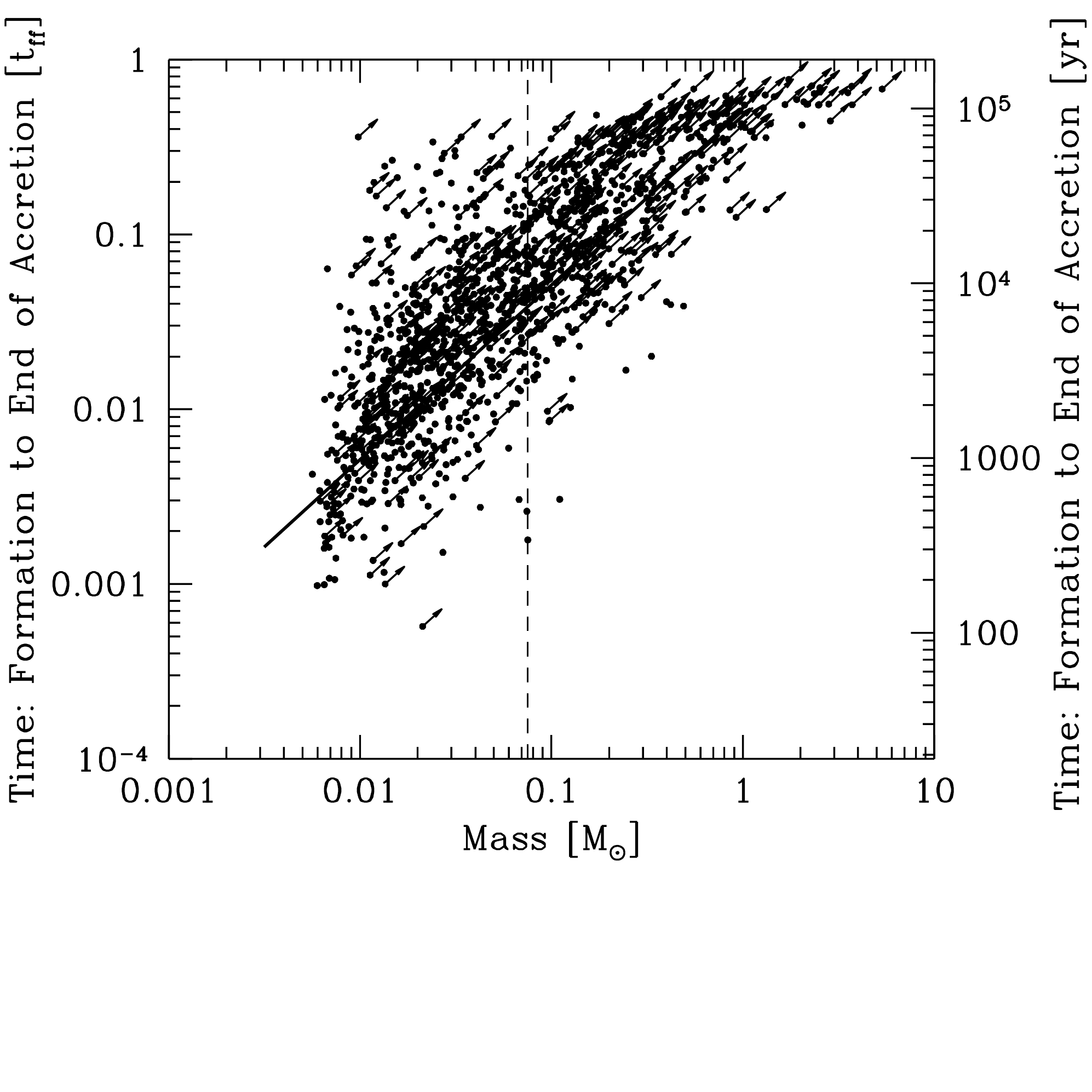}\vspace{-1.5cm}
\caption{The time between the formation of each object and the termination of its accretion or the end of the main calculation versus its final mass.  Objects that are still accreting significantly at the end of the calculation are represented with arrows.  As in BBB2003, BB2005, and B2005, there is a clear linear correlation between the time an object spends accreting and its final mass.  The solid line gives the curve that the objects would lie on if each object accreted at the mean of the time-averaged accretion rates. The accretion times are given in units of the $t_{\rm ff}$ on the left-hand axes and years on the right-hand axes. The vertical dashed line marks the star/brown dwarf boundary.
}
\label{acctime}
\end{figure}

BB2005 analysed the earlier calculation presented by BBB2003 and another calculation beginning with a denser cloud to determine the origin of the IMF in those calculations (see also B2005).  They found that the IMF resulted from competition between accretion and ejection.  There was no significant dependence of the mean accretion rate of an object on its final mass.  Rather, there was a roughly linear correlation between an object's final mass and the time between its formation and the termination of its accretion.  Furthermore, the accretion on to an object was usually terminated by a dynamical interaction between the object and another system, ejecting the object.  Thus, objects formed with very low masses (a few Jupiter masses) and accreted to higher masses until their accretion was terminated, usually, by a dynamical encounter.  This combination of competitive accretion and stochastic ejections produced the mass function.

\begin{figure}
\centering\vspace{-0.2cm}
    \includegraphics[width=8.4cm]{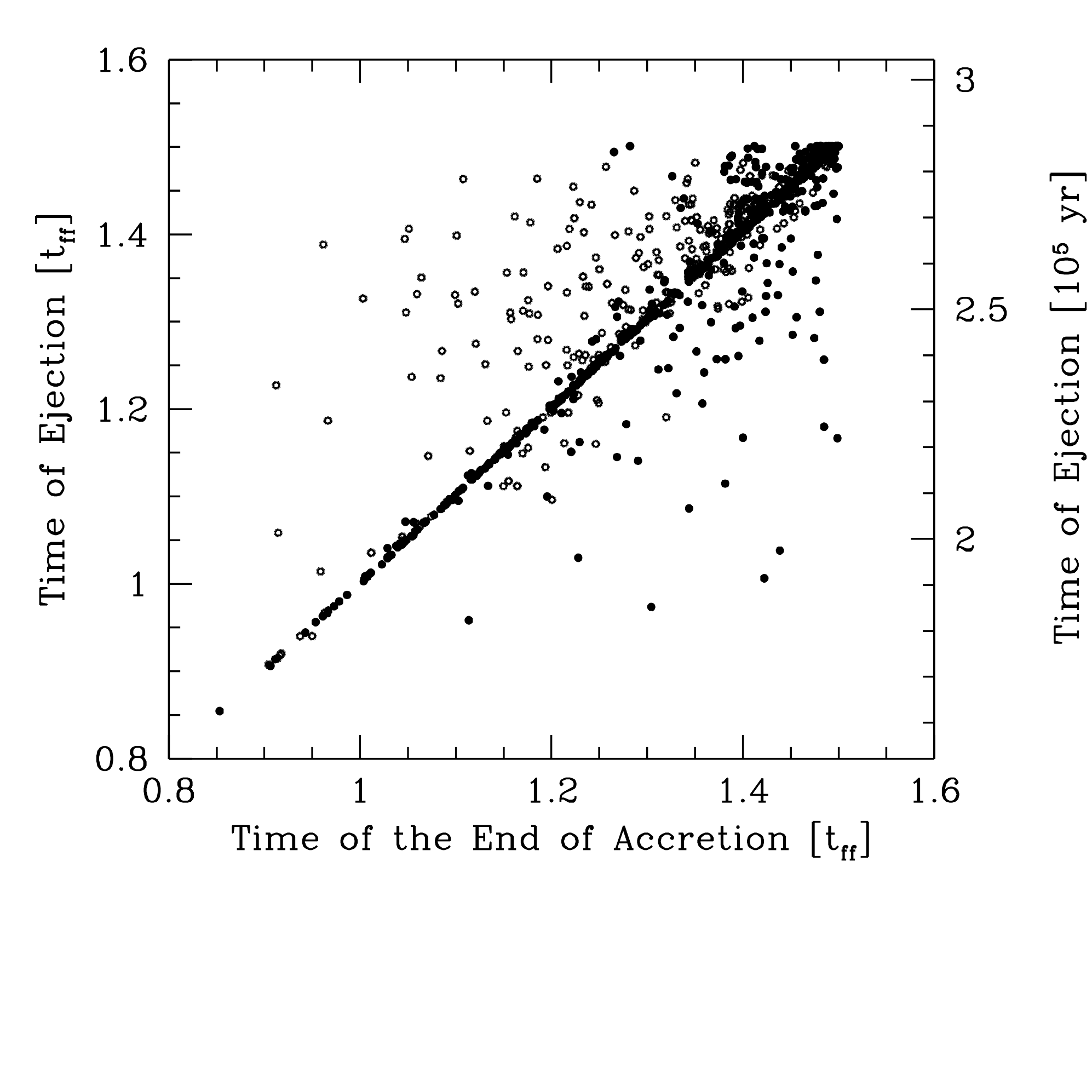}\vspace{-1.5cm}
\caption{For each single object that has stopped accreting by the end of the main calculation, we plot the time of the ejection of the object from a multiple system versus the time at which its accretion is  terminated.  As in the smaller calculations of BBB2003, BB2005, and B2005, these times are correlated, showing that the termination of accretion on to an object is usually associated with dynamical ejection of the object. Open circles give those objects where multiple `ejections' are detected by the ejection detection algorithm and, hence, the ejection time is ambiguous (see the main text).  Binaries have been excluded in the plot because it is difficult to determine when a binary has been ejected.
}
\label{ejtime_vs_acctime}
\end{figure}

In Figures \ref{accrate}, \ref{acctime}, and \ref{ejtime_vs_acctime}, we plot similar figures to those found in BB2005 and B2005.  These figures display the same trends as found by BB2005, but with a much greater statistical significance.  Figure \ref{accrate} gives the time-averaged accretion rates of all the objects formed in the main calculation versus the object's final mass.  The time-averaged accretion rate is the object's final mass divided by the time between its formation (i.e. the insertion of a sink particle) and the end of its accretion (defined as the last time its accretion rate drops below 10$^{-7}$ M$_\odot$/yr) or the end of the calculation.  As in BB2005, there is no dependence of the time-averaged accretion rate on an object's final mass, except that objects need to accrete at a rate at least as quickly as their final mass divided by their age (i.e., the lower right potion of Figure \ref{accrate} cannot have any objects lying in it).  This means that the most massive stars have higher time-averaged accretion rates than the bulk of the stars and VLM objects.  On the other hand, if the calculation were continued longer, objects that are accreting with lower time-averaged accretion rates could also reach high masses.  

The mean of the accretion rates is $1.02 \times 10^{-5}$ M$_\odot$/yr, which is within a factor of two of the mean accretion rates of the three calculations analysed by BB2005 and B2005.  Thus, the mean accretion rate does not depend significantly on cloud density (BB2005), the equation of state of high-density gas (B2005), or on the total mass of the gas cloud (this work).  The dispersion in the accretion rates is about 0.4 dex, also similar to the previous simulations.  Rather, the primary  determinant of the final mass of a star or brown dwarf is the period over which it accretes.  Figure \ref{acctime} very clearly shows the linear relation (with some dispersion) between the period of time over which an object accretes and it's final mass.

Finally, in Figure \ref{ejtime_vs_acctime}, for each object that has stopped accreting by the end of the main calculation (excluding the components of binaries), we plot the time at which the object undergoes an ejection versus the time that its accretion is terminated.  There is a very strong correlation between the two showing that accretion is usually terminated by a dynamical encounter with other objects, and confirming the results of BB2005 and B2005. We define the time of ejection of an object as the last time the magnitude of its acceleration drops below 2000 km~s$^{-1}$~Myr$^{-1}$ (or the end of the calculation). The acceleration criterion is based on the fact that once an object is ejected from a stellar multiple system, sub-cluster, or cluster through a dynamical encounter, its acceleration will drop to a low value. The specific value of the acceleration was chosen by comparing animations and graphs of acceleration versus time for individual objects. We exclude binaries because they have large accelerations throughout the calculation which frequently results in false detections of ejections.  Also, in Figure \ref{ejtime_vs_acctime}, we use two different symbols (filled circles and open circles).  For the former we are confident of the ejection time.  However, for those objects denoted by the open circles, we find that at least two `ejections' more than 2000 years apart have occurred.  These are usually objects that have had a close dynamical encounter with a multiple system that has put them into long-period orbits rather than ejecting them.  In these cases, we chose the `ejection' time closest to the accretion termination time but we use an open symbol to denote our uncertainty in whether or not we have identified the best time for the dynamical encounter.

In terms of raw results, we find that, excluding binaries, for 635 objects out of 899 (71\%) the accretion termination time and the ejection time are within 2000 years of each other.  If we also exclude those objects for which we are uncertain in our identifications of the ejection times as described above, we find 483 objects out of 592 (82\%) are consistent with ejection terminating their accretion.  These are probably lower limits in the sense that it is difficult to determine in an automated way the time at which an ejection occurs and an erroneous value is much more likely to differ from the accretion termination time by more than 2000 years than coincide with it.  In any case, it is clear that for the majority of objects their accretion is terminated by dynamical encounters with other stellar systems.

\subsection{Stellar cluster properties}

\begin{figure}
\centering
    \includegraphics[width=8.4cm]{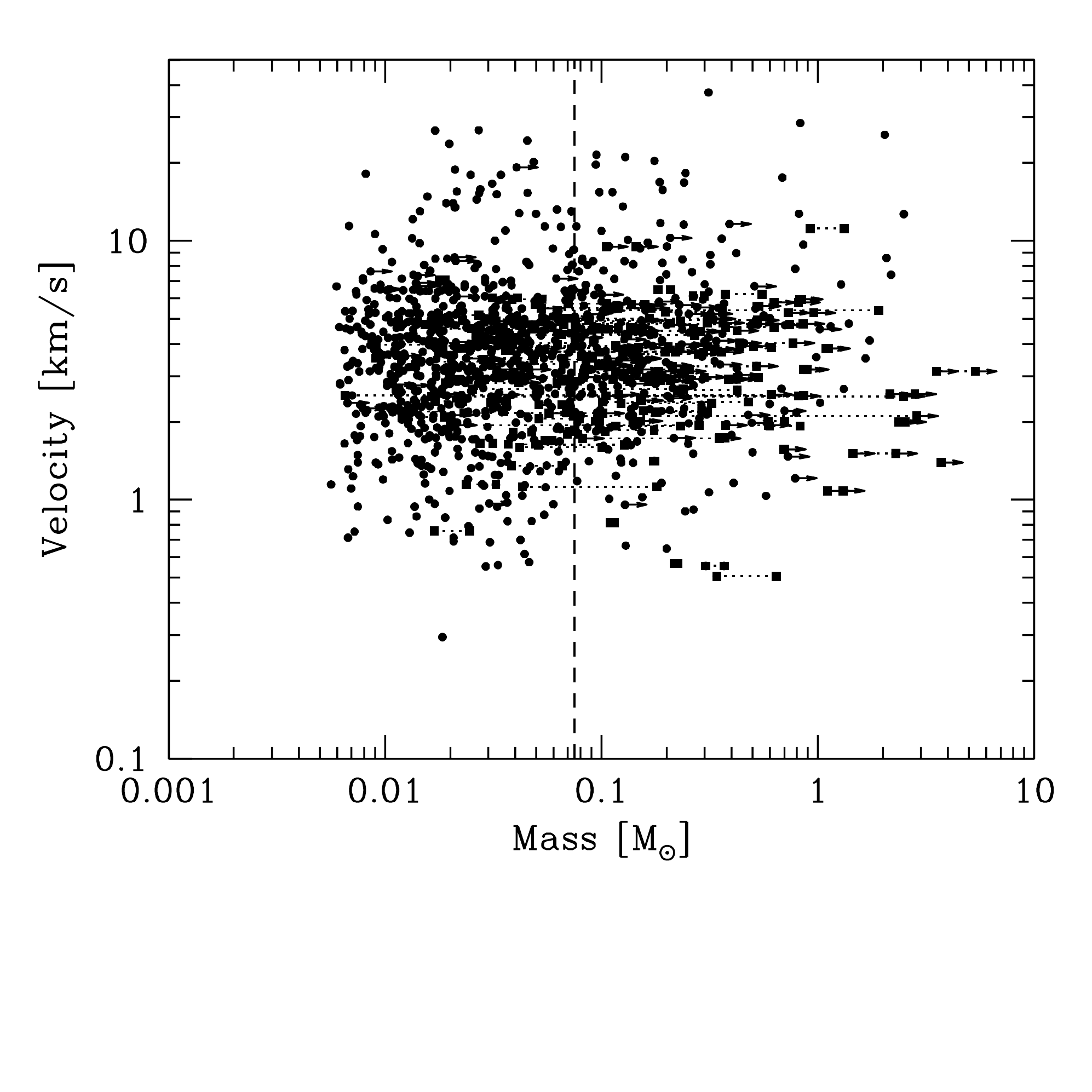}\vspace{-1.5cm}
\caption{The magnitudes of the velocities of each star and brown dwarf relative to the centre-of-mass velocity of the stellar system at the end of the main calculation.  For binaries, the centre-of-mass velocity of the binary is given, and the two stars are connected by dotted lines and plotted as squares rather than circles.  Objects still accreting at the end of the calculation are denoted by horizontal arrows.  The root mean square velocity dispersion for the association (counting each binary once) is 5.6 km~s$^{-1}$ (3-D) or 3.2 km~s$^{-1}$ (1-D).  There is a weak dependence of the velocity dispersion on mass with VLM objects having a slighly lower velocity dispersion than stars (see the main text).  Binaries are found to have a lower velocity dispersion than single objects of only 3.8 km~s$^{-1}$ (3-D).  The vertical dashed line marks the star/brown dwarf boundary.
}
\label{veldisp}
\end{figure}

\begin{table*}
\begin{tabular}{lcccccc}\hline
Quantity /  Distance range & $<1000$ AU & $1000-3000$ AU  & $3000-10^4$ AU  &  $1-3\times 10^4$ AU & $3-10\times 10^4$ AU & $>1\times 10^5$  AU \\ \hline

Median mass [M$_\odot$] & 0.18 & 0.024 & 0.035 & 0.056 & 0.054 & 0.045 \\
Upper quartile mass [M$_\odot$]  & 0.30 & 0.091 & 0.098 & 0.15 & 0.18 & 0.095 \\
Maximum mass [M$_\odot$] & 5.3 & 2.9 & 3.7 & 2.5 & 2.1 & 2.0 \\
Velocity dispersion [km/s] &    6.1 & 4.0 & 4.2 & 4.3 & 8.2 & 13.8 \\
Number objects &    8 & 56 & 569 & 408 & 172 & 41 \\
Number binaries &  2 & 8 &  68 & 55 & 13 & 0 \\
Binary fraction &   0.33 & 0.167 & 0.136 & 0.156 & 0.082 & 0.0 \\\hline
\end{tabular}
\caption{\label{tablecluster} Radial properties of the stellar cluster at the end of the main calculation.  The cluster is very compact, with a half-mass radius of 10,900 AU.  The radii containing 80\% and 90\% of the mass are 29,800 and 54,200 AU, respectively.  There is no evidence for radial mass segregation in terms of the median mass, the upper quartile mass, and the maximum mass, except in the inner 1000 AU.  In terms of the binary fraction and the stellar velocity dispersion, again there very centre of the cluster has a higher velocity dispersion and a higher binary frequency than the bulk of the cluster.  However, unlike the mass function, the velocity dispersion and binary fraction also differ in the outer regions of the cluster (the outer 20\% of the mass, beyond 3 half-mass radii).  The outer regions have a higher velocity dispersion and a lower binary fraction than the bulk of the cluster.}
\end{table*}

\begin{figure}
\centering
    \includegraphics[width=8.4cm]{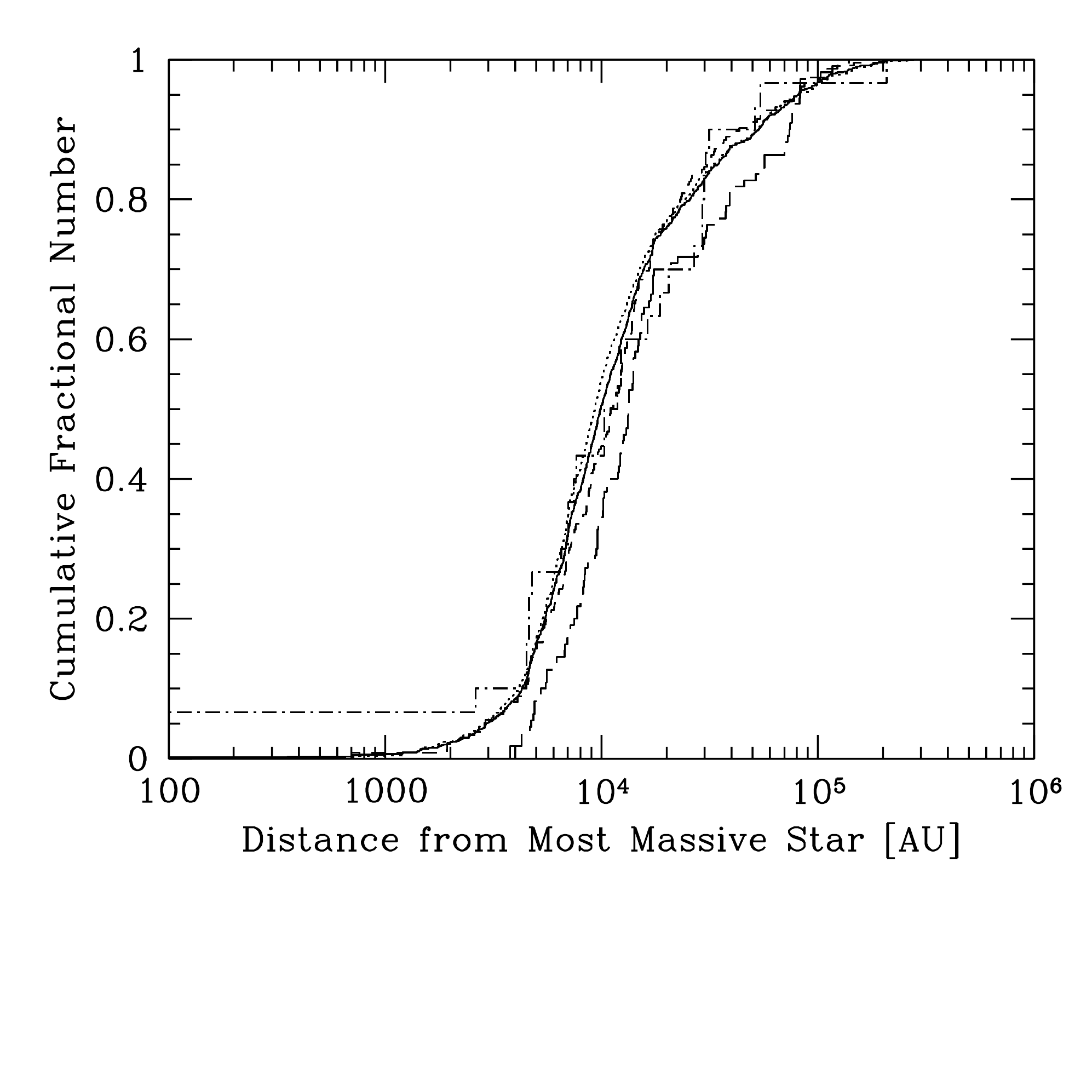}\vspace{-1.5cm}
\caption{The cumulative fractions of stars as a function of distance from the most massive star at the end of the main calculation.  The solid line gives the result for all stars, while the dotted, short-dashed, long-dashed, and dot-dashed give the cumulative distributions for the stellar mass ranges $M<0.1$, $0.1\leq M < 0.3$,  $0.3\leq M < 1.0$,  and $M \geq 1.0$ M$_\odot$, respectively.  There is no significant mass segregation observed.
}
\label{radialimf}
\end{figure}

\begin{figure}
\centering
    \includegraphics[width=8.4cm]{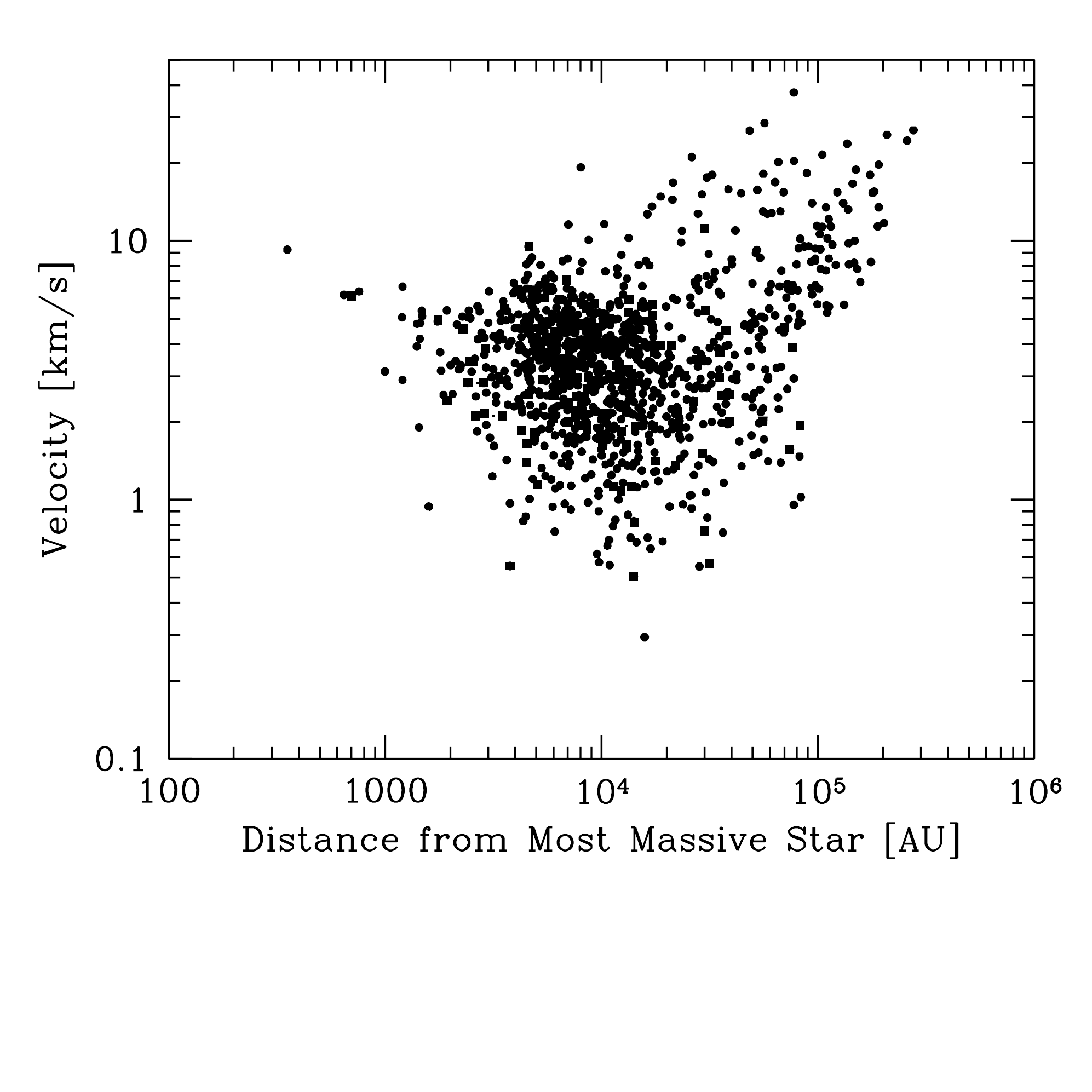}\vspace{-1.5cm}
\caption{For each star and brown dwarf, we plot the magnitude of its velocity relative to the centre-of-mass velocity of the stellar system versus its distance from the most massive star in the cluster at the end of the main calculation.  For binaries, the centre-of-mass velocity of the binary is given and the binary is plotted as a square rather than a circle.  The velocity dispersion clearly on depends radius, with the outer regions having a significantly larger velocity dispersion.  These outer objects have been ejected (see also Table \ref{tablecluster}).}
\label{veldispcentre}
\end{figure}

At the end of the main calculation, the bulk of the stars and brown dwarfs are contained within a single compact stellar cluster surrounded by a low density halo of objects (lower right panel of Figure \ref{mainevolution}).  The stellar cluster has a half-mass radius of only 10,900 AU (0.053 pc), ignoring the gas.  The radii containing 80\% and 90\% of the mass are 29,800 AU (0.14 pc) and 54,200 AU (0.26 pc), respectively.

In Figure \ref{veldisp}, we plot the magnitude of the velocity of every star or brown dwarf relative to the centre of mass of the stellar system at the end of the main calculation.  For binaries, we plot the two components with the centre of mass velocity of the binary using filled squares connected by a dotted line.  The overall root mean square (rms) velocity dispersion (counting each binary only once) is 5.6 km~s$^{-1}$ (3-D) or 3.2 km~s$^{-1}$ (1-D).  BBB2003, BB2005, and B2005, found no significant dependence of the velocity dispersion on mass.  Here, with a much larger sample of objects we find that stars tend to have a slightly higher dispersion than VLM objects.  The rms velocity dispersion of VLM systems is 5.4 km~s$^{-1}$ (3-D) while for the stars (masses $\geq 0.1$ M$_\odot$) the rms velocity dispersion is 6.9 km~s$^{-1}$ (3-D).  Binaries have a velocity dispersion of only 3.8 km~s$^{-1}$ (3-D), significantly lower than single objects.

Since this is the first hydrodynamical calculation to form a massive stellar cluster while simultaneously resolving brown dwarfs and binaries it is of interest to examine how the stellar properties vary with radius.  We define the cluster centre to be the location of the most massive star (5.3 M$_\odot$).  In Table \ref{tablecluster}, we present statistics on how the stellar masses, velocity dispersion, and binary fraction vary with radius from the cluster centre.  Note that for this table, we have defined the binary fraction as the number of binaries divided by the number of systems (single objects and binaries).  We do not make any attempt to identify triple or higher order systems.  Each binary is counted once and its centre-of-mass velocity is used when calculating the stellar velocity dispersions.

We find that within the radius containing 80\% of the mass (excluding the gas), there is little evidence of a radial variation in the stellar mass function (see Figure \ref{radialimf}), the velocity dispersion, or the binary fraction.  The exception may be the very centre of the cluster (within 1000 AU of the most massive star) where the median stellar mass, the upper quartile mass, the velocity dispersion, and the binary fraction are all higher than in the bulk of the cluster.  However, there are only 8 objects in this region so the statistical uncertainties are great.

In the periphery of the cluster containing 20\% of the stellar mass (perhaps better described as the halo) we do find statistically significant differences.  The mass function is still indistinguishable from the mass function found in the bulk of the cluster (the median mass, the upper quartile mass, and the maximum mass are all similar to those values found in the bulk of the cluster).  However, the velocity dispersion increases monotonically as the distance from the cluster centre increases (see Table \ref{tablecluster} and Figure \ref{veldispcentre}).  This is because only objects that have been ejected quickly can have made it out to these distances by the end of the calculation.  Also, the binary fraction decreases outside of the 80\%-mass radius.  It drops by a factor of two between the 10,000-30,000 AU ($1-3$ half-mass radii) radial bin and the 30,000-100,000 AU ($3-9$ half-mass radii) bin and there are no binaries (out of 41 objects) more than 100,000 AU ($>9$ half-mass radii) from the cluster centre.  Presumably, even though some binaries are ejected, they are less likely to be ejected than single objects and the likelihood of them surviving the ejection process decreases with increasing ejection velocity (since a closer dynamical encounter is required to achieve a higher ejection velocity).

Observationally, the best cluster to compare our results to is the Orion Nebula Cluster (ONC).  \citet{HilHar1998} examined its structure and dynamics.  They estimated the stellar mass to be $\approx 2\times 10^3$ M$_\odot$ and the half-mass radius to be $\approx 0.8$ pc, so the main simulation discussed here produces a cluster that is significantly less massive and more compact than the ONC.  Although the ONC is larger and more massive, it is probably at a similar stage of evolution as the main calculation when it is stopped in the sense that it does not contain significant substructure \citep*{BatClaMcC1998, ScaCla2002} and, if it was assembled from the merger of sub-clusters, the ONC's period of violent relaxation has ended.  By contrast, the $\rho$ Ophiuchi cloud contains a similar mass of stars and gas to the calculations presented here \citep{Bontempsetal2001}, but it is composed of many sub-clusters rather than a single large cluster.

\citeauthor{HilHar1998} investigated mass segregation in the ONC and found that within the half-mass radius there was evidence for general mass segregation with stars in various mass bins becoming more centrally concentrated with increasing stellar mass.  At larger radii, there was little evidence for mass segregation.  At the end of the main calculation, we find no significant mass segregation.  This is ironic since one of the main arguments usually advanced in favour of the competitive accretion model for star formation is that it naturally produces mass-segregated clusters \citep[e.g.][]{Bonnelletal1997, Bonnelletal2001a}.  The difference here is most probably that the stellar cluster existing at the end of the main calculation has just formed from the merger of 5 sub-clusters and even if these sub-clusters were mass segregated before their mergers it is going to take some time for the entire cluster to settle down again.  This does illustrate that competitive accretion does not necessarily produce clusters that are mass segregated throughout their entire formation process.  

\citet{Kohleretal2006} investigated binarity in the ONC.  They found that there was no significant dependence of the binary fraction on the distance from the cluster centre by comparing samples within $\approx 0.3$ pc (approximately 40\% of the half-mass radius) of the centre with observations between $0.7-1.8$ pc from the centre (approximately $1-2$ half-mass radii).  They stated that this was in contrast to the theory that the low binary frequency in the ONC compared to low-density star forming regions was due to dynamical disruption.  However, their result is consistent with our hydrodynamical simulation in that we also find no significant variation of binary fraction within 3 half-mass radii and binary disruption certainly occurs in the simulation.  Only outside of 3 half-mass radii does there appear to be a slow decline in binarity.  Needless to say, it would be interesting to try and detect a lower binary fraction or a higher velocity dispersion at distances more than 3 half-mass radii from the centre of the ONC to see whether the ONC displays variations like those apparent in the simulation.  However, this would presumably be very difficult given the low stellar density and the problems of determining membership so far from the cluster centre.

\subsection{Stellar encounters and disc sizes}

\begin{figure}
\centering
    \includegraphics[width=8.4cm]{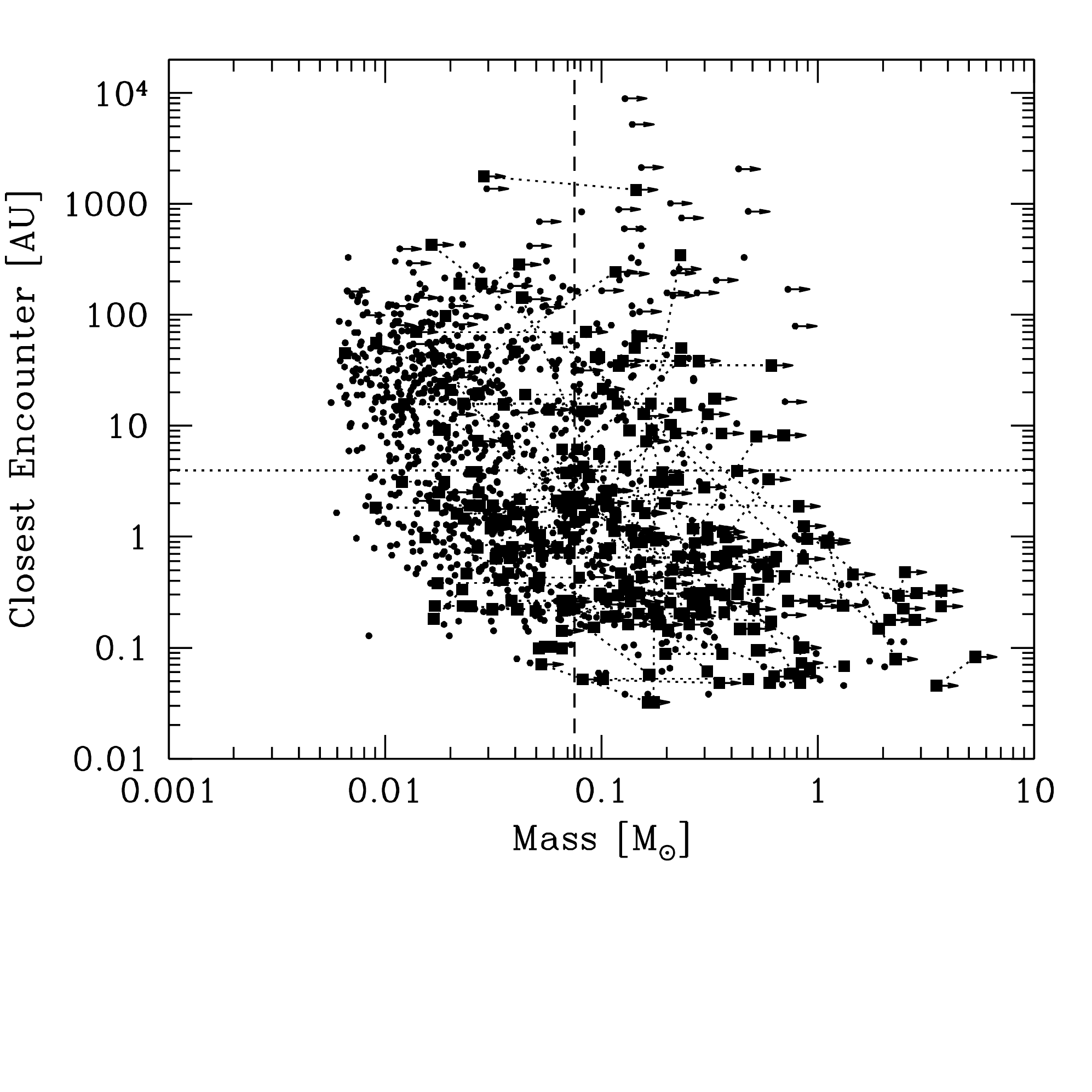}\vspace{-1.5cm}
\caption{The closest encounter distances of each star or brown dwarf during the main calculation versus the final mass of each object.  Objects that are still accreting significantly at the end of the calculation are denoted with arrows indicating that they are still evolving and that their masses are lower limits.  Binaries  are plotted with the two components connected by dotted lines and squares are used as opposed to circles.  Encounter distances less than 4 AU are upper limits since the point mass potential is softened within this radius.  The vertical dashed line marks the star/brown dwarf boundary.  The brown dwarfs in the top left corner of the figure that are still accreting formed shortly before the calculation was stopped are thus still evolving rapidly.  They may not end up as brown dwarfs.
}
\label{truncate}
\end{figure}

\citet{ReiCla2001} proposed that brown dwarfs may be formed from dynamical ejections of low-mass objects from accreting unstable multiple systems, thus terminating their accretion and fixing them at low masses.  \citet{BatBonBro2002a}, BBB2003, BB2005, and B2005 performed hydrodynamical simulations in which it was found that dynamical interactions were crucial in terminating accretion and setting an object's mass, but that this applied to stars as well as brown dwarfs (see also Section \ref{origin} of this paper).  Brown dwarfs were simply ejected soon after they had formed while those objects ending up as stars suffered ejections only after a longer period of accretion.

\citet{ReiCla2001} also speculated that if brown dwarfs formed via ejection, they might have smaller, lower-mass discs than stars.  BBB2003, BB2005, and B2005 found that discs around stars and brown dwarfs were frequently truncated by dynamical encounters.  However, some large discs were found to exist around both stars and brown dwarfs, while other stars and brown dwarfs had discs truncated to below the resolution limit of $\approx 10$ AU in their calculations.

In the calculations presented here, discs are resolved with radii down to $\approx 10$ AU in the main calculation and down to a few AU in the re-run calculation.  However, with SPH, the resolution length depends on density.  Thus, for example, more massive discs are better resolved than low-mass discs.  Furthermore, low-mass discs evolve much more quickly than high-mass discs due to the artificial viscosity present in the simulations (since the magnitude of the viscosity also depends on density).  Because of these numerical effects it is difficult to determine robustly the statistical properties of discs (e.g.\ their size and mass distributions).

By contrast, it is relatively simple to determine the closest dynamical encounter every star or brown dwarf has had during the calculation.  In Figure \ref{truncate}, we plot the distance of the closest encounter that every star and brown dwarf has had by the end of the main calculation.  As in the earlier papers, there is a wide range of closest encounter distances, but stars have generally had closer encounters than brown dwarfs.  However, this is somewhat misleading for several reasons.  First, as will be seen in the next section, multiplicity is a strong function of primary mass.  In Figure \ref{truncate} it clear that (close) binaries are responsible for many of the `closest encounters'.  Second, objects that are still accreting at the end of the calculation are still evolving and, since the mass of an object depends on its `age' more massive accreting objects are more likely to have had close encounters.  In particular, most objects with brown dwarf masses that are still accreting have formed shortly before the calculation was stopped.  They have not had much time for dynamical encounters to occur and may not end up as brown dwarfs.  Finally, BBB2003, BB2005, and B2005 found that many stars that had close encounters still had resolved discs at the end of their calculations because those discs formed from accretion {\em subsequent} to their closest dynamical encounter.

Despite these difficulties, if an object suffers a dynamical encounter that terminates its accretion this encounter will truncate any disc that is larger than approximately 1/2 of the periastron distance during the encounter \citep*{HalClaPri1996}.  Therefore, excluding binaries and objects that are still accreting, determining the distribution of 1/2 of the closest encounter distance should give us an indication of the disc size distribution around single objects that have reached their final masses.  Note that formally we have still included the wide components of triple and quadruple systems, but these constitute only 48 objects out of the 884 `single' non-accreting objects so should not adversely affect any conclusions.

\begin{figure}
\centering
    \includegraphics[width=8.4cm]{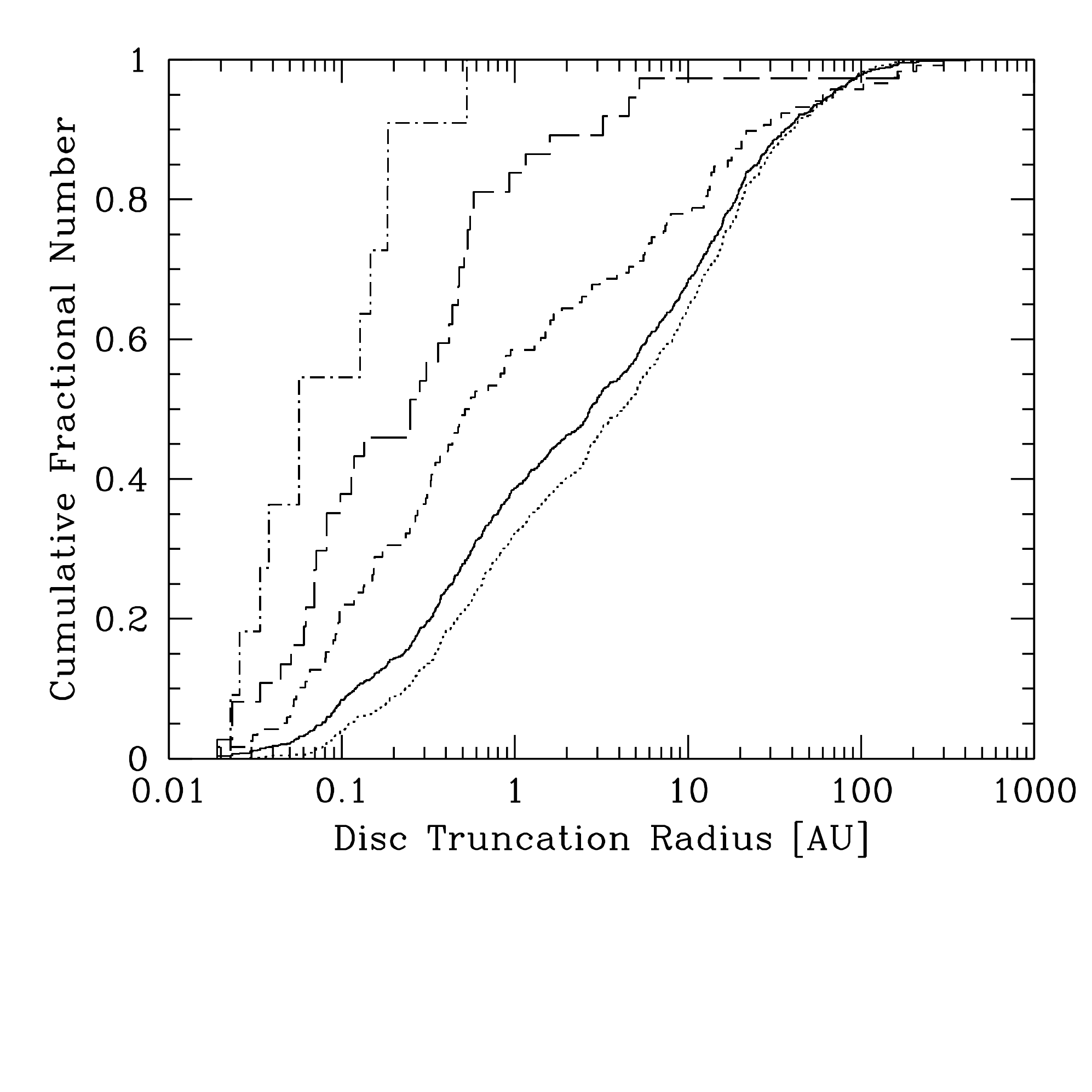}\vspace{-1.5cm}
\caption{Due to dynamical interactions, stars and brown dwarfs {\it potentially} have their discs truncated to approximately 1/2 of the periastron separation during the encounter (see also Figure \ref{truncate}).  At the end of the main calculation, we plot the cumulative fraction objects as a function of the potential truncation radius. We exclude binaries and any objects that are still accreting at the end of the calculation.  The solid line gives the result for all stars and brown dwarfs, while the dotted, short-dashed, long-dashed, and dot-dashed lines give the cumulative distributions for the mass ranges $M<0.1$, $0.1\leq M < 0.3$,  $0.3\leq M < 1.0$,  and $M \geq 1.0$ M$_\odot$, respectively.  More massive stars tend to have had closer encounters.
}
\label{cumtruncate}
\end{figure}

In Figure \ref{cumtruncate}, we plot the cumulative distributions of disc truncation radii (taken to be 1/2 of the closest encounter distance) for these objects.  The solid line gives the cumulative distribution for all 884 objects, while in the other distributions we break the sample into mass bins of $M<0.1$, $0.1\leq M < 0.3$,  $0.3\leq M < 1.0$,  and $M \geq 1.0$ M$_\odot$.  More massive stars tend to have had closer encounters and, thus, have smaller disc truncation radii.  The median truncation radius is two orders of magnitude larger for the VLM objects than for the solar-type stars.  In particular, we note that 10\% of VLM stars have truncation radii greater than 40 AU, while 1/3 have truncation radii greater than 10 AU.

{\em We emphasise that Figure \ref{cumtruncate} should be used with caution}.  First, the simulation presented here produces a very dense stellar cluster.  Disc truncation may be less important for setting disc sizes in a lower-density star-forming region.  Second, Figure \ref{cumtruncate} does {\it not} give a disc size distribution.  At best, it is a distribution of {\it lower limits} to disc sizes because of the fact that stars can suffer a close dynamical encounter, but then accrete more material from the molecular cloud and form a new disc.  This happens frequently in the simulation, especially for the higher mass stars.  The distribution is likely to be most useful for VLM objects because they tend to have their accretion terminated soon after they form by dynamical encounters and generally will not subsequently accrete significantly from the molecular cloud.

\citet*{ArmClaPal2003} considered the lifetimes of circumstellar discs surrounding young stars.  They obtained a good fit to the observed distributions of lifetimes with a $1\sigma$ dispersion of 0.5 dex in initial disc masses, with the exception of the $\approx 30$\% of young weak-lined T-Tauri stars (WTTS) that appeared to have lost their discs even with an age of 1 Myr.  There are two points of interest here.  First, we note that the dispersion of the time-averaged accretion rates for an object of a given final mass (Section \ref{origin} and Figure \ref{accrate}) is 0.4 dex in the main calculation (and similar values were obtained by BBB2005 and B2005).   This might naturally be expected to lead to the dispersion in disc masses that Armitage et al.\ required to explain the disc lifetime distributions.  Second, we find that many objects have had very close dynamical encounters.  For some objects, their closest encounters will be the one that ejects them from stellar group they are formed in.  Once they are ejected it is unlikely they will accrete a new disc.  Such objects might help to explain the observation that some WTTS appear to have lost their discs at a very young age \citep[see also][]{ArmCla1997}.

\subsection{Multiplicity as a function of primary mass}

\begin{table}
\begin{tabular}{lccccc}\hline
Mass Range ~ [M$_\odot$]& Single & Binary  & Triple & Quadruple  \\ \hline
\hspace{0.83cm}$M<0.01$       &      82      &     0     &      0      &     0    \\
$0.01\leq M<0.03$      &    348      &     8    &       1     &      0   \\
$0.03\leq M<0.07$      &    207      &   18    &       2     &      0   \\
$0.07\leq M<0.10$      &      78      &    6      &     1     &      2   \\
$0.10\leq M<0.20$      &      99     &     22    &       4     &      2   \\
$0.20\leq M<0.50$      &      59     &     23    &       5     &     10   \\
$0.50\leq M<0.80$      &     16      &     7      &     4     &      4   \\
$0.80\leq M<1.2$        &       7      &     3      &     3     &      3   \\
\hspace{0.83cm}$M>1.2$        &       9       &    3      &     3     &      4   \\ \hline
All masses                    &   905   &      90    &      23   &      25           \\ \hline
$ 0.10\leq M < 0.20$~ (no BD)     &      116    &      15    &       0     &      1   \\
$0.20\leq M<0.50$~ (no BD)     &      66     &     25     &      8      &     1   \\
$0.50\leq M<0.80$~ (no BD)     &      18     &     10     &      3      &     1   \\
$0.80\leq M<1.2$~ \hspace{0.08cm} (no BD)       &     8      &     5     &      3     &      0   \\
\hspace{0.83cm}$M>1.2$~ \hspace{0.13cm}(no BD)       &    12     &      4     &      3    &       0   \\ \hline
\end{tabular}
\caption{\label{tablemult} The numbers of single and multiple systems for different primary mass ranges at the end of the main calculation.  In the lower portion of the table, the numbers exclude brown dwarf ($M<0.075$ M$_\odot$ companions) to allow better comparison with the surveys of \citet{DuqMay1991} and \citet{FisMar1992} which were not sensitive to brown dwarfs (e.g. a solar-type star with any number of brown dwarf companions would be counted as a single solar-type star, while a solar-type star with a close brown dwarf companion and a wide M-star companion would be counted as a solar-type binary).}
\end{table}

\begin{figure*}
\centering
    \includegraphics[width=8.4cm]{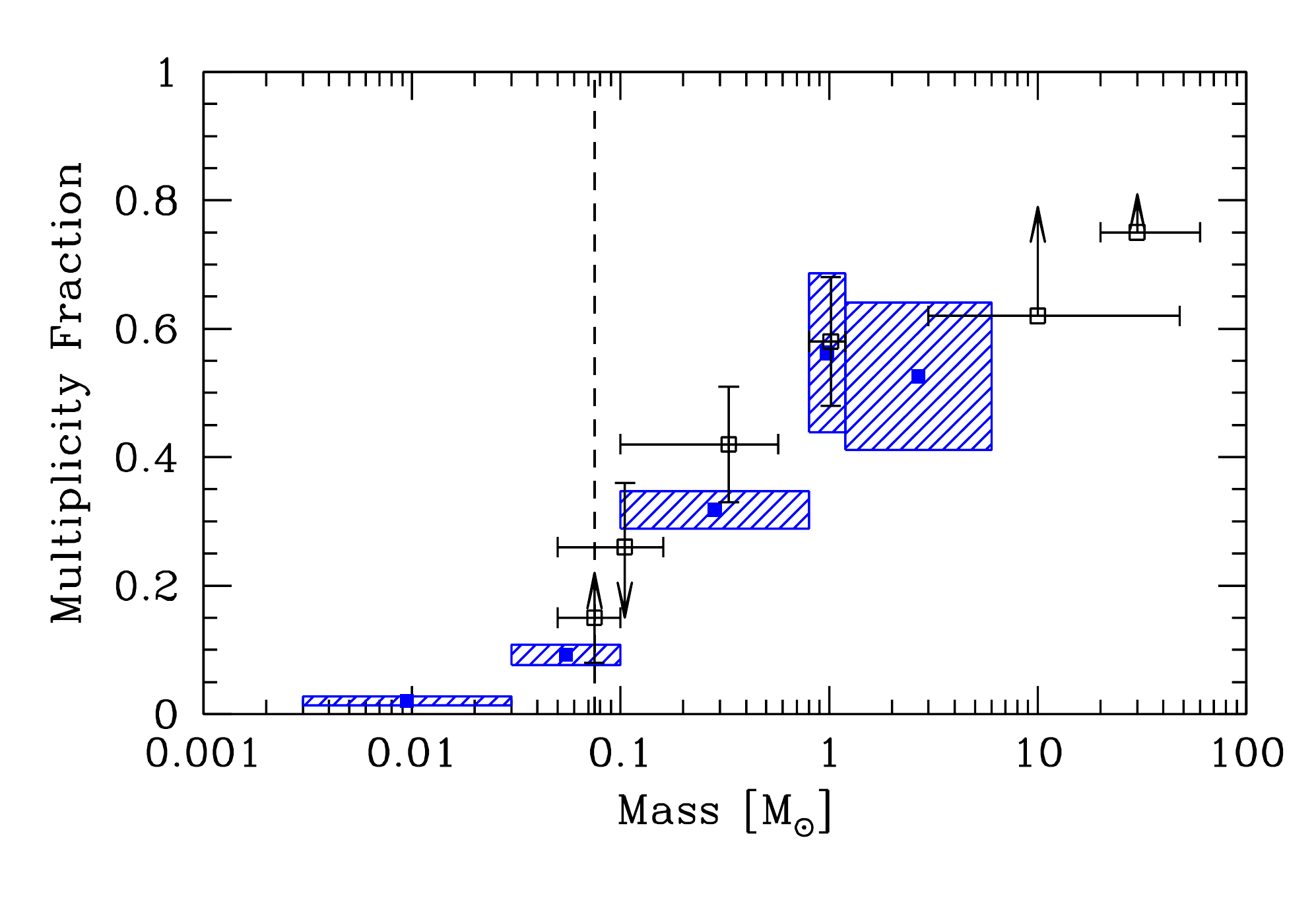}
    \includegraphics[width=8.4cm]{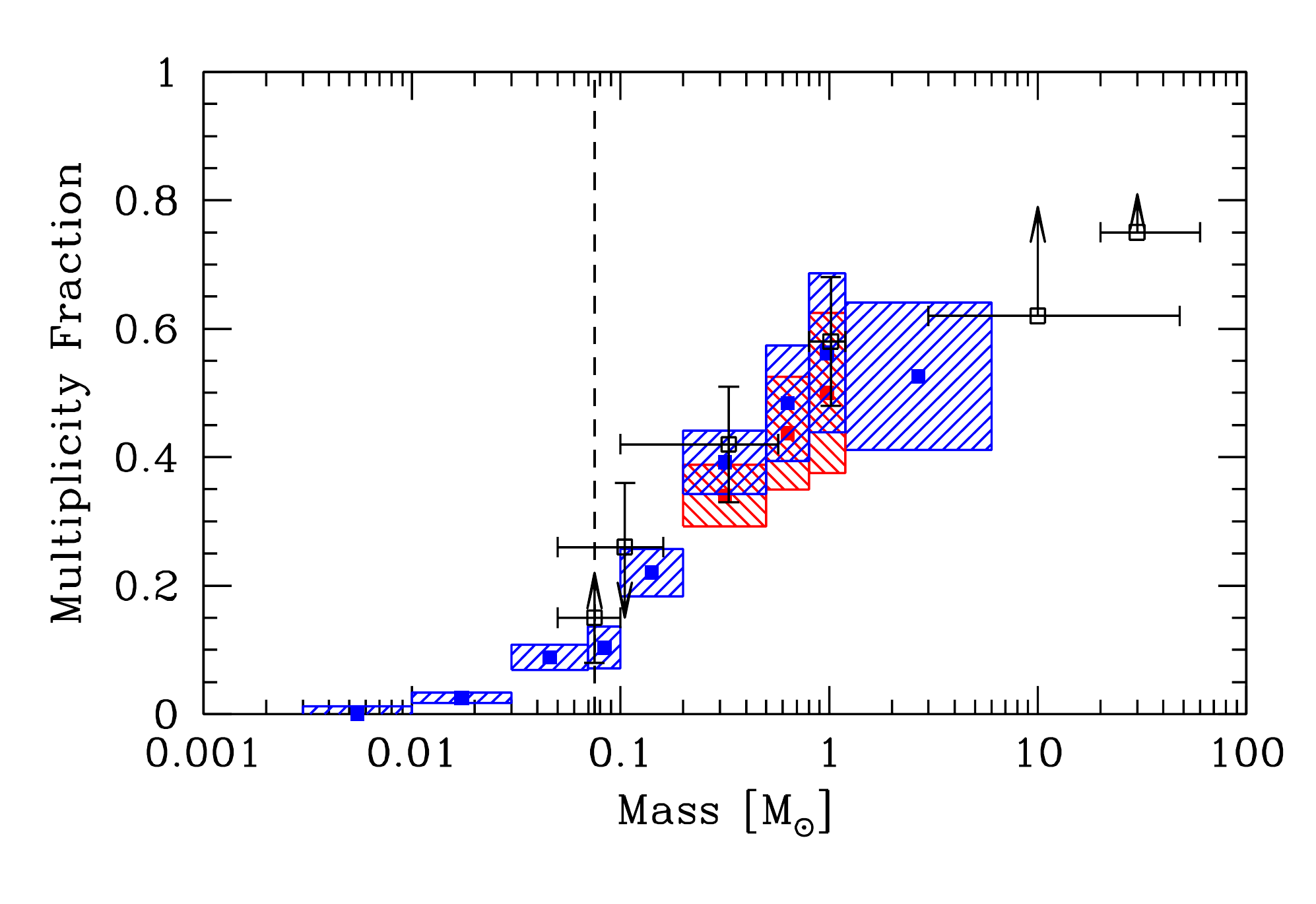}
\caption{Multiplicity fraction as a function of primary mass.  The left and right panels both give results from the main calculation, but different mass ranges are used for the low-mass stars.  On the right, the mass ranges are those given in the upper section of Table \ref{tablemult}, while on the left only three mass ranges are used for objects with masses $M<0.8$~M$_\odot$ (namely, $M<0.03$, $0.03\leq M < 0.1$, and $0.1 \leq M < 0.8$ M$_\odot$).  The blue filled squares surrounded by shaded regions give the results from the main calculation with their statistical uncertainties.  The open black squares with error bars and/or upper/lower limits give the observed multiplicity fractions from the surveys of Close et al. (2003), Basri \& Reiners (2006), Fisher \& Marcy (1992), Duquennoy \& Mayor (1991), Preibisch et al. (1999) and Mason et al. (1998), from left to right.  The red filled squares and associated shaded regions in the right panel give the multiplicity fractions excluding brown dwarf companions (masses $<0.075$ M$_\odot$) to allow better comparison with the surveys of Duquennoy \& Mayor and Fischer \& Marcy. The general trend of increasing multiplicity with primary mass is well reproduced by the main calculation.  Note that because the multiplicity is a steep function of primary mass it is important to ensure that similar mass ranges are used when comparing the simulation with observations. }
\label{multiplicity}
\end{figure*}

We turn now to the properties of the binary and higher-order multiple stars and brown dwarfs produced by the simulations.  The properties of multiple stellar systems have been investigated in the past through ensembles of small $N-$body \citep[e.g.,][]{McDCla1993, McDCla1995, SteDur1998,SteDur2003,HubWhi2005} or hydrodynamical \citep[e.g.,][]{Delgadoetal2004, GooWhiWar2004a, GooWhiWar2004b} simulations, with some of the observed trends in properties being reproduced depending on the input parameters.  However, this is the first time a large number of multiple stars and brown dwarfs has been produced from a single hydrodynamical simulation of star formation.  Although the calculation produces more brown dwarfs than is realistic, it is still of great importance to compare the multiple systems with observations.  It may be, for example, that precisely modelling the IMF requires radiative transfer to be included, but that some binary properties do not depend significantly on whether radiative transfer is included or not.  

Observationally, it is clear that the fraction of stars or brown dwarfs that are in multiple systems increases with stellar mass (massive stars: \citealt{Masonetal1998,Preibischetal1999}; intermediate-mass stars: \citealt{Patienceetal2002}; solar-type stars: \citealt{DuqMay1991}; M-dwarfs: \citealt{FisMar1992}; and very-low-mass stars and brown dwarfs: \citealt{Closeetal2003, Siegleretal2005,BasRei2006}).  It also seems that the multiplicity of young stars in low-density star-forming regions is somewhat higher than that of field stars \citep{Leinertetal1993, GheNeuMat1993,Simonetal1995,Ducheneetal2007}.  However, IC348 has a similar binary frequency to the field \citep{DucBouSim1999}.
In the Orion Nebula Cluster, \citet{Kohleretal2006} find that the binary frequency of low-mass stars is similar to that of field M dwarfs and lower than that of field solar-type stars, but that stars with masses $M>2$ M$_\odot$ have a higher binarity than stars with $0.1<M<2$ M$_\odot$ by a factor of 2.4 to 4.

To quantify the fraction of stars and brown dwarfs that are in multiple systems we use the multiplicity fraction, $mf$, defined as a function of stellar mass.  We define this as
\begin{equation}
mf = \frac{B+T+Q}{S+B+T+Q},
\end{equation}
where $S$ is the number of single stars within a given mass range and, $B$, $T$, and $Q$ are the numbers of binary, triple, and quadruple systems, respectively, for which the primary has a mass in the same mass range.  Note that this differs from the companion star fraction, $csf$, that is also often used and where the numerator has the form $B+2T+3Q$.  We choose the multiplicity fraction following \citet{HubWhi2005} who point out that this measure is more robust observationally in the sense that if a new member of a multiple system is found (e.g. a binary is found to be a triple) the quantity remains unchanged.  We also note that it is more robust for simulations too in the sense that if a high-order system decays because it is unstable the numerator only changes if a quadruple decays into two binaries (which is quite rare).  Furthermore, if the denominator is much larger than the numerator (e.g. for brown dwarfs where the multiplicity fraction is low) the production of a few single objects does not result in a large change to the value of $mf$.  This is useful because many of the systems in existence at the end of the calculations presented here may undergo further dynamical evolution.  By using the multiplicity fraction our statistics are less sensitive to this later evolution.

When analysing the simulations, some subtleties arise.  For example, many `binaries' are in fact members of triple or quadruple systems and some `triple' systems are components of quadruple or higher-order systems.  From this point on, unless otherwise stated, we define the numbers of multiple systems as follows.  The number of binaries excludes those that are components of triples or quadruples.  The number of triples excludes those that are members of quadruples.  However, higher order systems are ignored (e.g. a quintuple system may consist of a triple and a binary in orbit around each other, but this would be counted as one binary and one triple).  We need to stop counting larger and larger multiple systems at some point because in fact the simulation forms one large cluster to which many of the multiple systems are still bound when the calculation is finished (see Section \ref{calcmultiples} for a description of how we identify multiple systems).  We choose quadruple systems as a convenient point to stop as it is likely that most higher order systems would decay if the cluster was evolved for many millions of years.  The numbers of single and multiple stars produced by the main hydrodynamical calculation are given in Table \ref{tablemult} following these definitions.

In Figure \ref{multiplicity}, we plot the multiplicity fraction of the stars and brown dwarfs as a function of stellar mass for the main calculation, based on the numbers given in Table \ref{tablemult}.  In the left panel, we divide the objects into low-mass brown dwarfs (masses $<30$ Jupiter-masses or 0.03 M$_\odot$), VLM objects excluding the low-mass brown dwarfs (masses $0.03-0.10$ M$_\odot$), low-mass stars (masses $0.10-0.80$ M$_\odot$), solar-type stars (masses $0.80-1.20$ M$_\odot$), and intermediate mass stars (masses $>1.2$ M$_\odot$).  In the right panel, finer mass divisions are used for masses less than 0.8 M$_\odot$.  These divisions are chosen for comparison with various observational surveys.  In Figure \ref{multiplicity}, the filled blue squares give the multiplicity fraction while the surrounding blue hatched regions give the range in stellar masses over which the fraction is calculated and the $1\sigma$ (68\%) uncertainty on the multiplicity fraction (e.g.\ for solar-type primary stars, the multiplicity fraction is  $0.56\pm 0.12$).  The black open boxes and their associated error bars and/or upper/lower limits give the results from a variety of observational surveys (see the figure caption).  Finally, in the right panel, the filled red squares and their associated red hatched regions give the multiplicity fractions excluding brown dwarfs (masses less than 0.075 M$_\odot$).

\begin{figure*}
\centering
    \includegraphics[width=8.4cm]{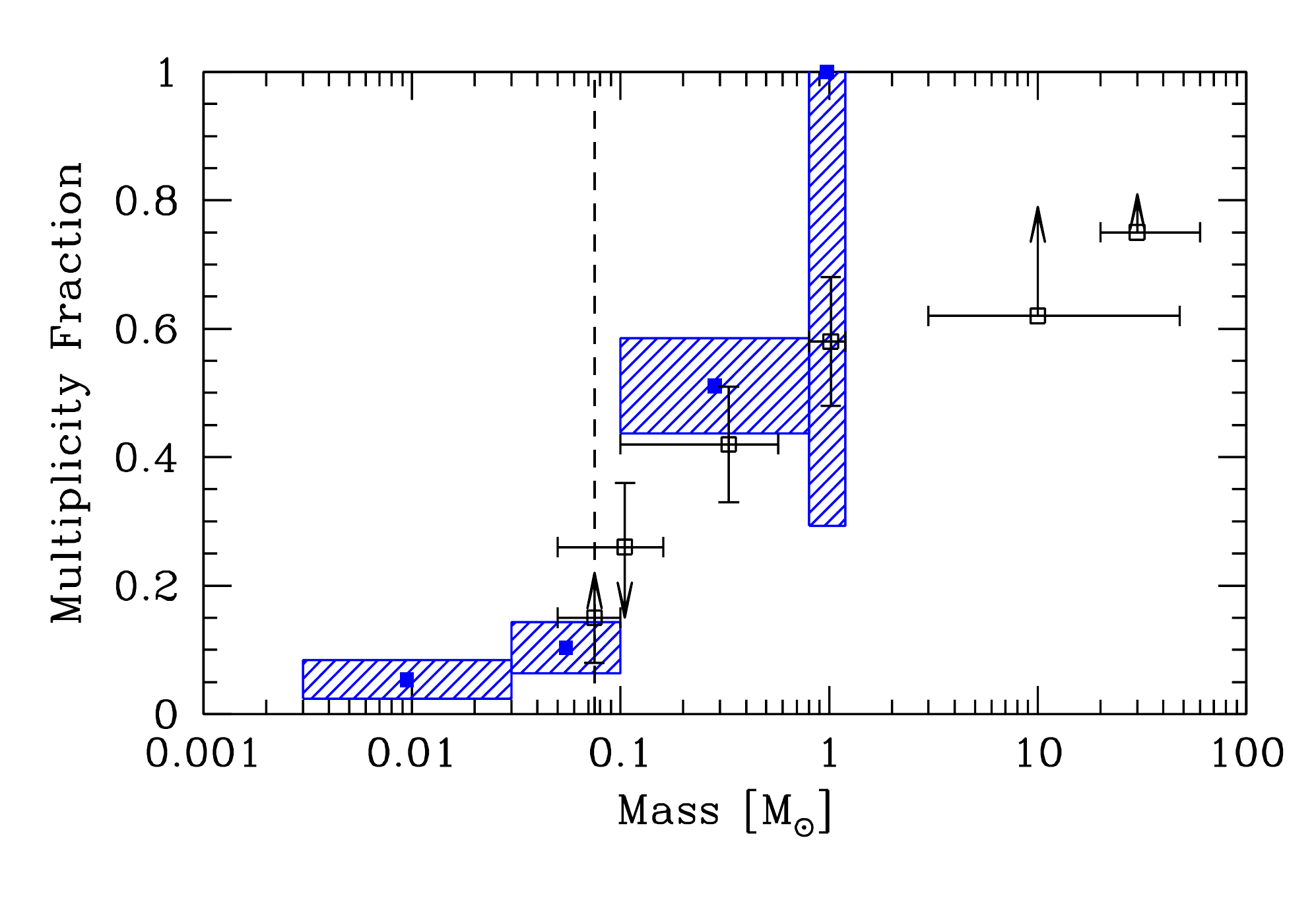}
    \includegraphics[width=8.4cm]{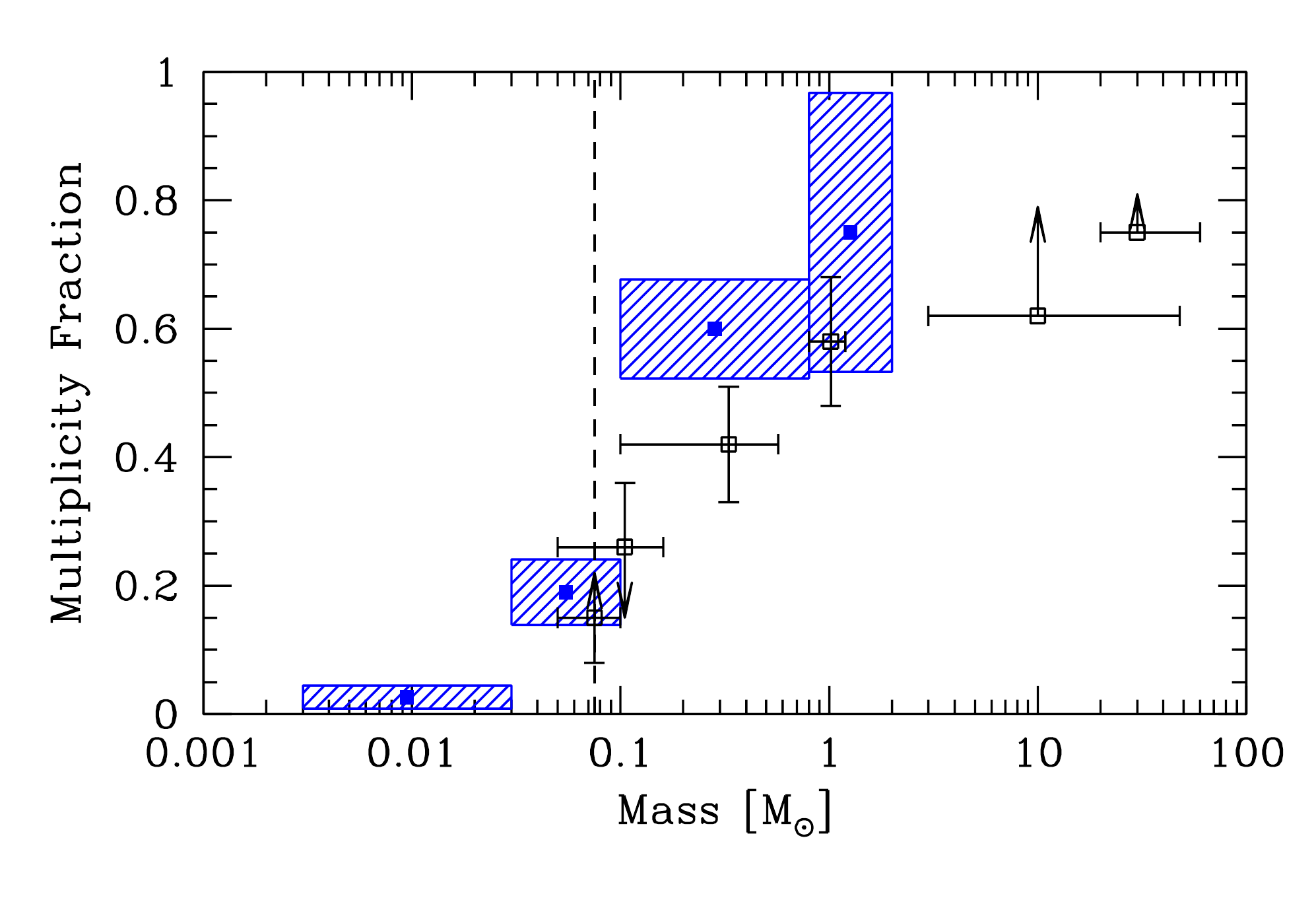}
\caption{Multiplicity fraction as a function of primary mass for the main calculation at $t=1.038t_{\rm ff}$ (left) and the re-run calculation at the same time (right).  The blue filled squares surrounded by shaded regions give the results from the calculations.  The open black squares with error bars and/or upper/lower limits give the observed multiplicity fractions from the surveys of Close et al. (2003), Basri \& Reiners (2006), Fisher \& Marcy (1992), Duquennoy \& Mayor (1991), Preibisch et al. (1999) and Mason et al. (1998), from left to right.  The multiplicities for primaries with masses in the range $0.03-0.8$ M$_\odot$ are higher in the re-run calculation in which the sink particles have smaller accretion radii and no gravitational softening.}
\label{multiplicitycomp}
\end{figure*}

The main hydrodynamical calculation clearly predicts that the multiplicity fraction strongly increases with increasing primary mass.  Furthermore, the values in each mass range are in reasonable agreement with observation.  There is excellent agreement for solar-type and low-mass stars.  For intermediate mass stars the statistics from the calculation are poor (and the observed value is also uncertain), while for VLM objects the hydrodynamical calculation gives a slightly lower prediction that the observations, but not unreasonably so.

In detail, we find:
\begin{description}
\item[{\bf Solar-type stars:}] \citet{DuqMay1991} find an observed multiplicity fraction of $0.58 \pm 0.1$.  The main calculation gives a multiplicity fraction of $0.56\pm 0.12$. However, this figure includes brown dwarf companions and Duquennoy \& Mayor's survey was not sensitive to brown dwarfs.  Excluding them, we obtain $0.50 \pm 0.13$ which is still in good agreement with the observed value.

\item[{\bf M-dwarfs:}] \citet{FisMar1992} find an observed multiplicity fraction of $0.42 \pm 0.09$.  In the mass range $0.1-0.8$ M$_\odot$ we obtain $mf=0.32 \pm 0.03$ which is slightly lower than the observed value, though still within the uncertainties.  However, in this mass range the multiplicity fraction changes quite rapidly with mass.  Fischer \& Marcy's sample contains stars with masses between 0.1 and 0.57 solar masses, but the vast majority have masses in the range $0.2-0.5$ M$_\odot$ whereas in the hydrodynamical simulation around half of the low-mass stars have masses less than 0.2 M$_\odot$.  In the $0.2-0.5$ M$_\odot$ mass range we obtain $mf=0.39 \pm 0.05$.  However, Fischer \& Marcy's survey was also not sensitive to brown dwarfs companions.  Removing these, we obtain $0.34\pm 0.05$.  This value is consistent with the observed value, lying well within the $1\sigma$ uncertainties.

\item[{\bf VLM objects:}]  There has been much interest in the multiplicity of VLM objects in recent years (\citealt{Martinetal2000, Martinetal2003, Closeetal2003, Closeetal2007, Gizisetal2003, Pinfieldetal2003, Bouyetal2003, Bouyetal2006, Siegleretal2003, Siegleretal2005, Luhman2004, MaxJef2005}; \citealt*{KraWhiHil2005, KraWhiHil2006}; \citealt{BasRei2006, Reidetal2006, Allenetal2007, Konopackyetal2007, Ahmicetal2007, Reidetal2008}; \citealt*{LawHodMac2008}; \citealt{Maxtedetal2008}).  For a recent review, see \citet{Burgasseretal2007}.  Over the entire mass range of $0.003-0.10$ M$_\odot$, we find a very low multiplicity of just $0.047\pm 0.008$.  We note the main calculation, which is essentially a larger version of the calculation reported in BBB2003, produces a VLM object multiplicity in agreement with the earlier, smaller calculations which gave $mf \approx 0.06$ (B2005).  However, in the earlier calculations it was impossible to sub-divide the VLM objects because of the small numbers.  As with the M-dwarfs, the multiplicity drops rapidly with decreasing primary mass and the {\it observed} VLM objects tend to have high masses.  The main calculation gives multiplicities of $0.22\pm 0.04$ for the mass range $0.1-0.2$ M$_\odot$, $0.10 \pm 0.03$ for the mass range $0.07-0.10$ M$_\odot$, $0.09\pm 0.02$ for the mass range $0.03-0.07$ M$_\odot$, $0.025 \pm 0.008$ for the mass range $0.01-0.03$ M$_\odot$, and $0.00 \pm 0.01$ for masses less than 0.01 M$_\odot$.  Therefore, to compare with observations it is very important to compare like with like.  The observed frequency of VLM binaries is typically found to be $\approx 15$\% \citep{Closeetal2003, Closeetal2007, Martinetal2003, Bouyetal2003, Gizisetal2003, Siegleretal2005, Reidetal2008}.  The surveys are most complete for binary separations greater than a couple of AU.  Recently \citep{BasRei2006} estimated the total frequency (including spectroscopic systems) to be $\approx 20-25$\%.  These surveys typically targeted primaries with masses in the range $0.03-0.1$ M$_\odot$, but most of these objects in fact have masses greater than 0.07 M$_\odot$.  Thus, the closest comparison with our calculation is our frequency of $0.10 \pm 0.03$ for the mass range $0.07-0.10$ M$_\odot$.  This is somewhat lower than the observed frequency (a factor of two at face value), but still in better agreement than that from the earlier simulations (B2005).  In the next section we show that decreasing the accretion radii of the sink particles increases the frequency of VLM binaries bringing them into good agreement with the observed value.  Thus, the main calculation produces a VLM binary frequency that is consistent with observations (at around the $2-3\sigma$ level), but it is lower and we attribute this to the effects of the sink particle approximation rather than a fundamental failing of the hydrodynamical star formation model.

\item[{\bf Low-mass brown dwarfs:}]  The frequency of low-mass binary brown dwarfs (primary masses less than 30 Jupiter masses) is observationally unconstrained.  {\it We predict that the multiplicity continues to fall as the primary mass is decreased as described above.}  Even if our predicted multiplicities are under-estimated by a factor of two or even three due to the effects of sink particles, we would predict that the binary frequency in the mass range $0.01-0.03$ M$_\odot$ is $\lsim 7$\%.  Companions to brown dwarfs with masses less than 10 Jupiter-masses should be exceptionally rare ($\lsim 3$\%).

\end{description}

\subsubsection{The dependence of multiplicity on sink particle approximations}

As with the IMF, the question arises of how dependent these results are on the use of sink particles.  In particular, in the main calculation, binaries cannot have separations smaller than 1 AU (due to the gravitational softening) and the sink particle accretion radius removes all gas within 5 AU of the sink particle, presumably affecting close dynamical interactions between protostellar objects.  This is likely to have a severe effect on the properties of short period binaries.  As mentioned above and will be seen in more detail in Section 3.3, this particularly affects VLM binaries whose median separation in the main calculation (and observationally) is less than 10 AU. 

In Figure \ref{multiplicitycomp}, we compare the multiplicity fractions produced by the main calculation (left) and the re-run calculation (right) at the end time of the re-run calculation ($t_{\rm ff}=1.038$).  The first point to note is that the fractions given by the main calculation at 1.038 $t_{\rm ff}$ and at 1.50 $t_{\rm ff}$ are the same within the statistical uncertainties.  Therefore, we conclude that the fractions do not evolve significantly with time (though their mass ratios and separations might -- see Sections \ref{sec:separations} and \ref{sec:massratios}).  There are few stars with masses greater than 0.8 M$_\odot$ at the earlier time because they have not yet had time to accrete to high masses.  Thus, the multiplicity fractions of solar-type and intermediate mass stars are poorly defined.  However, for low-mass stars, the fractions are $0.51\pm  0.07$ and $0.32 \pm 0.03$ respectively, which lie within $2\sigma$ of each other.  For VLM systems the fractions are $0.10\pm 0.04$ and $0.092\pm 0.016$, respectively.  For low-mass brown dwarfs, the fractions are $0.054\pm 0.030$ and $0.021\pm 0.007$, respectively.

We now compare the fractions give by the main calculation and the re-run calculation which has smaller sink particle accretion radii (left and right panels of Figure \ref{multiplicitycomp}).  The multiplicity fractions are greater in the re-run calculation for VLM objects and low-mass stars, but not for the low-mass brown dwarfs.  An increase in the multiplicity fractions for small sink particles is what we might expect since binaries can become tighter (due to the absence of gravitational softening) and dissipative processes can play a role on smaller scales (due to the smaller accretion radii of only 0.5 AU). Low-mass stars in the re-run calculation have a multiplicity of $0.60\pm 0.08$, which differs by $\approx 0.6 \sigma$ from the main calculation at the same time.  VLM binaries have a multiplicity of $0.19\pm 0.05$.  This is $1 \sigma$ higher than the main calculation at the same time.  Finally, low-mass brown dwarfs have a multiplicity of $0.026\pm 0.018$ which differs by $0.6\sigma$ from the main calculation at the same time.

Clearly, even with such large numbers of objects, statistical uncertainties still make comparison of the results difficult.  However, the indication is that decreasing the sizes of the sink particles increases the multiplicity fractions, at least for the mass range $0.03-0.80$ M$_\odot$.  In particular, decreasing the sizes of the sink particles maintains the good agreement with observations for solar-type stars and low-mass stars, and improves the agreement for VLM objects.  The multiplicity of $19\pm 5$\% for the mass range $0.03-0.10$ M$_\odot$ is in excellent agreement with the typically observed value of $\approx 15$\% \citet{Closeetal2003} and the upper limit of $20-25$\% estimated by \citet{BasRei2006}.  

In summary, it seems that {\it purely hydrodynamical simulations of star formation using sink particles can reproduce the observed multiplicities of solar-type stars, low-mass stars, and VLM objects}.  The results appear to depend slightly on the sink particle assumptions, with smaller sink particles generally leading to slightly higher multiplicities and better agreement with observations.

\subsubsection{Star-VLM binaries}

We turn now to the issue of VLM/brown dwarf companions to stars.  As in the previous section, we do not consider brown dwarf companions as such, rather we consider VLM companions ($< 0.1$ M$_\odot$) to stars ($\geq 0.1$ M$_\odot$).   The main calculation produced 26 stellar-VLM binaries out of 290 stellar systems, a frequency of $9.0\pm 1.6$\%.  For the vast majority of these stellar-VLM binaries, the star is a low-mass star: 14 of the primaries have masses between $0.1-0.2$ M$_\odot$, 7 have primary masses in the range $0.2-0.5$ M$_\odot$, and 3 have primary masses between 0.5 and 0.8 M$_\odot$.   However, within the statistical uncertainties, the frequency of VLM companions is not found to depend on primary mass.  Even around solar-type and intermediate mass stars we find VLM companions, but the statistics are very poor with only two out of the 35 systems with primary masses greater than 0.8 M$_\odot$ being star/VLM binaries ($6\pm 4$\%).

Although there is no statistically significant dependence of the frequency of such systems on primary mass, the separation distributions are very different.  For primaries with masses of $0.1-0.2$ M$_\odot$, the semi-major axes of all but 3 of the 14 systems are less than 30 AU.  The other three all have semi-major axes greater than 1000 AU.  This separation distribution is very similar to the VLM and brown dwarf binaries discussed in Section \ref{sec:separations}.  For the 7 primaries with masses of $0.2-0.5$ M$_\odot$, three have VLM companions within 10 AU, there is one at 49 AU, and the remaining three have wide companions (greater than 1000 AU).  The VLM companions of the 3 primaries with masses of $0.5-0.8$ M$_\odot$ have semi-major axes between 27 and 65 AU.  Finally, the 2 star/VLM binaries with primary masses greater than 0.8 M$_\odot$ both have semi-major axes greater than 1000 AU.  Thus, {\it the typical separation of star/VLM binaries seems to increase strongly as the mass of the primary increases}.   

In addition to the star/VLM binaries, there are four triple systems consisting of a star with two VLM companions and eight quadruple systems that contain at least one star/VLM pair.  In all but three of these 12 systems the widest orbit has a semi-major axis in the range $50-500$ AU.  The remaining three systems have very wide outer orbits ($>1000$ AU).

There has been much discussion over the past decade of the observed ``brown dwarf desert" for close brown dwarf companions solar-type stars \citep[frequency $\approx 1$\%;][]{MarBut2000, GreLin2006} and how this changes for wider separations and different primary masses.  \citet{McCZuc2004} found that the frequency of wide brown dwarfs to G, K, and M stars between 75-300 AU was 1\%$\pm$1\%.  The frequencies of wide brown dwarf companions to A and B stars \citep{KouBroKap2007}, M dwarfs \citep{Gizisetal2003}, and other brown dwarfs appears to be similarly low, although the frequency of wide binary brown dwarfs may be higher when they are very young \citep{Closeetal2007}.  Our results are consistent with these observations in the sense that we do not find brown dwarf companions to solar-type stars in close orbits (frequency $\lsim 8$\% at the 95\% confidence level), but that VLM companions exist orbiting stars and brown dwarfs with a wide range of masses.  Our results are also in good agreement with surveys of VLM objects that are frequently found to have companions, but where their separations are usually less than $\approx 20$ AU \citep{Closeetal2003, Closeetal2007,Allenetal2007}.  It would be of great interest to map out the separation distributions of VLM companions over a wide range of primary masses.  From the results of the main calculation {\it we predict that the frequency of star-brown dwarf systems should not depend greatly on primary mass, but that the typical star-brown dwarf binary separation should increase monotonically from $\lsim 10$ AU for primary masses less than 0.2 M$_\odot$ to $\sim 50$ AU for primary masses $\sim 0.4$ M$_\odot$ and to $> 100$ AU for solar-type stars}.

\begin{figure*}
\centering
    \includegraphics[width=8.4cm]{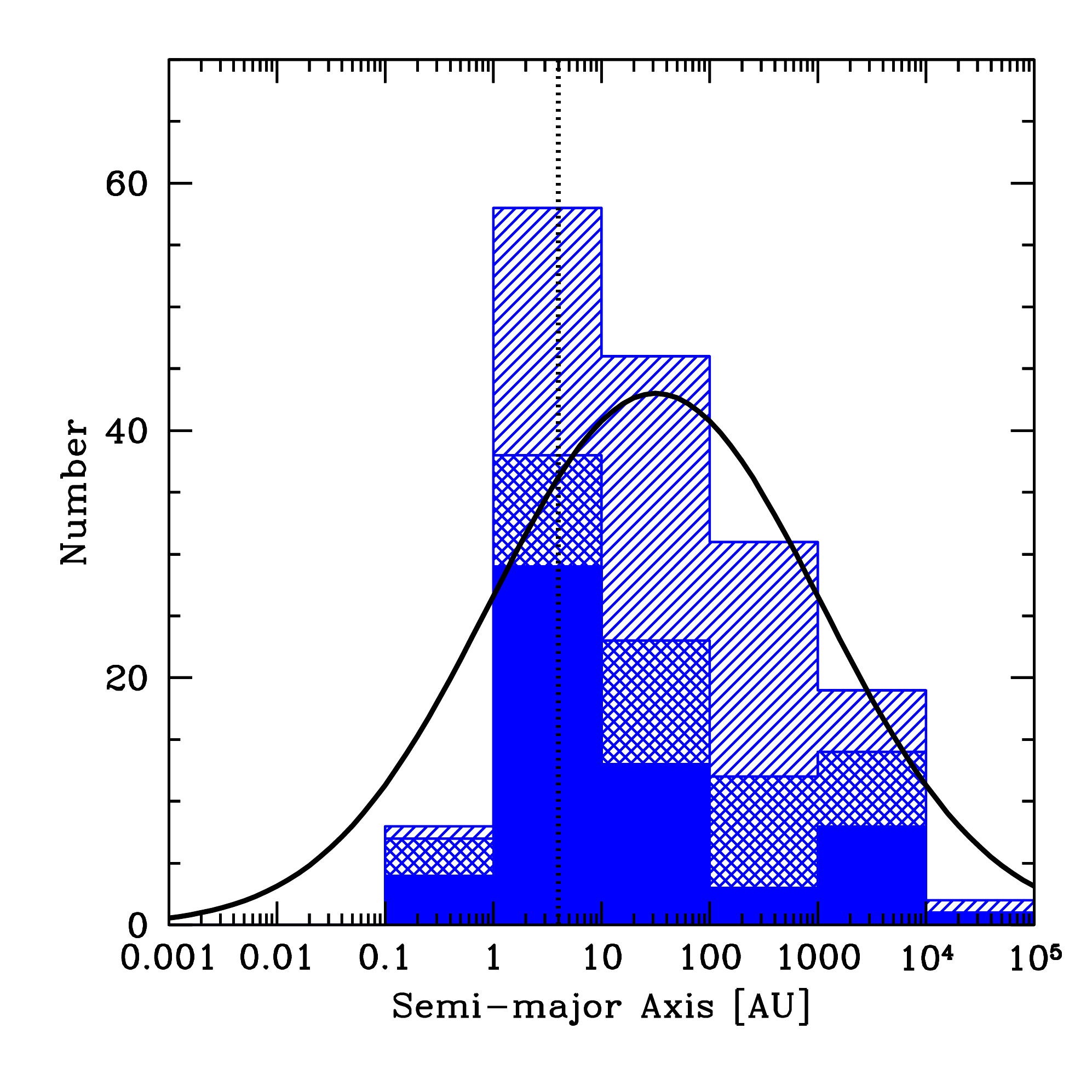}  %Uses factor of 43.
    \includegraphics[width=8.4cm]{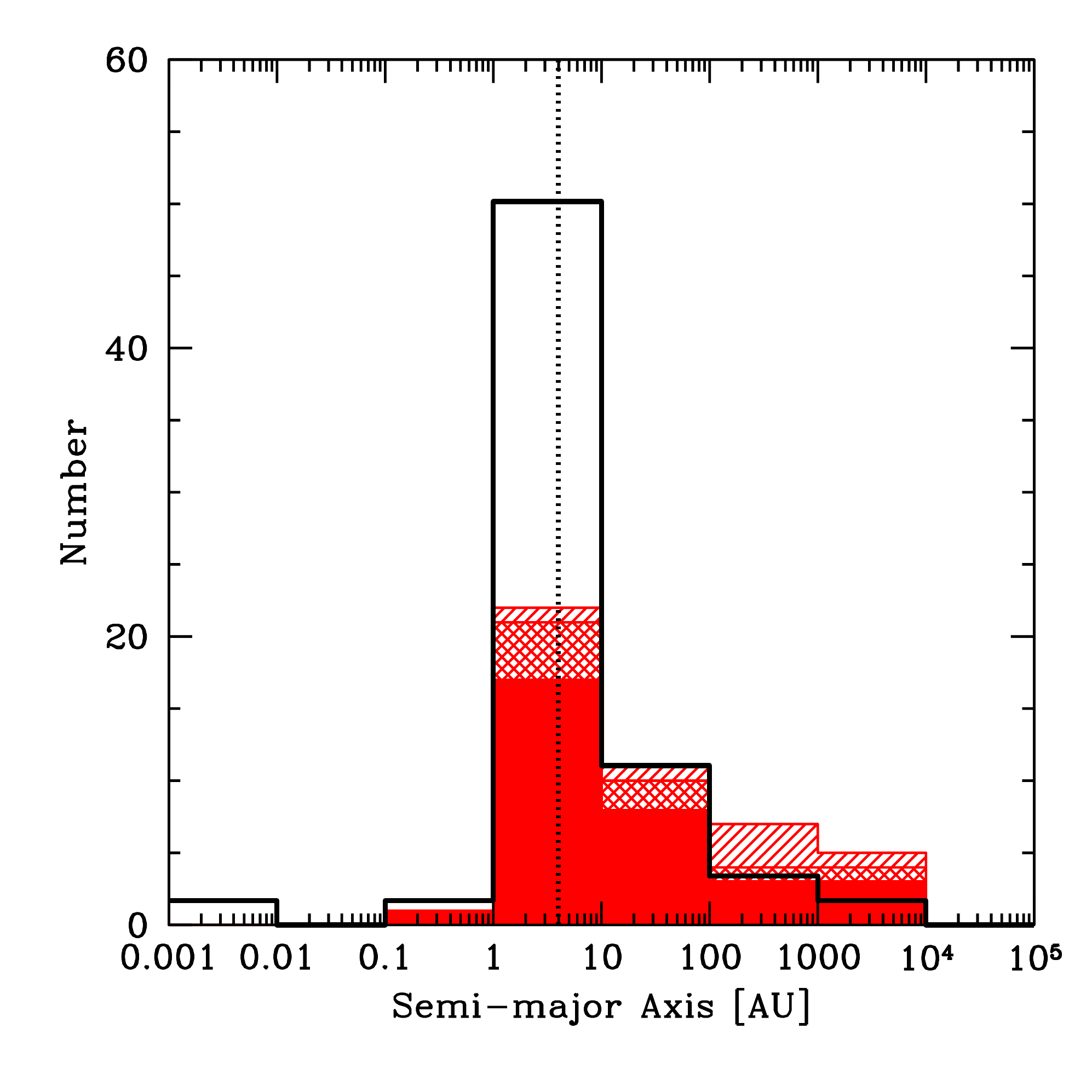} % Uses factor of 0.85
\caption{The distributions of separations (semi-major axes) of multiple systems with stellar (left) and VLM (right) primaries produced by the main calculation.  The solid, double hashed, and single hashed histograms give the orbital separations of binaries, triples, and quadruples, respectively (each triple contributes two separations, each quadruple contributes three separations).  In the stellar graph, the curve gives the G-dwarf separation distribution (scaled to match the area) from Duquennoy \& Mayor (1991).  In the VLM systems graph, the open black histogram gives the (scaled to match the number in the 10-100 AU range) separation distribution of the known very-low-mass multiple systems maintained by N. Siegler at http://vlmbinaries.org/ (last updated on February 4th, 2008).  The vertical dotted line gives the resolution limit of the calculations as determined by the gravitational softening and accretion radii of the sink particles.}
\label{separations}
\end{figure*}

\begin{figure*}
\centering
    \includegraphics[width=8.4cm]{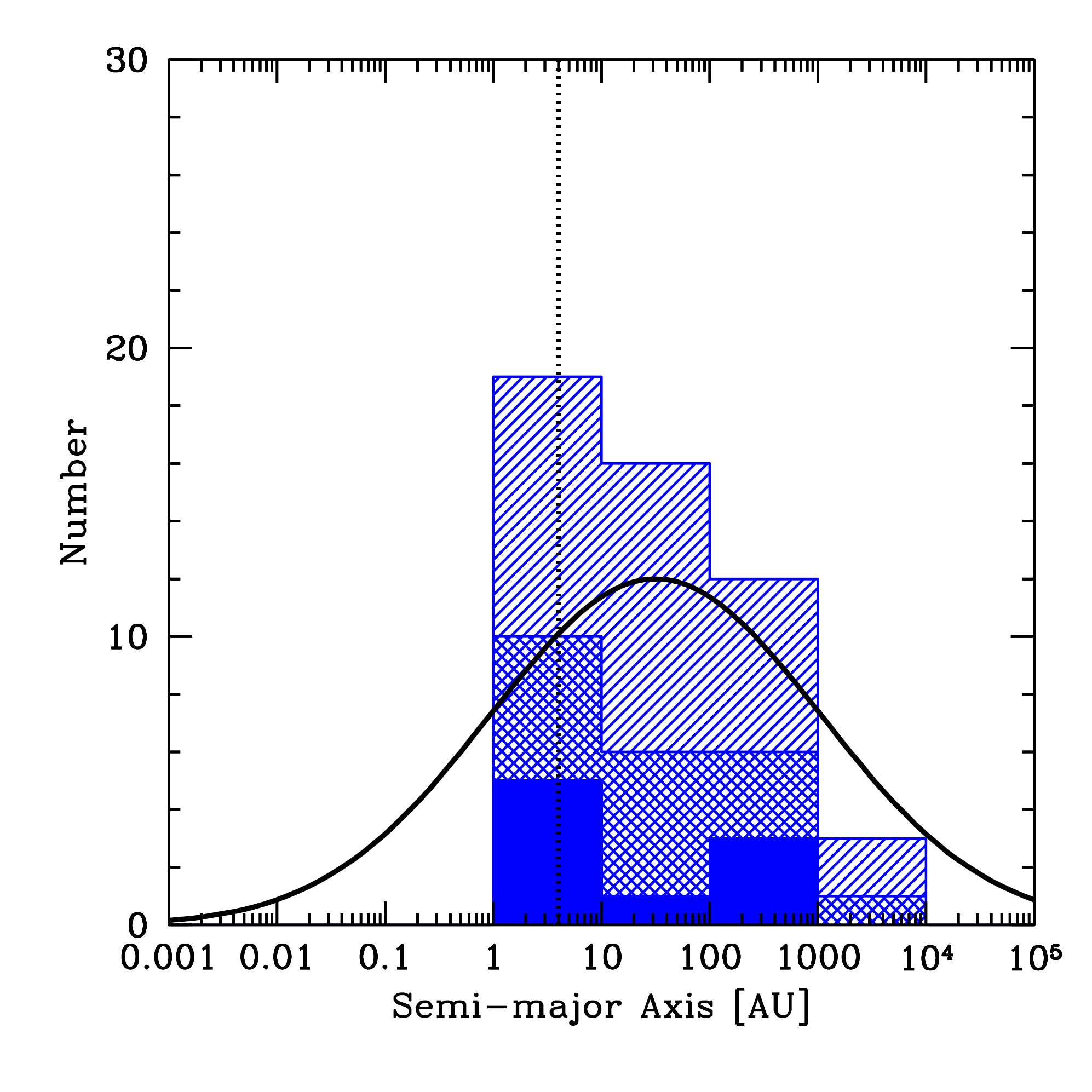} % Uses factor of 12.0
    \includegraphics[width=8.4cm]{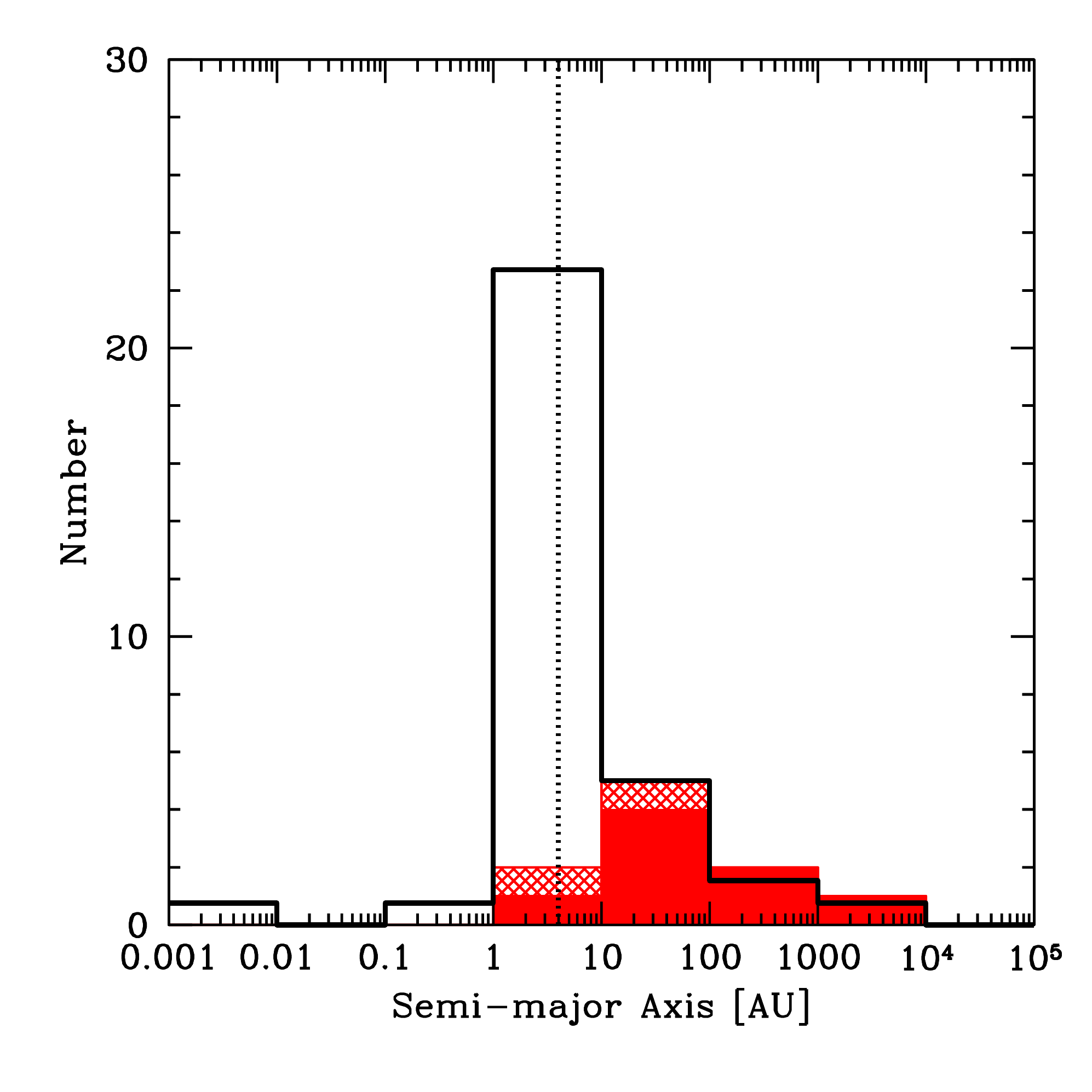} % Uses 0.385
    \includegraphics[width=8.4cm]{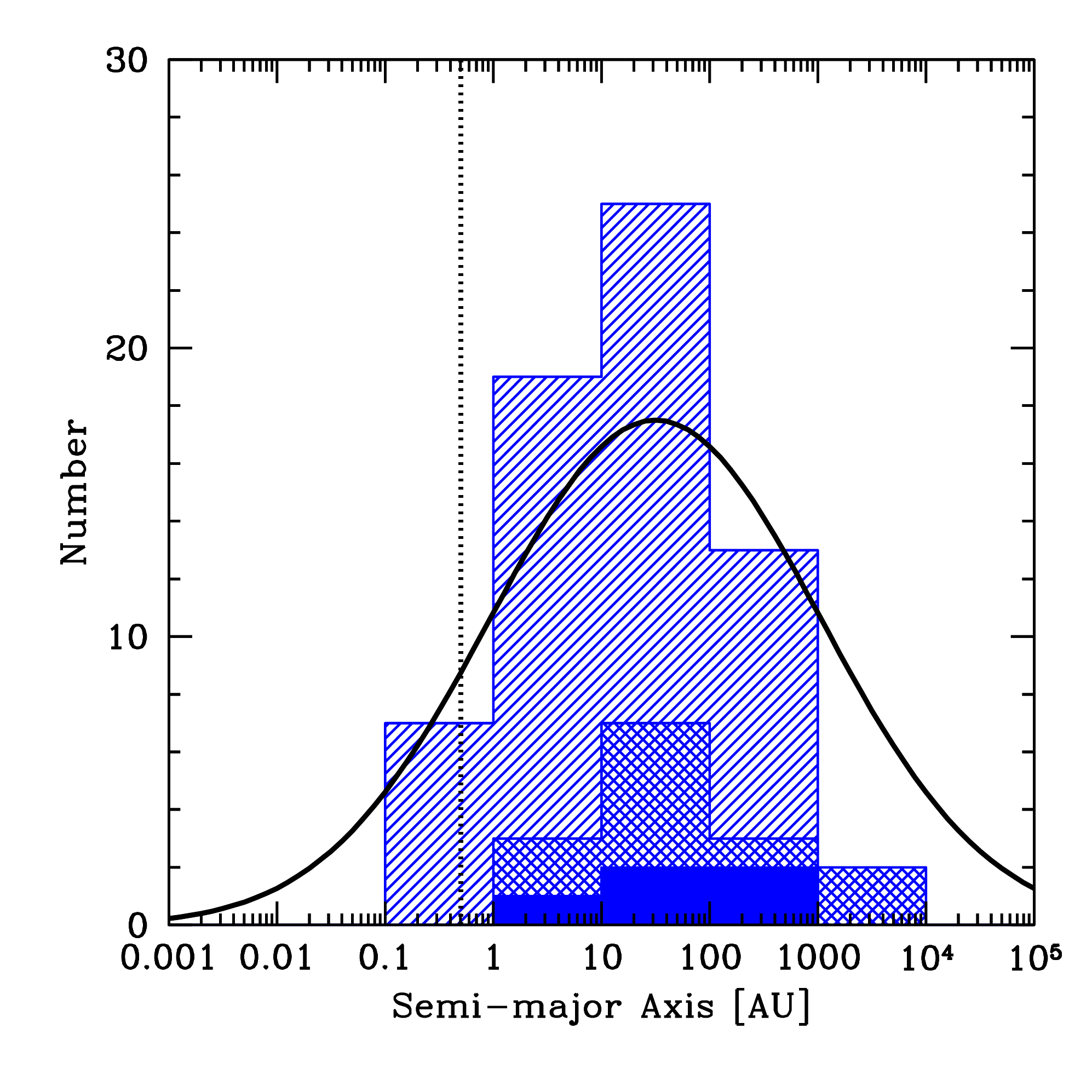} % Uses 17.5
    \includegraphics[width=8.4cm]{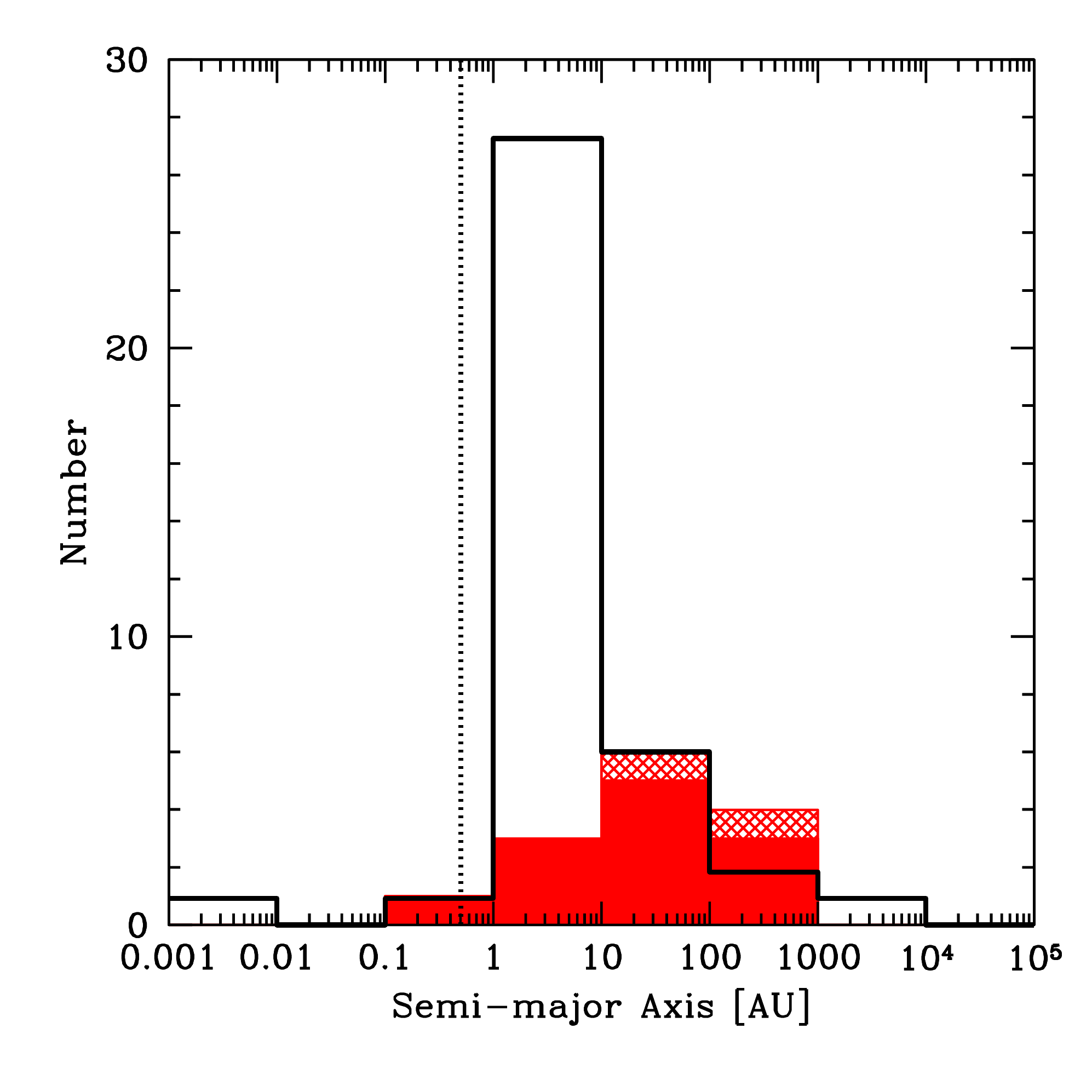} % Uses 0.462
\caption{The same as Figure \ref{separations} but the separation (semi-major axis) distributions are given at $t=1.038 t_{\rm ff}$ for the main calculation (top) and the re-run calculation which uses sink particles with small accretion radii (0.5 AU) and without gravitational softening (bottom).  As expected, reducing the length-scales of the sink particle accretion radii and gravitational softening produces a higher fraction of small-separation multiple systems.  In addition, the `pile up' of stellar system separations in the $1-10$ AU bin (top left) disappears when smaller separations are allowed (bottom left), recovering a bell-shaped distribution more similar to the observed Duquennoy \& Mayor (1991) distribution for solar-type primaries.}
\label{separations:re-run}
\end{figure*}

\subsubsection{The frequencies of triple and quadruple systems}
\label{freq_high_order}

Consulting Table \ref{tablemult}, we find that the main calculation produced 905 single stars/brown dwarfs, 90 binaries, 23 triples and 25 quadruples.  This gives an overall frequency of triple and quadruple systems of only $2.3\pm 0.5$\% and $2.5\pm 0.5$\%, respectively.  These are upper limits because some of these systems may be disrupted if the calculation were followed longer.

Although the overall frequencies are low, it is clear from the table that the frequencies of high-order multiples depend strongly on primary mass.  For VLM primaries, the frequencies of triple/quadruple systems range from $3.4 \pm 2.0$\% for the mass range $0.07-0.10$ M$_\odot$ to $0.9\pm 0.6$\% for $0.03-0.07$ M$_\odot$ and much less than 1\% for lower primary masses.  For low-mass M-stars in the range $0.10-0.20$ M$_\odot$ the frequency of triples/quadruples is $5\pm 2$\%.
For M-stars with masses in the range $0.20-0.50$, the frequency of triples/quadruples is $15 \pm 4$\% while for solar-type and intermediate mass stars the frequency is $\approx 37 \pm 12$\%.

How do these frequencies compare with observations?  \citet{FisMar1992} find 7 triples and 1 quadruple amongst 99 M-star primaries giving a frequency of $8\pm 3$\%.  As mentioned earlier, Fischer and Marcy's survey was not sensitive to brown dwarf companions and most of their M-stars had masses in the range $0.2-0.5$ M$_\odot$.  Excluding brown dwarfs from the multiple statistics, we find a frequency of $9\pm 3$\% for this stellar mass range, in excellent agreement.  \citet{DuqMay1991} found 7 triples and 2 quadruples from their 164 solar-type primaries giving a frequency of $5\pm 2$\%.  For solar-type stars (excluding brown dwarf companions), we find a frequency of $18\pm 10$\%.  The large uncertainty in our result makes comparison difficult for the solar-type stars, but our result is not unreasonable, especially given the fact that Duquennoy \& Mayor admit that they are likely to have missed some high-order multiple systems.

In summary, our frequencies of triples/quadruples are consistent with current observational surveys, though more robust statistics from observations, particularly for VLM objects, and improved statistics from the simulations, particularly for intermediate-mass stars, are obviously desireable.

\subsection{Separation distributions of multiples}
\label{sec:separations}

\begin{figure*}
\centering
    \includegraphics[width=5.8cm]{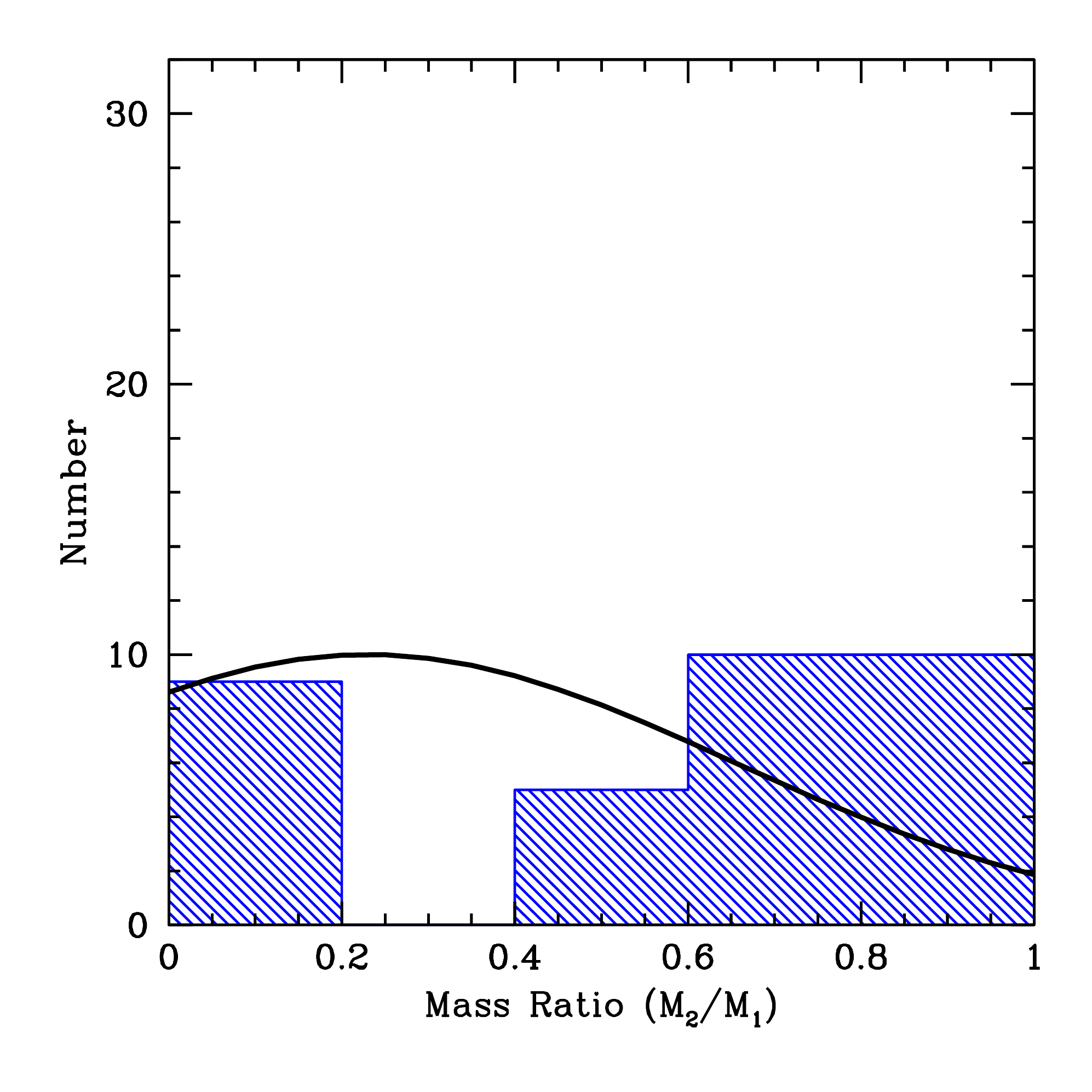}
    \includegraphics[width=5.8cm]{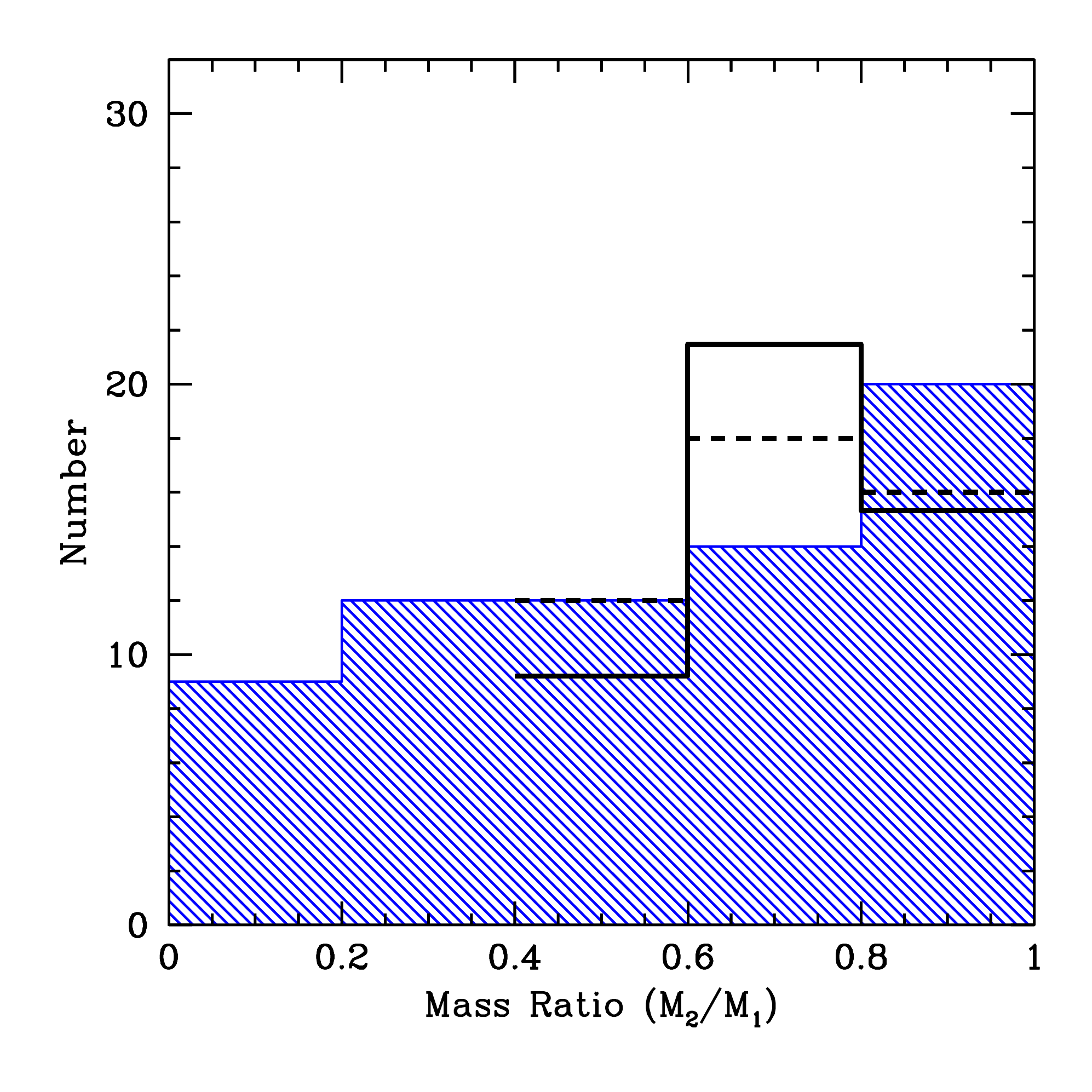}
    \includegraphics[width=5.8cm]{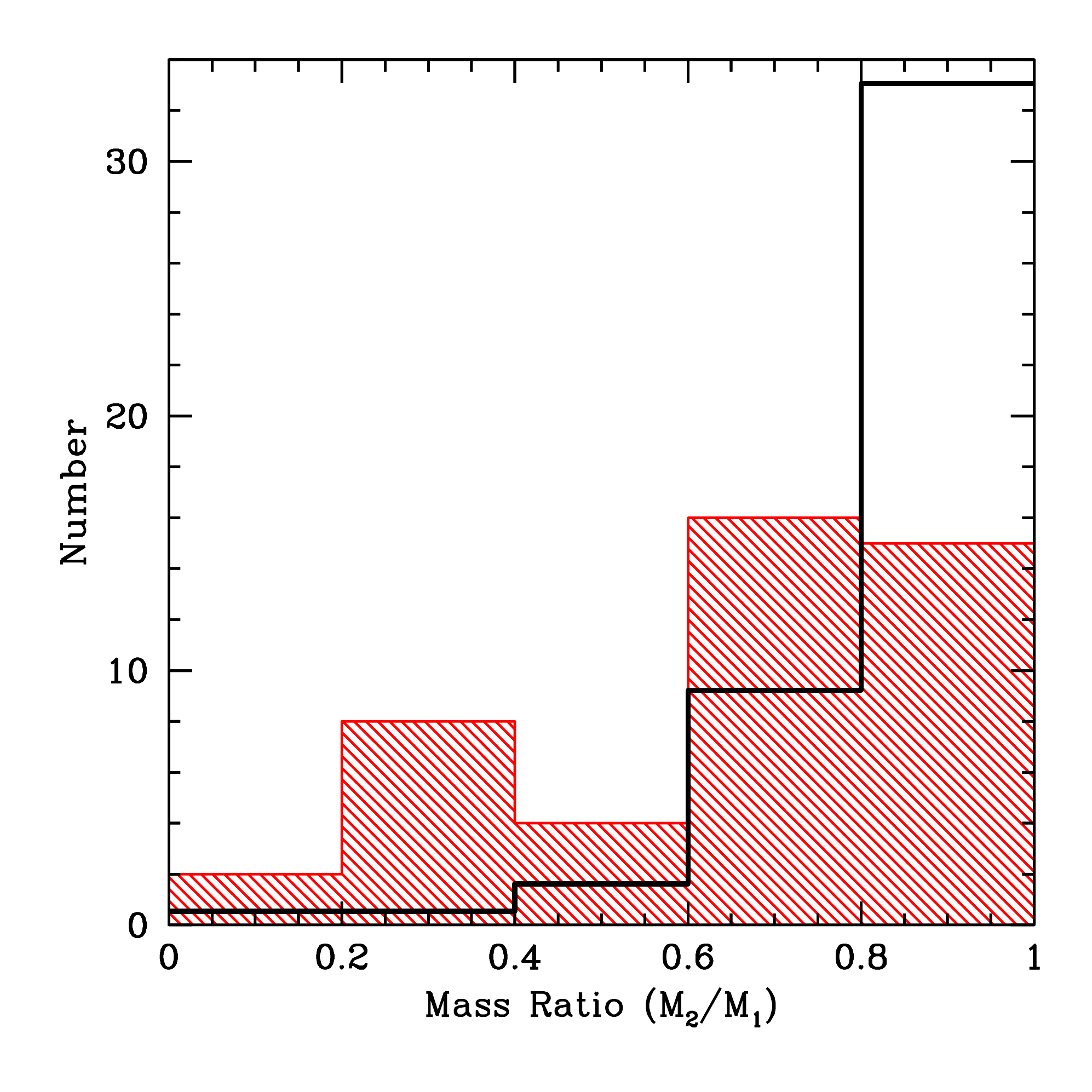}
\caption{The mass ratio distributions of binary systems with stellar primaries in the mass ranges $M_1>0.5$ M$_\odot$ (left) and $M_1=0.1-0.5$ M$_\odot$ (centre) and VLM primaries (right; $M_1<0.1$ M$_\odot$) produced by the main calculation.  The solid black lines give the observed mass ratio distributions of \citet{DuqMay1991} for G dwarfs (left), \citet{FisMar1992} for $M_1=0.3-0.57$ M$_\odot$ (centre, solid line) and $M_1=0.2-0.57$ M$_\odot$ (centre, dashed line), and of the known very-low-mass binary systems maintained by N. Siegler at http://vlmbinaries.org/ (right).  The observed mass ratio distributions have been scaled so that the areas under the distributions ($M_2/M_1=0.4-1.0$ only for the centre panel) match those from the simulation results.  The VLM binaries produced by the simulation are biased towards equal masses when compared with M dwarf binaries (primary masses in the range $M_1=0.1-0.5$ M$_\odot$).  71\% of the VLM binaries have $M_2/M_1>0.6$ while for the M dwarf binaries the fraction is only 51\%. }
\label{massratios}
\end{figure*}

\begin{figure*}
\centering
    \includegraphics[width=8.4cm]{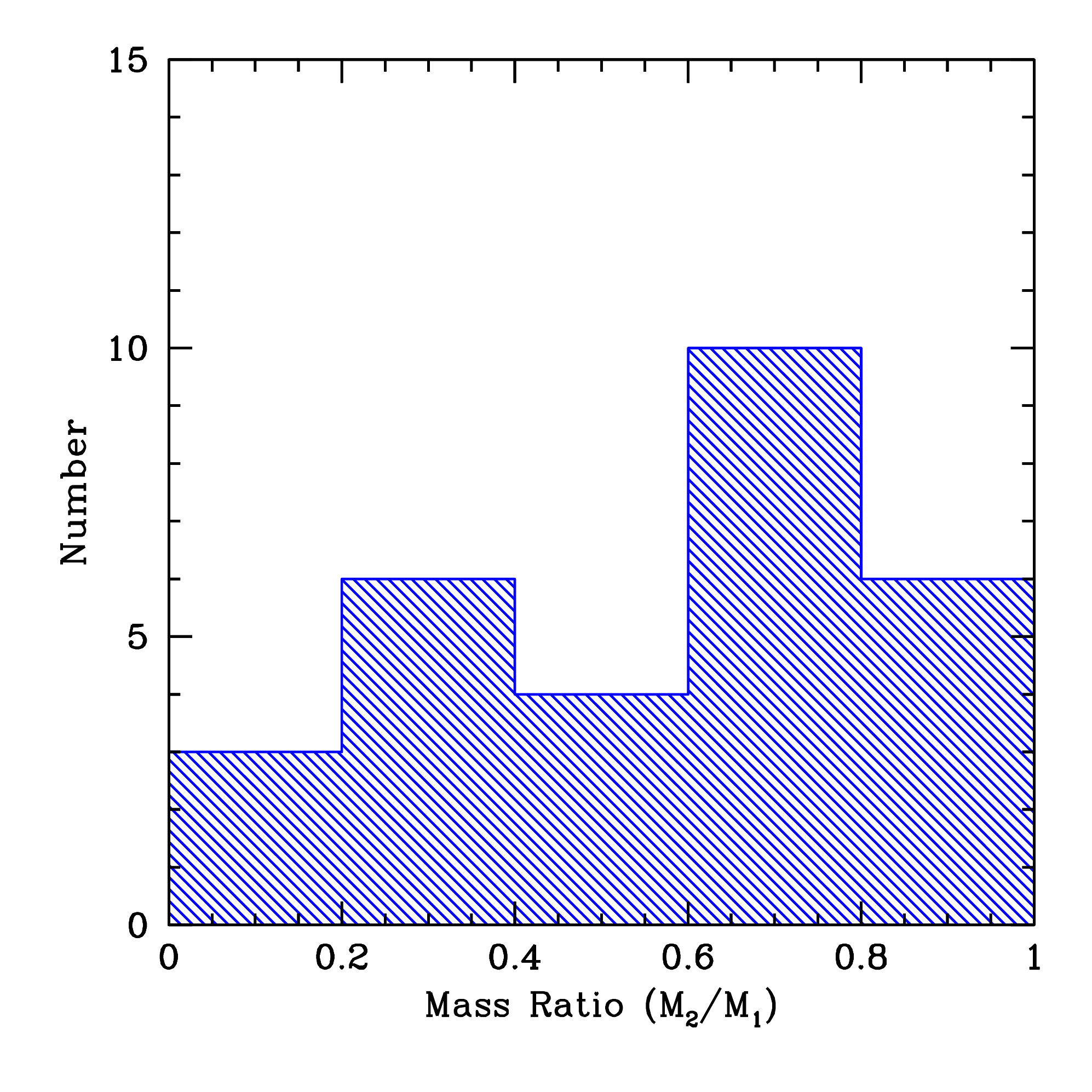}
    \includegraphics[width=8.4cm]{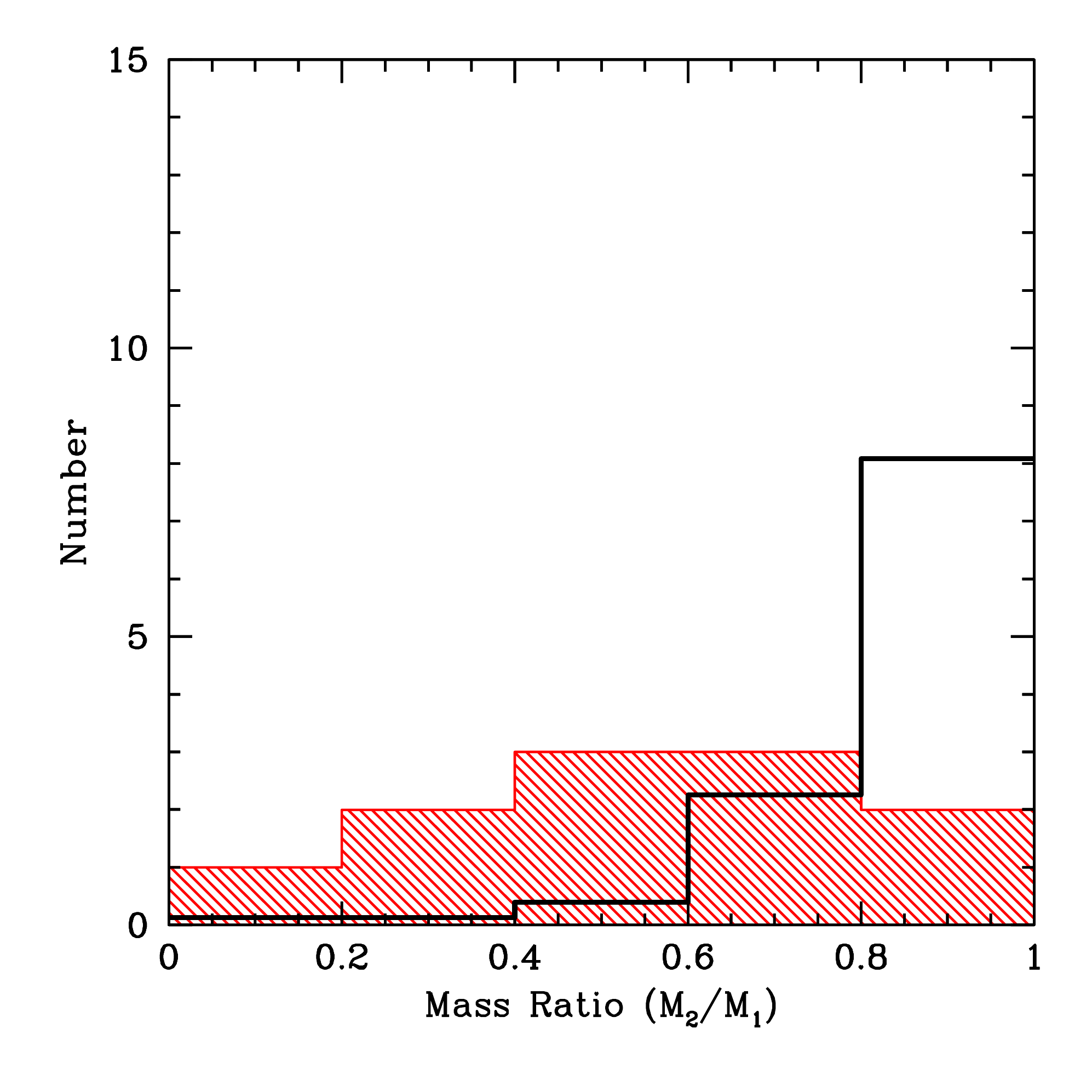}
    \includegraphics[width=8.4cm]{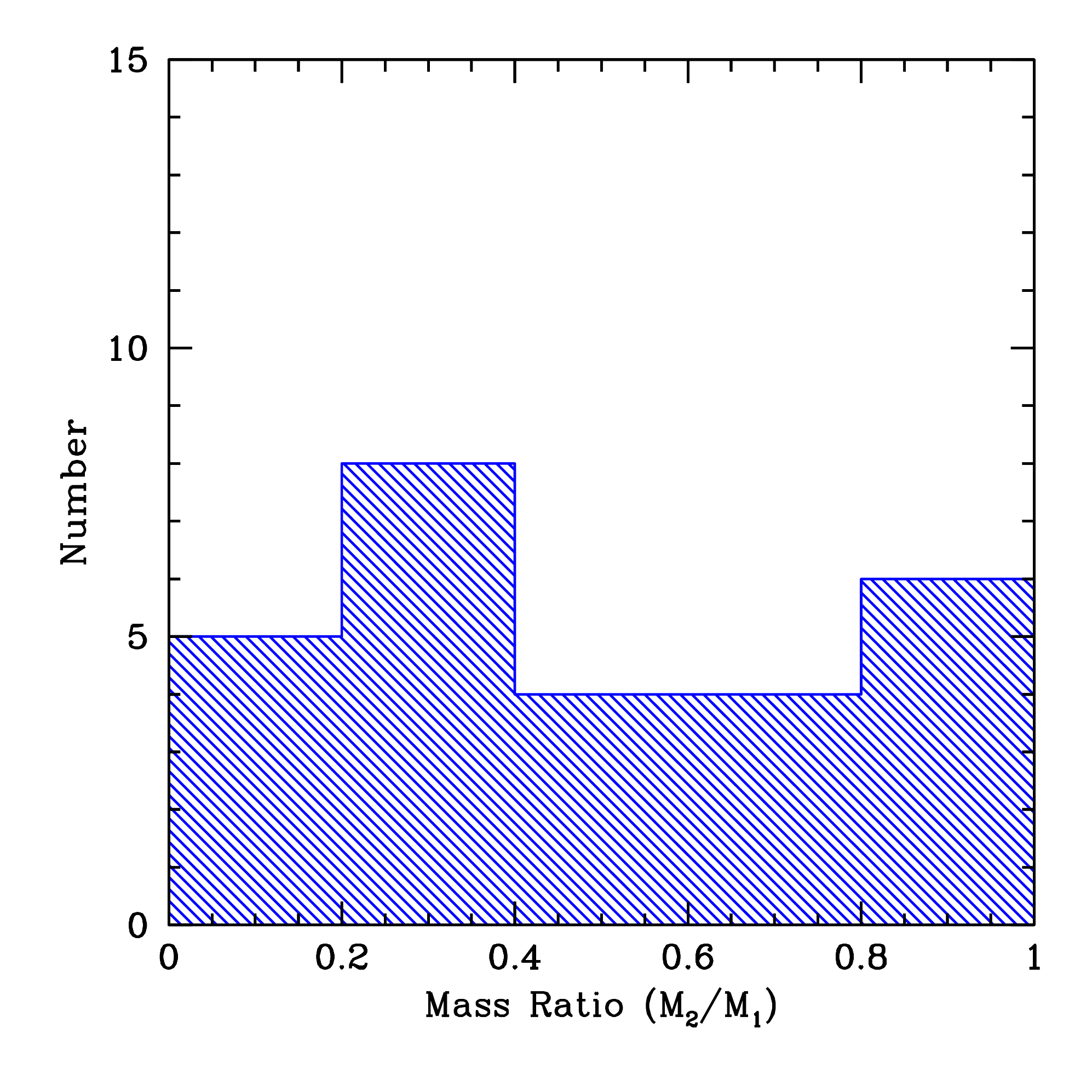}
    \includegraphics[width=8.4cm]{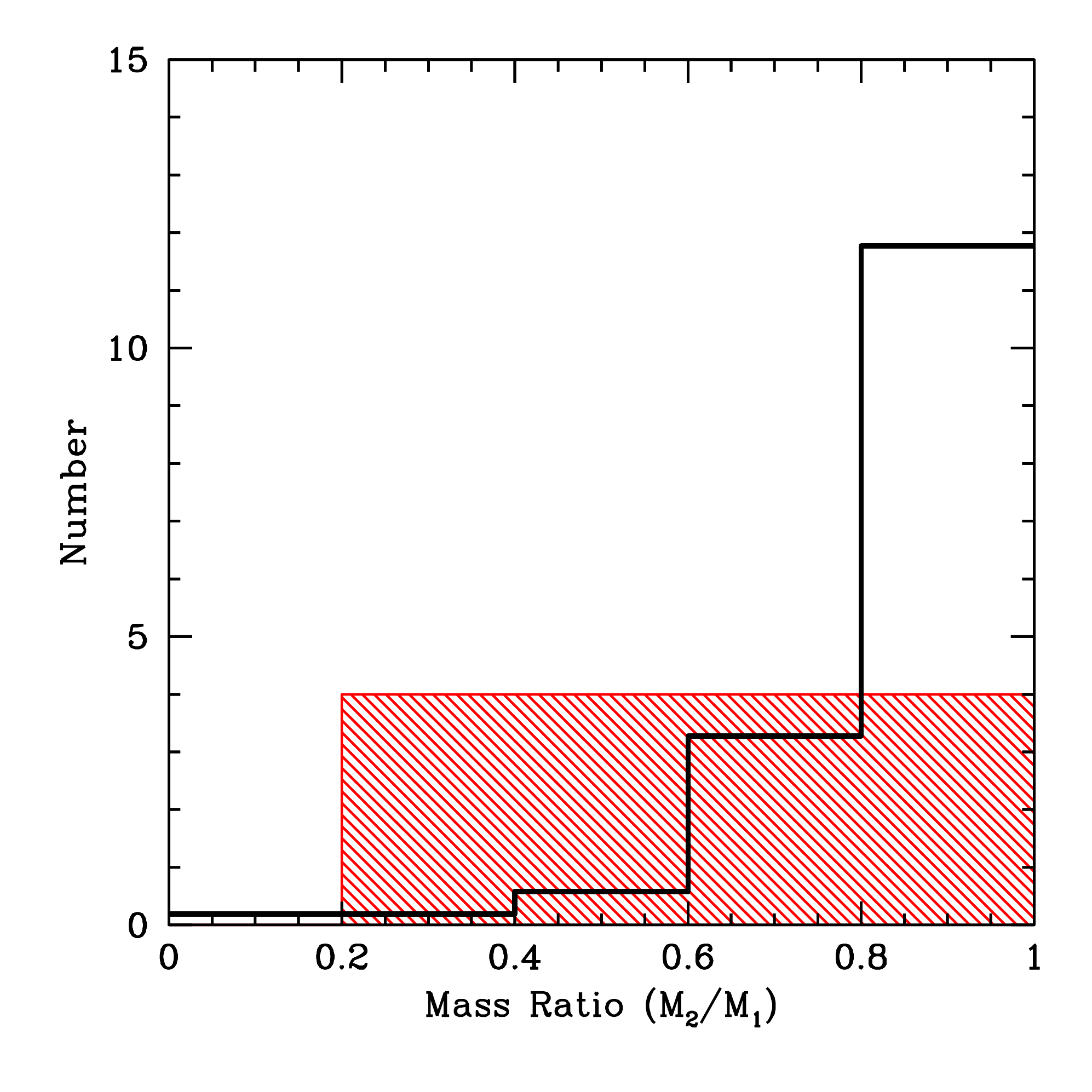}
\caption{The mass ratio distributions of binary systems with stellar ($M_1>0.1$ M$_\odot$; left) and VLM (right) primaries produced by the main calculation (upper panels) and re-run calculation (lower panels), both at $t=1.038t_{\rm ff}$.  In the VLM graphs, the open black histogram gives the mass ratio distribution of the known VLM multiple systems maintained by N. Siegler at http://vlmbinaries.org/  (scaled to match the total number) .  The frequency of VLM binaries is higher in the re-run calculation, but the mass ratio distributions of both stars and VLM objects are indistinguishable given the small number statistics.  Comparing the VLM panels with that in Figure \ref{massratios} there is evidence that the VLM binaries begin with more uniform mass ratio distributions and evolve towards equal-masses as the main calculation proceeds.}
\label{massratios:re-run}
\end{figure*}

With 58 stellar and 32 VLM binaries we can, for the first time, study the properties of a reasonably large sample of binary systems formed in a single star cluster. The main calculation also produced 19 stellar and 4 VLM triple systems and 23 stellar and 2 VLM quadruple systems.

Observationally, the median separation of binaries is found to depend on primary mass.  \citet{DuqMay1991} found that the median separation of solar-type binaries was $\approx 30$ AU.  \citet{FisMar1992} found indications of a smaller median separation of $\approx 10 $ AU for M-dwarf binaries.  Finally, VLM binaries are found to have a median separation $\lsim 4$ AU \citep{Closeetal2003, Closeetal2007, Siegleretal2005}, with very few VLM binaries found to have separations greater than 20 AU, particularly in the field \citep{Allenetal2007}.  Most recently, \citet{Closeetal2007} estimated that young VLM objects have a wide ($>100$ AU) binary frequency of $\sim 6$\%$\pm$3\% for ages less than 10 Myr, but only 0.3\%$\pm$0.1\% for field VLM objects.

Unfortunately, in the main calculation the gravitational force between sink particles is softened when they approach within 4 AU with the maximum acceleration, and hence the minimum binary separation, occurring at 1 AU.  Furthermore, gas within 5 AU of a sink particle is accreted, meaning that dissipative interactions with the gas are omitted on these scales.  These numerical approximations necessarily affect the formation of the multiple systems.  In the re-run calculation, no gravitational softening is applied and binaries with separations as small as 0.02 AU could be produced.  However, the sink particles still accrete gas within 0.5 AU which is likely to affect binary formation and smaller numbers of multiple systems are produced in the re-run calculation giving poorer statistics.

In Figure \ref{separations} we present the separation (semi-major axis) distributions of the stellar (primary masses greater than 0.10 M$_\odot$) and VLM multiples.  These distributions are compared with the surveys of \citet{DuqMay1991}, \citet{FisMar1992} and the listing of VLM multiples maintained by N.\ Siegler at http://vlmbinaries.org/, respectively.  The filled histograms give the separations of binary systems, while the double hashed region adds the separations from triple systems (two separations for each triple, determined by sub-dividing the triple into a binary with a wider companion), and the single hashed region includes the separations of quadruple systems (three separations for each quadruple which may be composed of two binary components or a triple with a wider companion).

We find that in the main calculation the median separation (including separations from binary, triple, and quadruple systems) increases with increasing primary mass.  The stellar systems have a median separation of 26 AU while the VLM systems have a median separation of 10 AU.  These values are in reasonable agreement with the observed values mentioned above, and the shapes of the separations distributions for stellar and VLM primaries are satisfactory (at least beyond 10 AU).  However, it is also clear from Figure \ref{separations} that the resolution limits imposed by the sink particle approximations (vertical dotted lines) almost certainly affect the distributions since the peaks of both the stellar and VLM distributions occur in the $1-10$ AU separation bin.

To investigate the effects of the sink particle approximations on the distributions, in Figure \ref{separations:re-run}, we display the stellar and VLM separation distributions from the re-run calculation ({\it lower} panels) and the main calculation at the same time ($t=1.038 t_{\rm ff}$; upper panels).  As expected, reducing the sink particle accretion radii and gravitational softening produces closer multiple systems.  The effect on the stellar distribution is particularly pleasing in that the separation distribution becomes more bell-like and the peak occurs in the 10-100 AU bin (rather than the 1-10 AU bin) which is well separated from the resolution limit (vertical dotted line).

More VLM multiple systems are formed in the re-run calculation and there are more with separations $<10$ AU.  Of even more interest is the fact that, at $t=1.038 t_{\rm ff}$, the median separations of the VLM multiples in the main calculation and the re-run calculation are similar to each other and similar to the stellar multiples, but much larger than at the end of the main calculation ($\approx 30$ AU at early times, but $\approx 10$ AU at the end of the main calculation).  Admittedly, the smaller numbers of VLM multiples at early times means that the uncertainties are large.  However, this indicates that {\it VLM systems may form with reasonably wide separations and evolve to smaller separations}.  We note that at $t=1.038t_{\rm ff}$ two thirds of the VLM multiples in the main calculation and more than 80\% of those in the re-run calculation are still accreting (and, thus, still evolving) whereas at the end of the main calculation all but 1 VLM multiple has ceased accreting.  \citet{BatBonBro2002b} discuss how close binaries (separations less than 10 AU) are formed from wider systems in the BBB2003 calculation through a combination of dynamical encounters with other protostars, their interactions with circumbinary and circumtriple discs, and accretion.  Since the main calculation is simply a larger version of BBB2003's calculation, it is probable that such evolution is also occurring here.  The possibility of VLM binaries undergoing evolution has also been suggested observationally.  \cite{Closeetal2007} and \cite{Burgasseretal2007} suggest that young wide VLM binaries are disrupted, leading to the observed paucity of old wide VLM systems.  They also find evidence that a higher proportion of young VLM systems may have unequal-mass components than for older systems (see also the next section).

\subsection{Mass ratio distributions of binaries}
\label{sec:massratios}

Along with the separation distributions of the multiple systems we can investigate the mass ratio distributions.  In this section we only consider binaries, but we include binaries that are components of triple and quadruple systems.  A triple system composed of a binary with a wider companion contributes the mass ratio from the binary, as does a quadruple composed of a triple with a wider companion.  A quadruple composed of two binaries orbiting each other contributes two mass ratios - one from each of the binaries.

Observationally, the mass ratio distribution of binaries also is found to depend on primary mass.  \citet{DuqMay1991} found that the mass ratio distribution of solar-type binaries peaked at $M_2/M_1 \approx 0.2$.  \citet{Halbwachsetal2003} found a bi-modal distribution for spectroscopic binaries with primary masses in the mass range $0.6-1.9$ M$_\odot$ and periods $\lsim 10$ years with a broad peak in the range $M_2/M_1=0.2-0.7$ and a peak for equal-mass systems (so-called twins; \citealt{Tokovinin2000b}).  They also noted that the frequency of twins was higher for periods $<100$ days, though this is not relevant for the calculations presented here since they do not probe such short periods.   \citet{Mazehetal2003} found a flat mass ratio distribution for spectroscopic binaries with primaries in the mass range $0.6-0.85$ M$_\odot$.   \citet{FisMar1992} also found a flat mass ratio distribution in the range $M_2/M_1 = 0.4-1.0$ for M-dwarf binaries with all periods.  Finally, VLM binaries are found to have a strong preference for equal-mass systems \citep{Closeetal2003, Siegleretal2005,Reidetal2006}.

In Figure \ref{massratios}, we present the mass ratio distributions of the stars with masses $\geq 0.5$ M$_\odot$ (left panel), M-dwarfs with masses $0.1\leq M<0.5$ M$_\odot$ (centre panel), and VLM objects (right panel).  We compare the M-dwarf mass ratio distribution to that of \citet{FisMar1992}, and the higher mass stars to the mass ratio distribution of solar-type stars obtained by \citet{DuqMay1991}.  The VLM mass ratio distribution is compared with the listing of VLM multiples maintained by N.\ Siegler at http://vlmbinaries.org/.  

We find that in the main calculation the ratio of near-equal mass systems to systems with dissimilar masses decreases going from VLM objects to M dwarfs in a similar way to the observed mass ratio distributions, but that the trend is not as strong as in the observed systems.  Specifically, 71\% of the VLM binaries have $M_2/M_1>0.6$ while for primary masses $0.1-0.5$ M$_\odot$ the fraction is only 51\%.  The stellar mass ratio distribution is consistent with Fischer \& Marcy's distribution.  The VLM binaries, although biased towards equal-mass systems, are not as strongly biased as is observed.  However, currently there is no volume-limited sample for VLM systems and systems with more equal-mass components are easier to detect so the degree to which the observed mass ratio distribution might be affected by selection effects is not yet clear.

What is clear, however, is that the mass ratios of binaries with primary masses greater than 0.5 M$_\odot$ do not agree with Duquennoy \& Mayor's mass ration distribution.  Of the 34 binaries, only 10 have mass ratios less than $M_2/M_1=0.5$.

In Figure \ref{massratios:re-run}, we display the stellar (primary masses $>0.1$ M$_\odot$) and VLM mass ratio distributions from the re-run calculation ({\it lower} panels) and the main calculation at the same time ($t=1.038 t_{\rm ff}$; upper panels).  The stellar mass ratio distributions are not significantly different from each other or from Figure \ref{massratios}.  However, the VLM binary mass ratio distributions at early times (for both the main and re-run calculations) are flatter than that obtained at the end of the main calculation.  {\it Again, this implies that the properties of the VLM binaries evolve.}  Both the apparent evolution of VLM binary separations and mass ratios are consistent with the evolution discussed by \citet{BatBonBro2002b}.  Dynamical exchange interactions between binaries and single objects tend to produce more equal-mass components, as does accretion of gas from circumbinary discs or the accretion of infalling gas with high specific angular momentum.  Thus, the apparent evolution of both the VLM binary separations and mass ratios may be due to evolution during their formation.

\subsubsection{Mass ratio versus separation}

\begin{figure}
\centering
    \includegraphics[width=8.4cm]{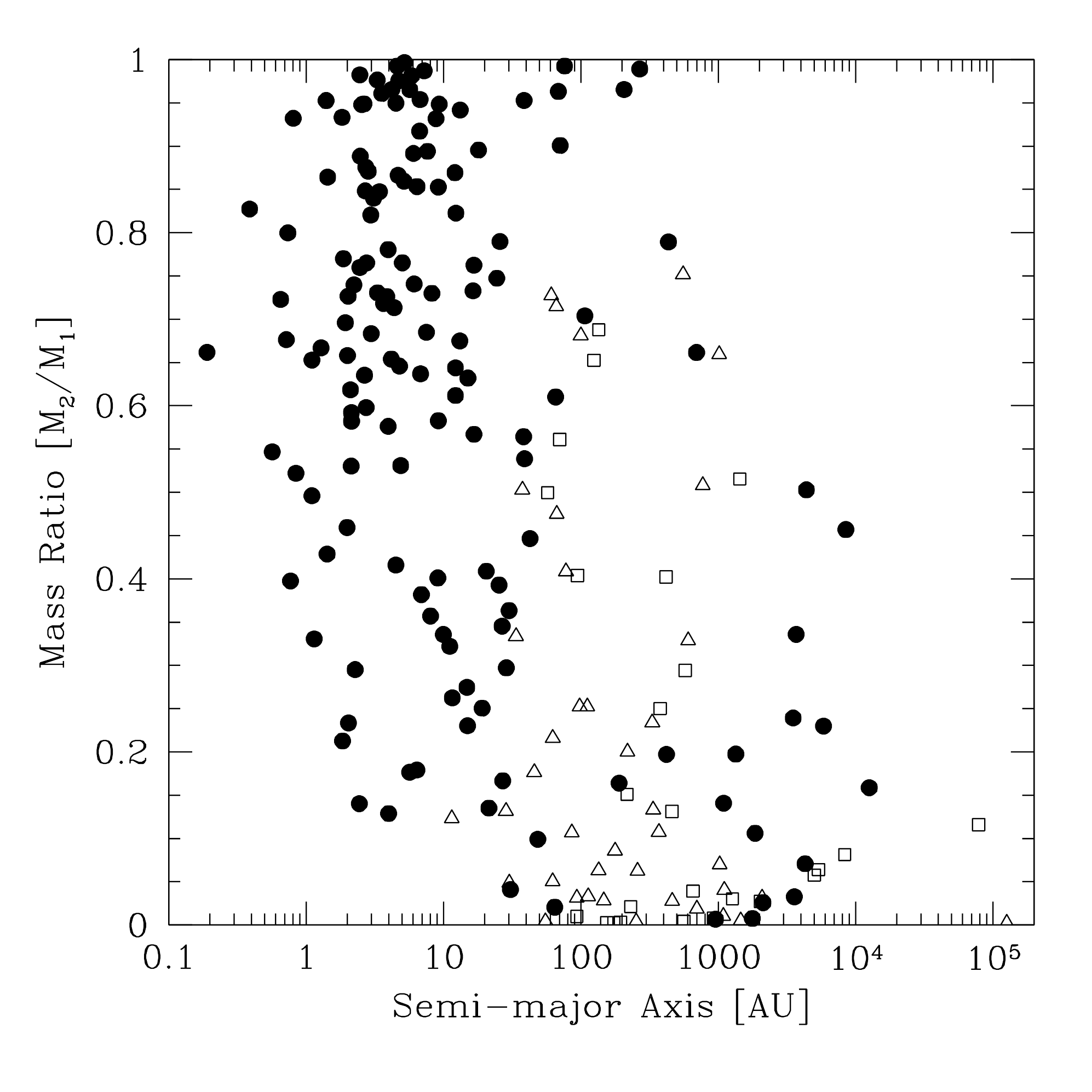}
\caption{The mass ratios of binaries (filled circles), triples (open triangles), and quadruples (open squares) as a function of semi-major axis for the main calculation.  For triples, the mass ratio compares the mass of the widest component to the sum of the masses of the two closest components.  For quadruples involving a two binary components, the mass ratio is between the two binaries, and for quadruples involving a triple, the mass ratio is between the mass of the fourth component and the triple.  All mass ratios are defined to be $\leq 1$. There is a clear relationship between mass ratio and separation with closer binaries having a greater fraction of near equal-mass systems.}
\label{a_q}
\end{figure}

In Figure \ref{a_q}, we plot mass ratios against separation (semi-major axis) for the binaries, triples, and quadruples at the end of the main calculation.  Note that for this figure we include systems that are sub-components of higher-order systems.  Thus, the closest two objects in a triple also appear in the plot as a binary.  Similarly, for quadruples consisting of two binary sub-components, each of the binaries appears in the plot and for each of the quadruples that involves a triple system, the triple appears in the plot.

There is clearly a relation between mass ratio and separation for the binaries with closer systems having a preference for equal masses.  The median mass ratios for {\it binary} separations in the ranges $1-10$, $10-100$, $100-1000$ and $1000-10^4$ AU are $M_2/M_1=0.74, 0.57, 0.68, 0.17$, respectively.  Including the mass ratios of triples and quadruples (as defined in the caption of Figure \ref{a_q}), these median values become 0.74, 0.41, 0.15, and 0.07, respectively.  The median mass ratio for triples is 0.11 and the median mass ratio for quadruples is 0.07.  However, the quadruples include those composed of two binaries and those composed of a triple and a fourth wide component.  The mass ratios of the latter tend to be much lower than those of the former.  There are 8 quadruples composed of two binaries and 16 composed of triples and a fourth component.  The median mass ratios for these two sub-samples are 0.45 and 0.03 respectively.  There are also only 11 (out of 40) triples composed only of stars (as opposed to containing VLM objects).  For these, the median mass ratio is 0.48.  All but one of the quadruple systems contains at least one VLM object.

A trend of more unequal mass binaries with increasing separation is expected from the evolution of protobinary systems accreting gas from an envelope \citep{Bate2000}.  Furthermore, dynamical interactions between binaries and single stars tend to tighten binaries at the same time as increasing the binary mass ratio through exchange interactions.

Observationally, closer binaries are found to have a higher fraction of `twins' \citep{Tokovinin2000b, Soderhjelm1997, Halbwachsetal2003}.  \citet{Tokovinin2000b} found evidence for the frequency of twins falling off for orbital periods greater than 40 days, but \citet{Halbwachsetal2003} found that the fraction of near equal-mass systems ($M_2/M_1>0.8$) is always larger for shorter period binaries than longer period binaries regardless of the dividing value of the period (from just a few days up to 10 years).  However, despite the fact that the fraction of twins decreases with increasing separation, the mass ratio distributions of both short-period and long-period binaries appear to have a peak at $M_2/M_1=1$ \citep[e.g.][]{Tokovinin2000b, Halbwachsetal2003, Soderhjelm2007}.   These observed relations are in qualitative agreement with the decreasing median mass ratio with increasing separation discussed above.  In Figure \ref{a_q}, we also note that although there is a higher fraction of twins at small separations, there are still some wide twins (separations $30-300$ AU).  

For stellar triple and quadruple systems, \citet{Tokovinin2008} reports that triples are observed to have a median mass ratio of 0.39 independent of the outer orbital period while quadruples involving two binary sub-components have a similar median mass ratio of $\approx 0.45$, but there may be a dependence on the outer orbital period.  The median mass ratio of the triples systems from the main calculation are in agreement with observations, as long as we only consider the triples containing stellar components (no VLM components).  This is consistent with the observational sample, but it does raise the question of how many triple systems containing VLM components exist in reality.  Similarly, the median mass ratio of quadruples containing two binary sub-systems is in good agreement with observations, but all but one of the systems from the main calculation includes a VLM object whereas the observational sample is dominated by stellar-only systems.  It is also interesting to note that quadruples composed of a triple and a wide fourth component out number quadruples composed of two binaries by 2:1 in the main calculation.  \cite{Tokovinin2000a} finds roughly equal numbers of such quadruples.  However, if the wide components of quadruples containing triples as sub-components typically have low masses this could be attributed to observational bias.

For the binaries, the clear trend of decreasing mass ratio with separation may go some way to explaining the apparent deficit of unequal-mass binaries with primary masses greater than 0.5 M$_\odot$ in the main calculation (left panel of Figure \ref{massratios}).  It is clear from Figures \ref{separations} and \ref{a_q} that the main calculation does not produce many wide pure binaries -- most of the wide systems are triples or quadruples and the binaries components within them necessarily have smaller separations than the wide tertiary or quartic components.  Since the mass ratio distributions in Figure \ref{massratios} only contain binary mass ratios an unequal-mass visual binary may in fact be composed of an undetected close binary and a wider companion.  However, while an observer of the system would include the unequal mass ratio of the wide system, only the mass ratio of the close binary component would be included in a mass ratio distribution like Figure \ref{massratios}.

Therefore, one way to reconcile the main calculation with observations may be to include the mass ratios of tertiary and quartic components.  The problem with this is there is no unique way to do this -- should the mass ratio of a triple be simply the ratio of the total mass of the binary to the third component?  Should an attempt be made to model the luminosities of the two stars in the binary?  What if the ratio of the two separations is small so that if an observer identified it as a binary they would also have been likely to separate it into a triple?  Furthermore, \citet{DuqMay1991} actually found a rather low frequency of triple and higher-order systems anyway, so perhaps the question of how to treat these higher order systems is not important.  On the other hand, discussion continues as to how many triples and quadruples were missed by this and other surveys.

For the moment, we conclude that the main calculation appears to under-produce unequal-mass solar-type binaries compared with observations.  However, this may at least be partially reconciled if many of the observed binaries are in fact higher-order systems or, alternately, if the mass ratios of tertiary and quartic components from the main calculation are included in the statistics.  There is much less of a difference between observations and the main calculation for binaries with M dwarf primaries or VLM binaries simply because 
\begin{description}
\item[a)] the frequency of higher-order systems decreases rapidly with decreasing primary mass (Section \ref{freq_high_order}) so the issue of how to treat higher-order systems does not arise, and
\item[b)] the typical separation of binaries decreases with decreasing primary mass (Section \ref{sec:separations}) so the wider systems that tend to have more unequal masses are much less frequent for low primary masses.
\end{description}

\subsection{Orbital eccentricities}

\begin{figure}
\centering
    \includegraphics[width=7.2cm]{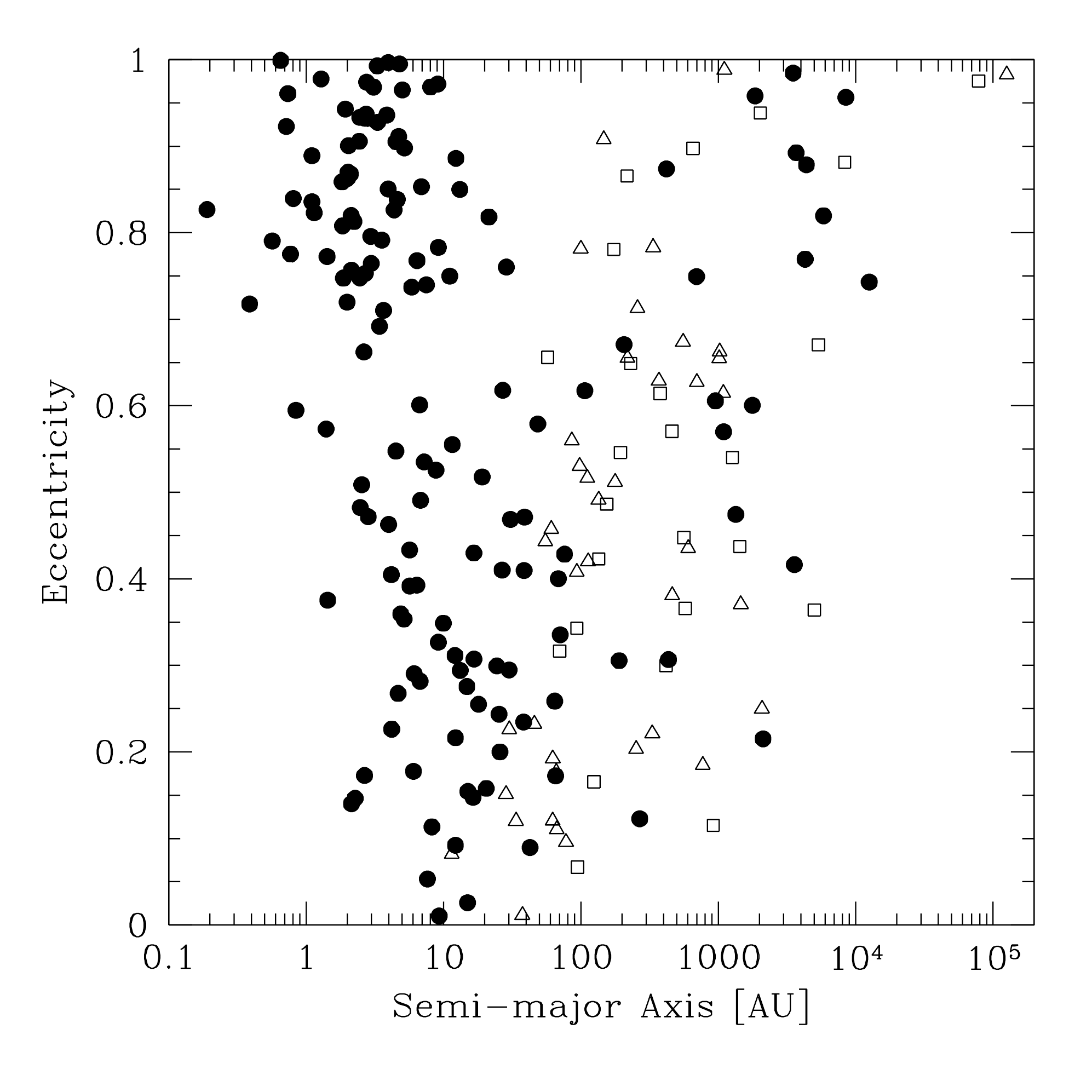}\vspace{-0.8cm}
    \includegraphics[width=7.2cm]{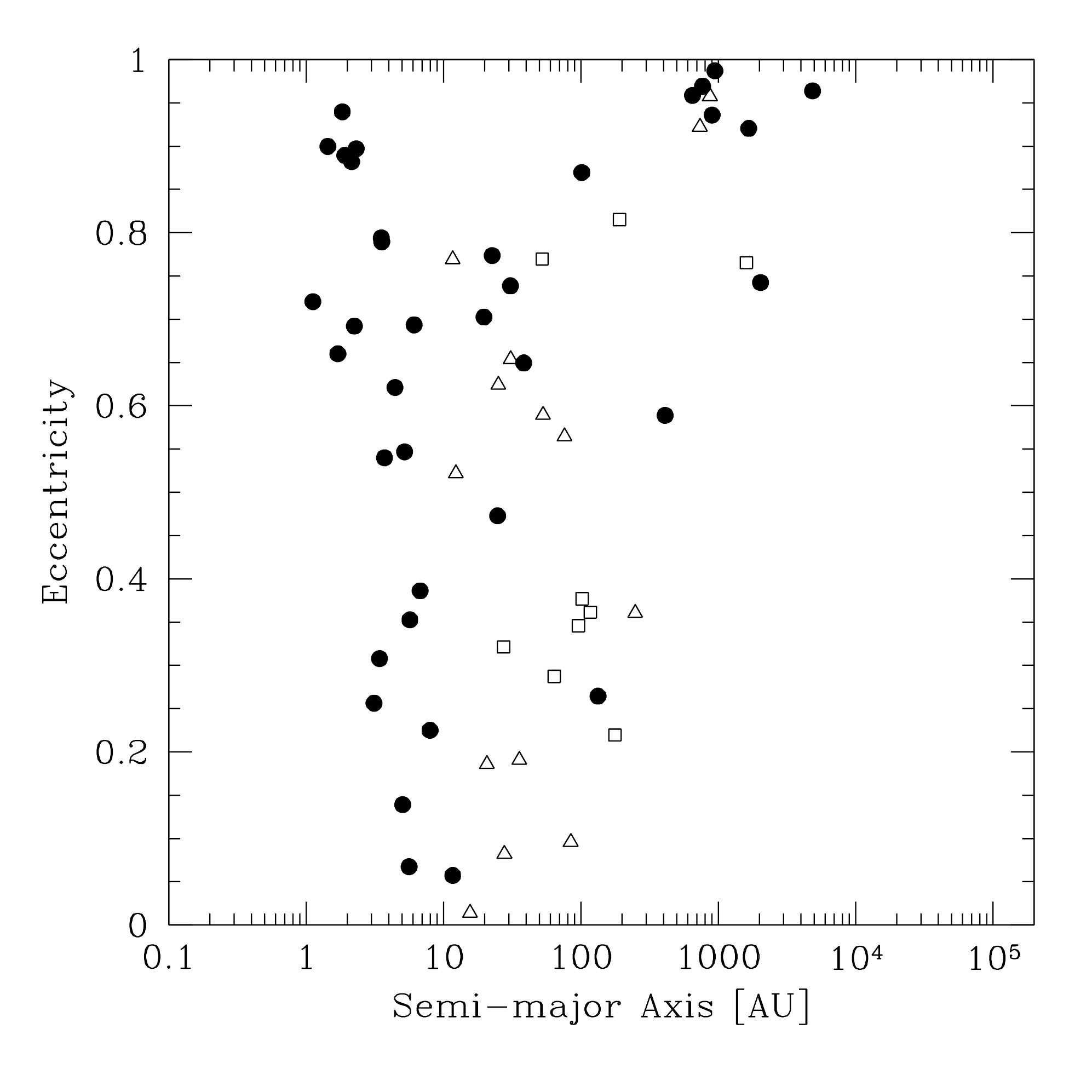}\vspace{-0.8cm}
    \includegraphics[width=7.2cm]{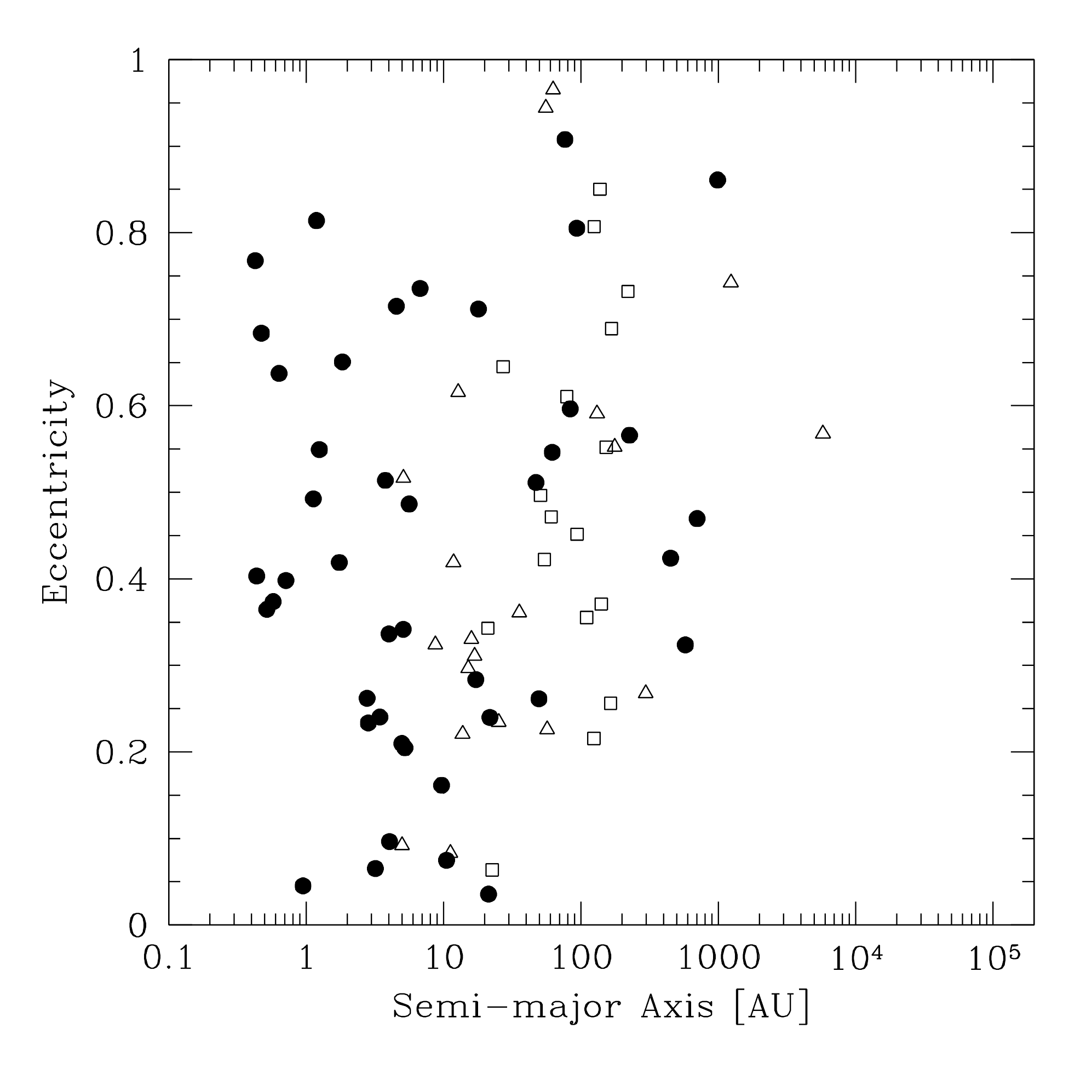}\vspace{-0.3cm}
\caption{The eccentricity distribution of binary (filled circles), triple (open triangles), and quadruple orbits (open squares) as a function of semi-major axis for the main calculation at the end (top panel) and at $t=1.038t_{\rm ff}$ (centre panel) and for the re-run calculation (lower panel).  The distribution at the end of the main calculation looks reasonable except for the group of binaries with semi-major axes less than $\sim 10$ AU and eccentricities $e\gsim 0.7$.  These systems would presumably have smaller eccentricities if the gas dynamics inside 5 AU of each sink particle were modelled.  This is tested by comparing the main calculation with the re-run calculation at $t=1.038t_{\rm ff}$ (the lower two panels).  As expected, although the main calculation still has a group of highly-eccentric close binaries, these systems are absent in the re-run calculation.
}
\label{eccentricity}
\end{figure}

Observationally, there is observed to be an upper envelope to binary eccentricities at periods less than  a few years \citep{DuqMay1991, Halbwachsetal2003}.  However, the main calculation does not allow us to probe such small separations.  Observations also indicate that eccentricities $e<0.1$ are rare for periods greater than $\approx 100$ days (separations $\gsim 1$ AU).  Finally, \citet{Halbwachsetal2003} find that the eccentricities of so called `twins' (binaries with mass ratios $M_2/M_1>0.8$) with periods greater than $\approx 10$ days (the tidal circularisation radius) are lower than for more extreme mass ratio systems.

In the upper panel of Figure \ref{eccentricity} we plot the eccentricities versus semi-major axes of the orbits of the binaries, triples and quadruples from the main calculation.  The distribution of eccentricities looks reasonable for separations greater than 10 AU.  In particular, of the 122 orbits with separations greater than 10 AU there are only 7 orbits with $e<0.1$ and these all have separations between 10 and 100 AU.  

However, there appears to be a strong excess of systems with $e>0.7$ and separations less than 10 AU.  This is almost certainly an artifact introduced by the sink particle approximation.  The absence of gas closer than 5 AU from a sink particle means that dissipative interactions between binary stars and the gas orbiting them are absent.  In Figure \ref{eccentricity}, we also plot eccentricity versus semi-major axis for the orbits of binaries, triples, and quadruples from the main calculation (middle panel) and re-run calculation (lower panel) at $t=1.038 t_{\rm ff}$.  The re-run calculation has no indication of the excess population at separations less than 10 AU and $e>0.7$, whereas even at this early time the main calculation has 5 binaries with separations less than 10 AU and $e>0.8$.  Thus, as expected, reducing the sink particle accretion radii allows dissipative interactions between sink particles on smaller scales and brings the calculations into better agreement with the observed eccentricity distributions.  The mean eccentricity of the systems in the re-run calculation is $\langle e \rangle=0.44$ for the binaries only and $\langle e \rangle=0.45$ if the orbits of the triples and quadruples are also taken into account.  The mean binary eccentricity is in good agreement with the observed mean eccentricities of long-period binaries \citep[periods $P\gsim 300$ days:][]{DuqMay1991,Halbwachsetal2003}.

\begin{figure}
\centering
    \includegraphics[width=7.5cm]{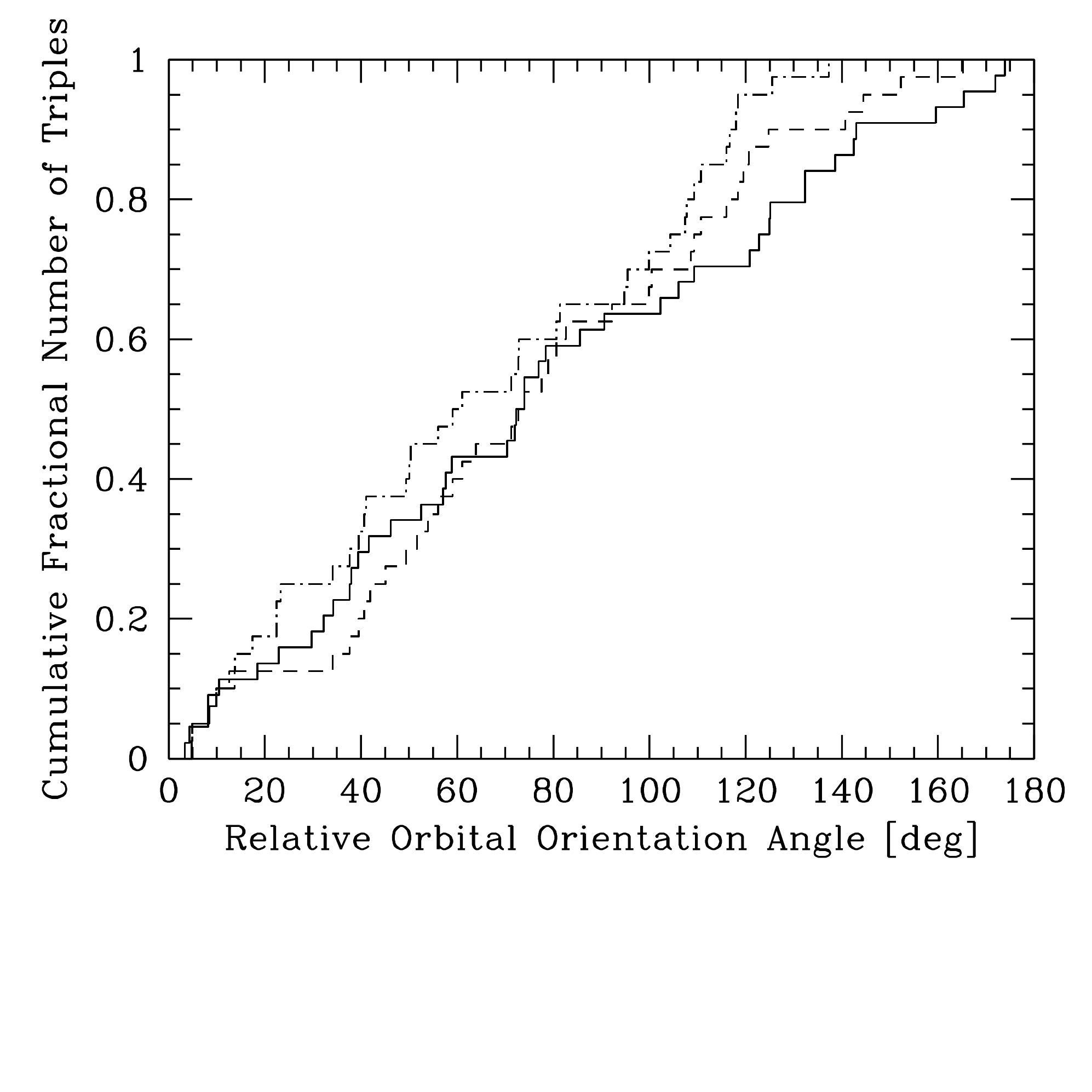}\vspace{-1.5cm}
    \includegraphics[width=7.5cm]{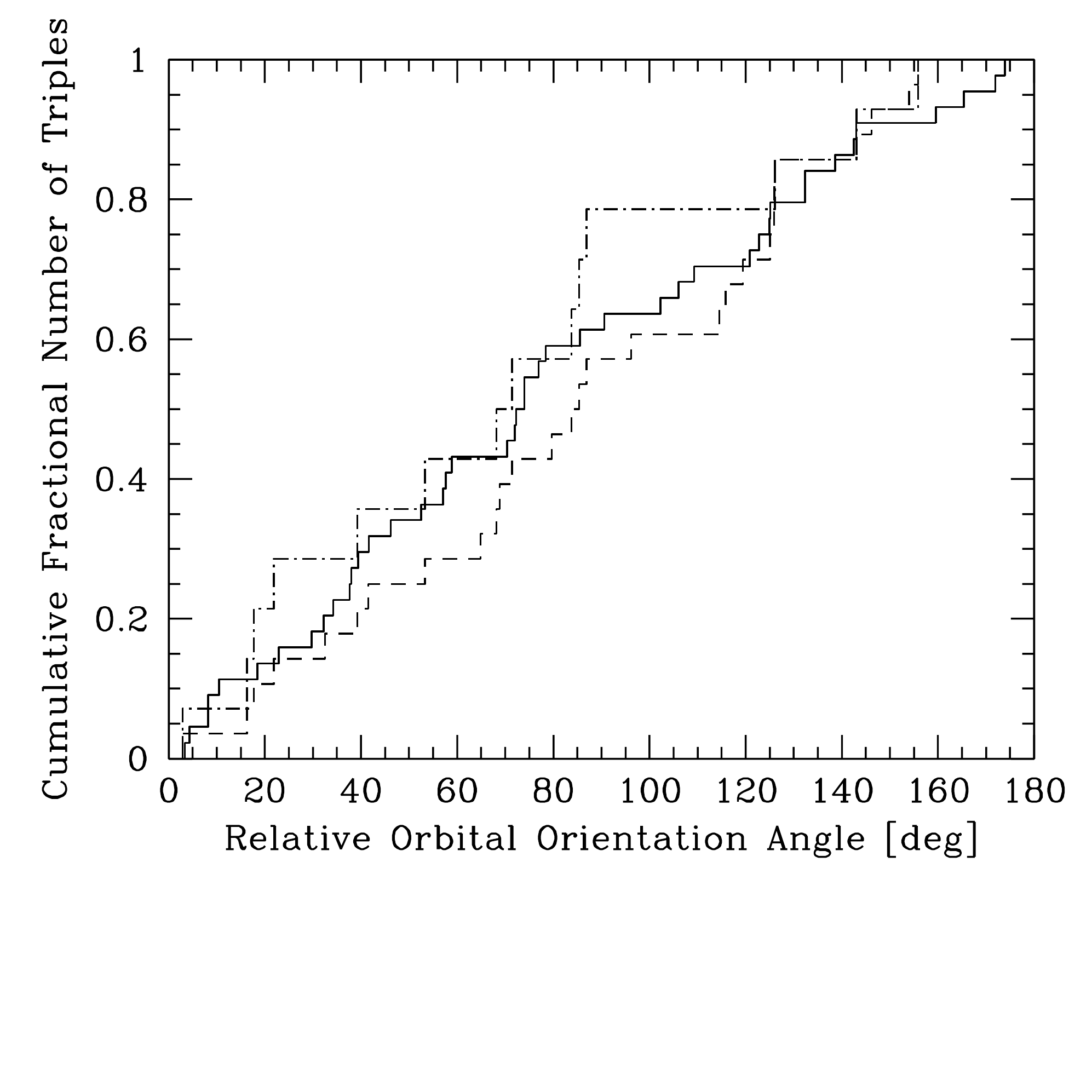}\vspace{-1.5cm}
    \includegraphics[width=7.5cm]{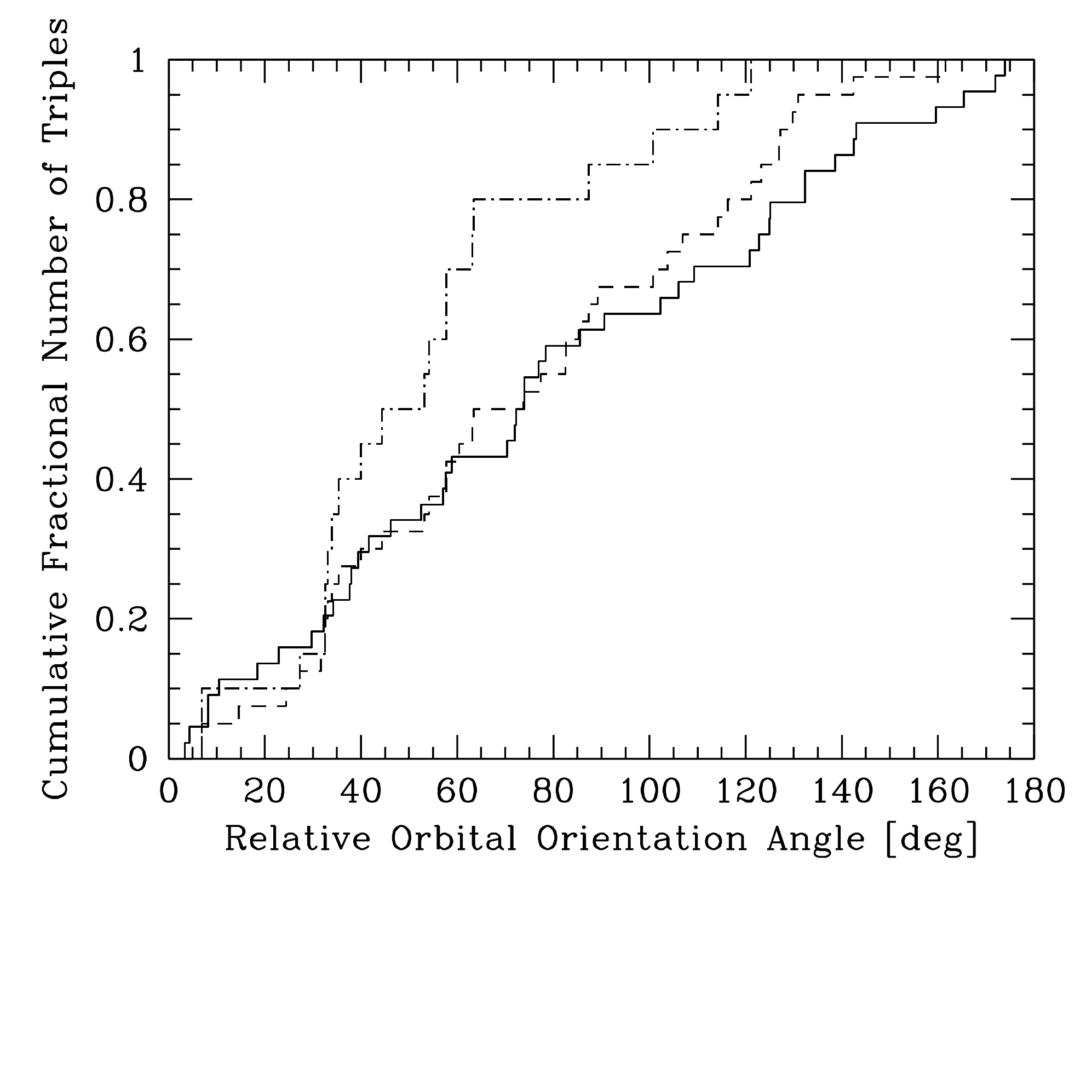}\vspace{-1.0cm}
\caption{The cumulative fraction of triples as a function of their relative orbital orientation angles at the end of the main calculation (top) and at $t=1.038 t_{\rm ff}$ for the main calculation (centre) and the re-run calculation (bottom).  In each case, the solid line gives the observed distribution of orientation angles including the $\cos(\Omega_1-\Omega_2)$ ambiguity \citep{SteTok2002}, the dot-dashed line gives actual the result from the simulation, and the dashed line gives the simulation result including the ambiguity present in the observed values. All simulated distributions are consistent with the observed distribution.  When the simulated distributions include the angle ambiguity the probabilities that they are drawn from the same population as the observed systems are 54\%, 72\%, and 66\%, respectively.  Even when the actual simulated distributions are compared with the observed distribution the probabilities are 14\%, 88\% and 3.5\%, respectively. }
\label{triples}
\end{figure}

We have also examined the dependence of the eccentricity on the mass ratio (for binary orbits only, but including binaries that are also components of higher order system) to see whether there is any sign of the tentative correlation between mass ratio and eccentricity found by \citet{Halbwachsetal2003}. For the main calculation the median eccentricity of binaries with mass ratios $M_2/M_1<0.8$ is $e=0.74$ (100 orbits) while for $M_2/M_1>0.8$ the median is $e=0.55$ (46 orbits).  Excluding orbits with separations less 10 AU (since they likely have high eccentricities due to the absence of dissipation on small scales)
 the median eccentricity of binaries with mass ratios $M_2/M_1<0.8$ is $e=0.47$ (47 orbits) while for $M_2/M_1>0.8$ the median is $e=0.37$ (10 orbits).  For $M_2/M_1>0.9$ the median is only $e=0.34$ (7 orbits).  For the re-run calculation, the statistics are that the median binary eccentricity for mass ratios $M_2/M_1<0.8$ is $e=0.45$ (33 orbits) while for $M_2/M_1>0.8$ the median is $e=0.39$ (10 orbits) and $e=0.36$ for $M_2/M_1>0.9$ (only 5 orbits).  Thus, in all cases, {\it we find evidence for a link between mass ratio and eccentricity such that `twins' have lower eccentricities, as is observed, though the effect is quite weak}.

\subsection{Relative alignment of orbital planes for triples}

For a hierarchical triple system there are two orbital planes, one corresponding to the short-period orbit and one to the long-period orbit.  There are many reasons why the inclinations of the orbital planes may not be randomly distributed relative to one another.  For example, if the triple system forms from the fragmentation of a disc around an initially single object, the orbital planes would be expected to be nearly coplanar.  If a triple system forms from a flatten core it may have preferentially aligned orbital planes.  If a triple system forms with initially non-coplanar orbital planes and subsequently accretes a lot of mass this may drive its orbital planes into closer alignment.  On the other hand, if a triple system forms from capture of a single object by a binary the orbital planes may be very misaligned.  Similarly, the wide tertiary in an initially aligned triple system may be perturbed by a passing object resulting in misaligned orbits.

Observationally, it is difficult to determine the relative orientations of the two orbits of a triple system due to the number of quantities that must be measured to fully characterise the orbits.  In particular, the relative angle between the two orbital angular momentum vectors is given by
\begin{equation}
\cos \Phi = \cos i_1 \cos i_2 + \sin i_1 \sin i_2 \cos(\Omega_1 - \Omega_2),
\end{equation}
where $i_1$ and $i_2$ are the orbital inclinations and $\Omega_1$ and $\Omega_2$ are the position angles of the lines of nodes.  The latter are only known with $180^\circ$ ambiguity unless the ascending node is identified by radial velocities.  Because for most observed triple systems the sign of the $\cos(\Omega_1 - \Omega_2)$ term is not known, there are two possible values of $\Phi$.  On the other hand, the mean value of $\Phi$ can be measured simply from knowledge of the number of co-rotating and counter-rotating systems \citep{Worley1967, Tokovinin1993, SteTok2002}.  These facts are important when we come to compare our results with observations below.  

The first studies \citep{Worley1967,vanAlbada1968} of the relative orbital orientations of triple systems found a small tendency towards alignment of the angular momentum vectors of the orbits.  Of 54 systems with known directions of the relative motions, 39 showed co-revolution and 15 counter-revolution resulting in a mean relative inclination angle of $\langle\Phi\rangle\approx 50^\circ$.  For 10 visual systems with known orbits, 5 systems were found to have $\Phi<90^\circ$, 2 had $\Phi>90^\circ$ and 3 were ambiguous.   \citet{Fekel1981} examined 20 systems with known orbits and periods of less than 100 years (for the wide orbit).  He found that 1/3 had non-coplanar orbits.  Finally, \citet{SteTok2002} performed the most detailed study to date.  From 135 visual triple systems for which the relative directions of the orbital motions are known they found $\langle\phi\rangle=67^\circ\pm 9^\circ$ and this result was also consistent with 22 systems for which the orbits were known.  They also found a tendency for the mean relative orbital angular momentum angle to increase with increasing orbital period ratio (i.e.\ systems with more similar orbital periods tend to be more closely aligned).

\begin{figure*}
\centering
    \includegraphics[width=8.4cm]{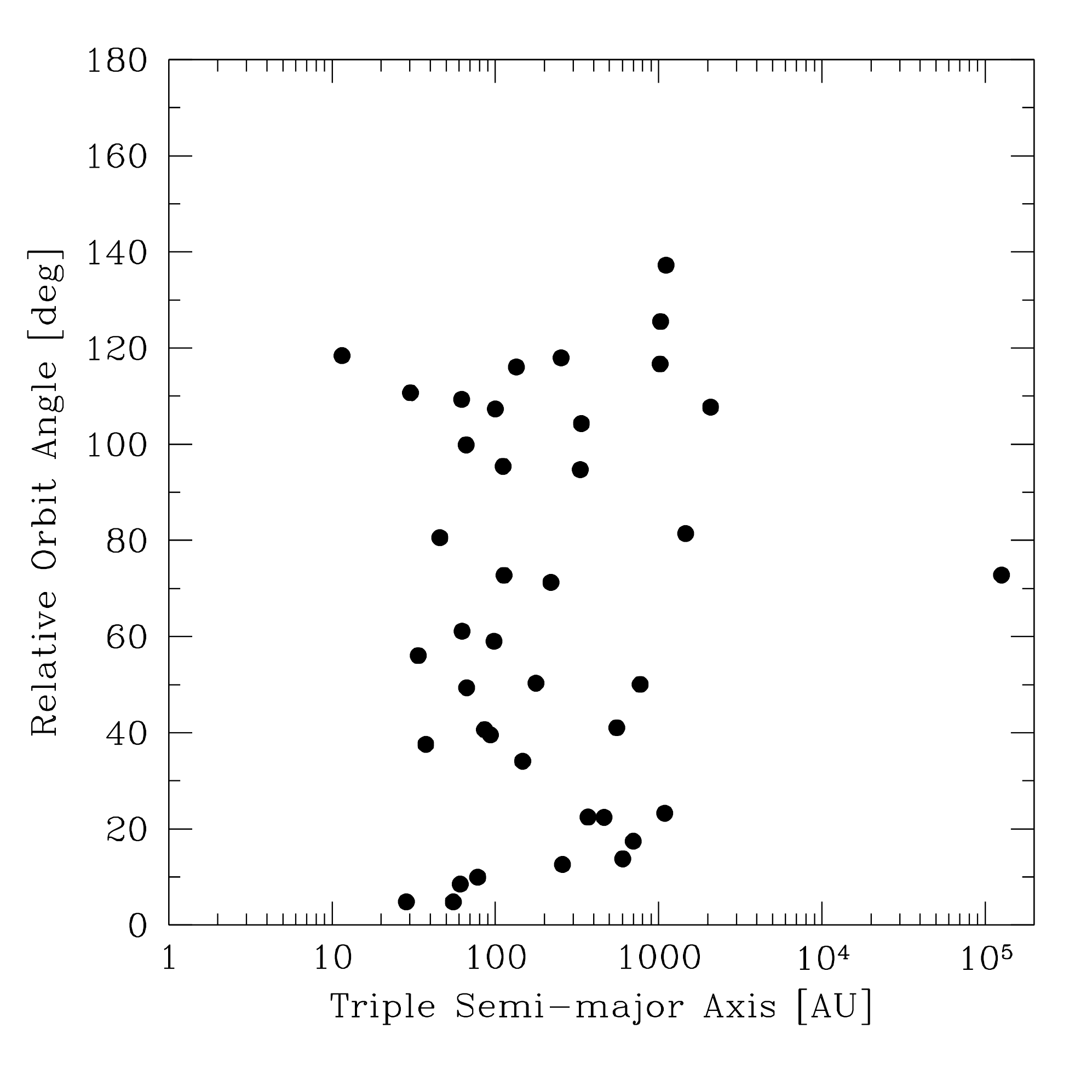}
    \includegraphics[width=8.4cm]{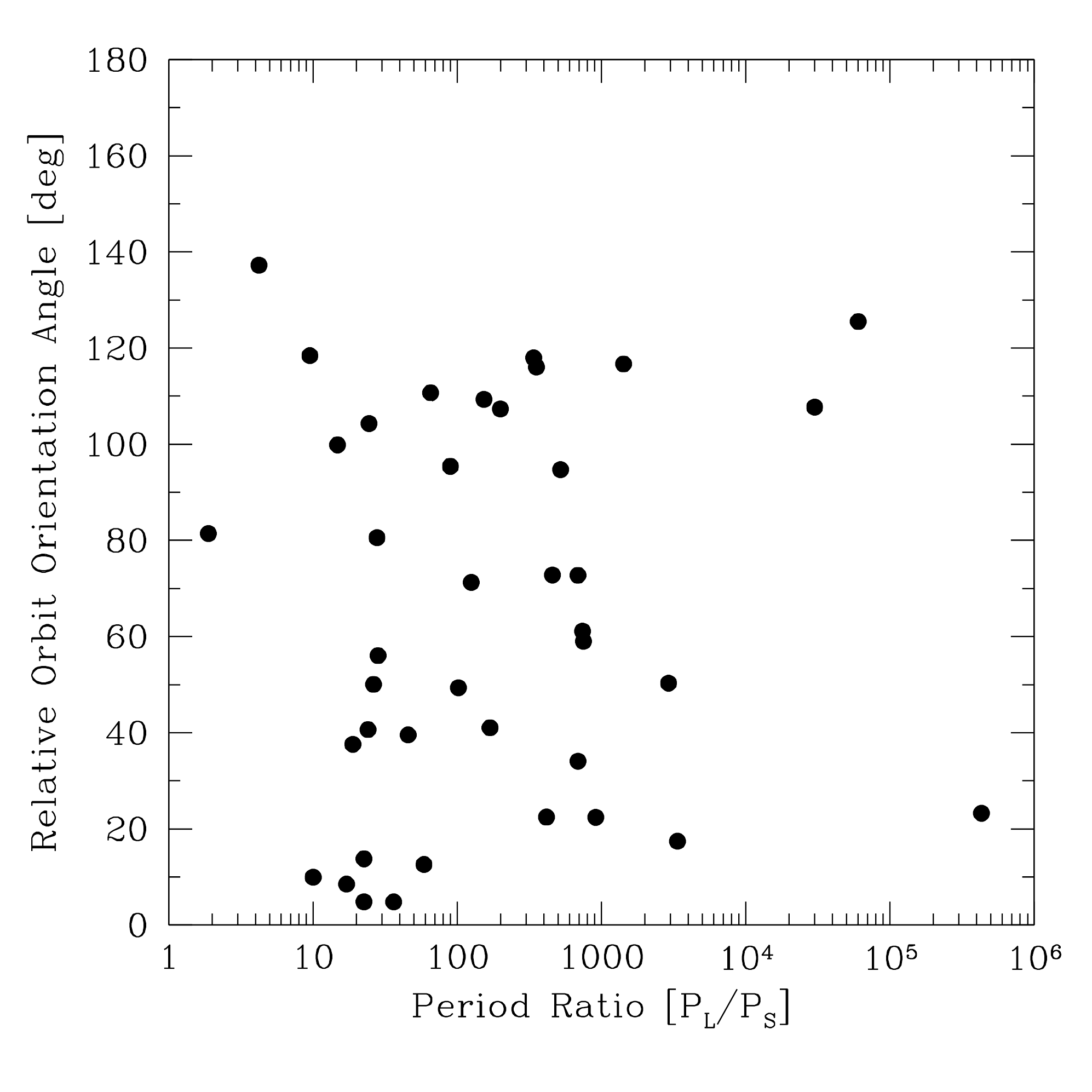}
\caption{The relative inclinations of the two orbital planes for the 40 triple systems produced by the main calculation (including those that are sub-components of quadruples). We give plots of the relative orbital orientation angle versus the semi-major axis of the third component (left) and versus the period ratio of the long and short period orbits (right).  There are no triples with relative orbital angles $>140^\circ$.   There is also the hint of an excess of systems with relative orbital angles less than $\approx 20^\circ$ for systems with period ratios less than 100.  Note that the two systems with period ratios $P_{\rm L}/P_{\rm S}<5$ are still dynamically unstable and would certainly undergo further evolution.}
\label{triples:a_periodratio}
\end{figure*}

At the end of the main calculation there are 40 triple systems (17 of these are sub-components of quadruple systems).  The mean relative orientation angle of the these systems is $\langle\Phi\rangle=65^\circ\pm 6^\circ$, in very good agreement with the observed value mentioned above.  This indicates that both the observed and simulated triple systems have a small tendency towards orbital coplanarity.  The re-run calculation and the main calculation at $t=1.038 t_{\rm ff}$ formed 20 and 14 triples with $\langle\Phi\rangle=53^\circ\pm 7^\circ$ and $\langle\Phi\rangle=69^\circ\pm 13^\circ$, respectively.  In Figure \ref{triples}, we compare the cumulative distributions of the orbital orientation angles for the triple systems of \citet{SteTok2002} with those formed by the main calculation at the end and at $t=1.038 t_{\rm ff}$, and by the re-run calculation.  The observational results (solid lines) include two angles for each observed triple system due to the ambiguity described above.  For the simulation results, we plot two cumulative distributions, one with the actual angles (dot-dashed lines) and one with two angles (dashed lines) for each triple (the true angle and the other possible angle allowed by reversing the rotation of one of the orbits).  {\it The observed and simulated distributions are in good agreement} when the angle ambiguity is included, but even without including the angle ambiguity the simulations are consistent with the observations.

\begin{figure}
\centering
    \includegraphics[width=8.4cm]{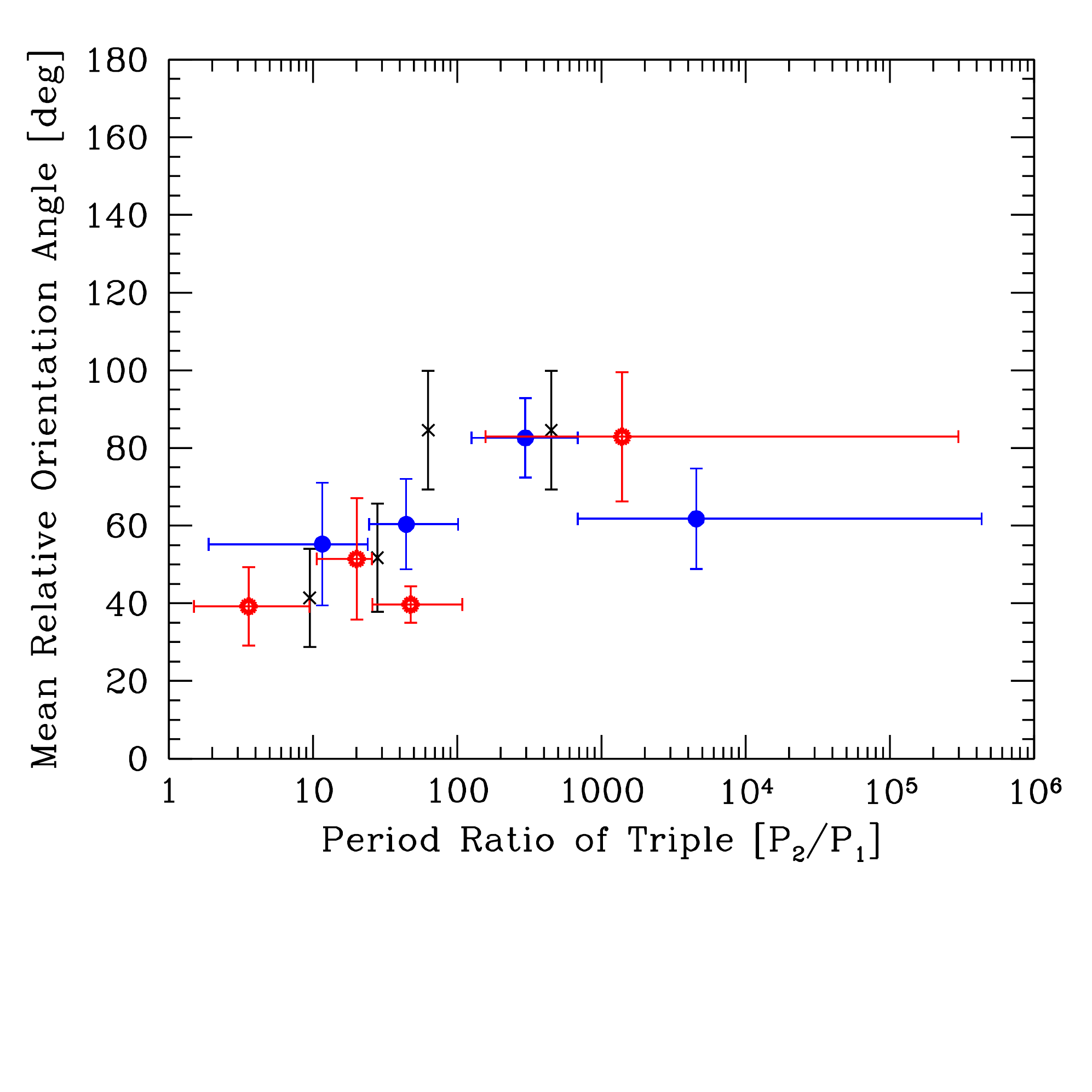}\vspace{-1.5cm}
\caption{The mean relative orbital orientation angle for triple systems. The blue filled circles give the results at the end of the main calculation with their statistical uncertainties.  The red open circles give the results from the re-run calculation.  The main calculation has not formed enough triple systems at $t=1.038 t_{\rm ff}$ to enable meaningful data to be plotted at the earlier time.  The black crosses give the observed mean angles from the Mulitple Star Catalogue as calculated by \citet{SteTok2002}.  The calculations are consistent with the observations and hint at an increasing mean orientation angle with increasing period ratio, but they are also consistent with a mean orientation angle that is independent of the period ratio.}
\label{mean_orientation}
\end{figure}

In Figure \ref{triples:a_periodratio} we plot the relative orbital orientation angle of the 40 triple systems as functions of the semi-major axis of the wide orbit and the ratio of the two orbital periods.  There is no clear correlation between the orbital orientation angle and semi-major axis or period ratio, or indeed on other quantities such as primary mass or the eccentricity of the long-period orbit.  However, although the triples are formed with a wide range of relative orbital inclinations, the absence of any angles greater than 140 degrees seems to be significant.  This implies that the triple systems are not formed purely by the capture of a third component.  We also note that there appears to be a small collection of 4 nearly coplanar triples with wide semi-major axes less than 100 AU, or 6 nearly coplanar triples with period ratios of less than 100.  This is intriguing, but unfortunately is not statistically significant.  

As mentioned above, \citet{SteTok2002} found a tendency for the mean relative orbital orientation angle to increase with increasing period ratio.  In Figure \ref{mean_orientation} we reproduce their observed results and plot the results from the main calculation and the re-run calculation.  Here we have performed averages over four groups of 10 (5 for the re-run calculation) triples, sorted by period ratio.  Our results are consistent with the observed values and there maybe a hint of a dependency on the period ratio, but our results are also consistent with no dependence.  Better statistics are required for both the simulations and observations to validate this trend.

\subsection{Relative alignment of discs and orbits}

Finally, we consider the relative alignment of the spins of the sink particles in binary systems.  Unfortunately there is not a direct analogy with real binary systems in this case because the sink particles are larger than stars and yet smaller than a typical disc.  The orientation of the sink particle spin thus represents the orientation of the total angular momentum of the star and the inner part of its surrounding disc.  This distinction is important because during the formation of an object the angular momentum usually varies with time as gas falls on to it from the turbulent cloud.  Thus, the orientation of the sink particle frequently differs substantially from the orientation of its resolved disc (if one exists) and, furthermore, the orientations of both the sink particles and their discs change with time while the object continues to accrete gas.

\begin{figure*}
\centering
    \includegraphics[width=8.4cm]{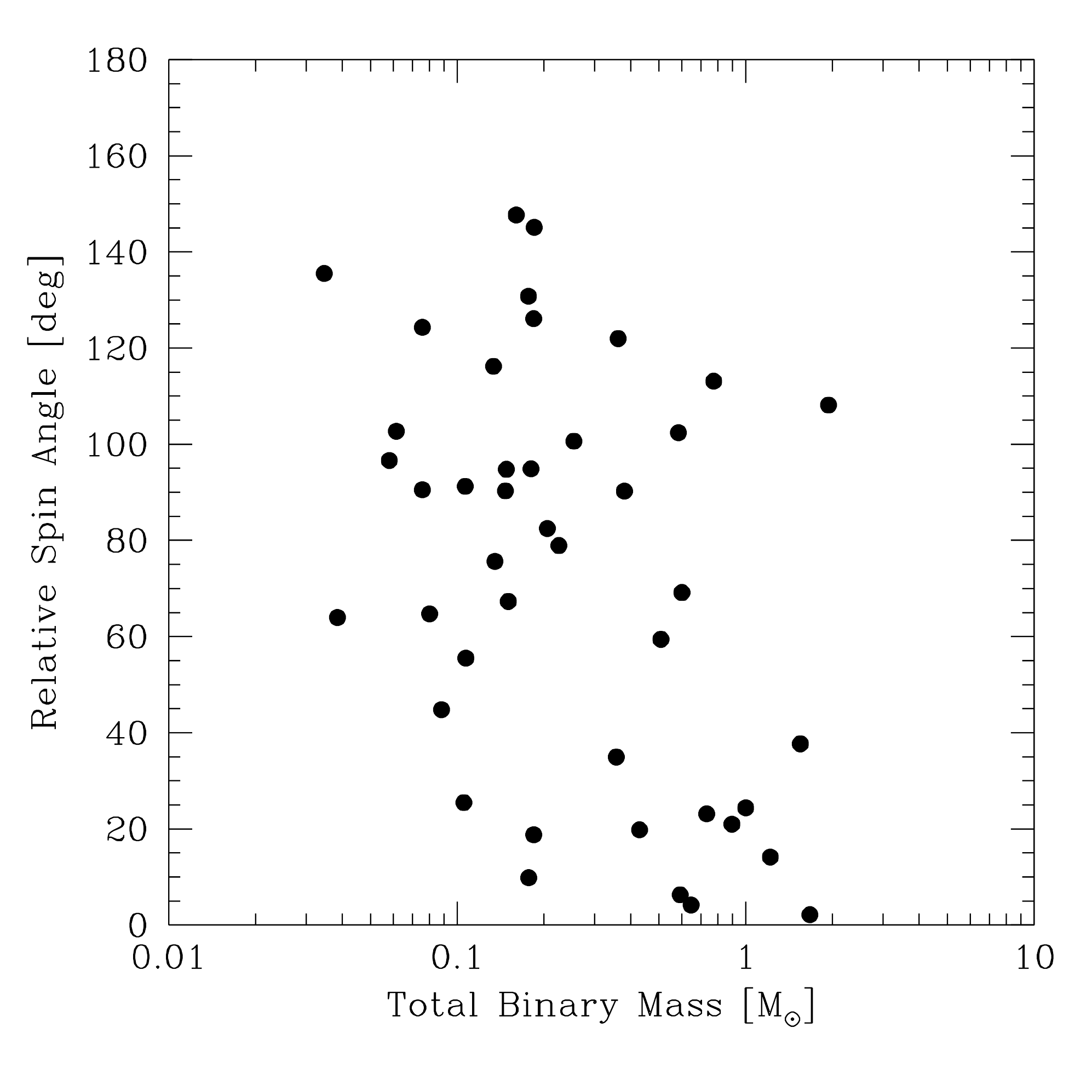}
    \includegraphics[width=8.4cm]{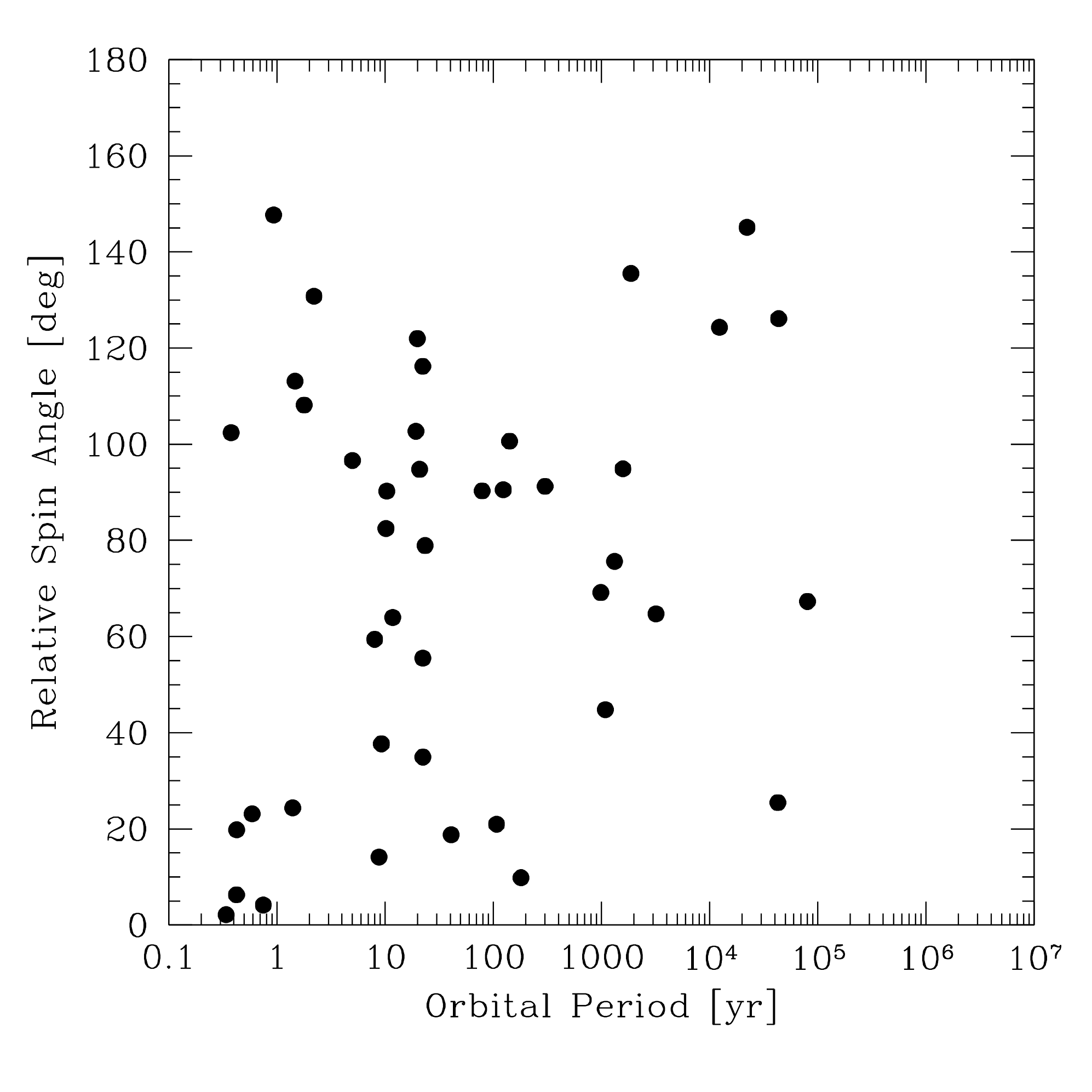}
\caption{The relative inclinations of the rotation axes of the sink particles (modelling stars and their inner discs) of the binary systems produced by the re-run calculation (including those that are sub-components of triples and quadruples). There is an excess of nearly aligned systems with a high total mass and/or orbital periods less than a few years.  The main calculation also shows a slight tendency for high-mass binaries to have aligned rotation axes, but it is not statistically significant (see the main text). 
}
\label{spinspin}
\end{figure*}

Observationally, \citet{Weis1974} found a tendency for alignment between the stellar equatorial planes and orbital planes among primaries in F star binaries, but not A star binaries.  The orbital separations were mainly in the $10-100$ AU range.  Similarly, \citet{Guthrie1985} found no correlation for 23 A star binaries with separations 10-70 AU.  Most recently, \citet{Hale1994} considered 73 binary and multiple systems containing solar-type stars and found evidence for approximate coplanarity between the orbital plane and the stellar equatorial planes for binary systems with separations less than $\approx 30$ AU and apparently uncorrelated stellar rotation and orbital axes for wider systems.  For higher-order multiple systems, however, non-coplanar systems were found to exist for both wide and close orbits.  Hale found no evidence to support a difference dependent on spectral type, eccentricity or age.  In terms of circumstellar discs, there is evidence for misaligned discs from observations of misaligned jets from protostellar objects \citet*{DavMunEis1994}, inferred jet precession \citep{Eisloffeletal1996,Davisetal1997}, and direct observations \citet{Koresko1998, Stapelfeldtetal1998}.  However, these are not statistically useful samples.  Finally, \citet*{MonMenDuc1998}, \citet*{DonJenMat1999}, \citet{Jensenetal2004}, \citet*{WolSteHen2001}, and \citet*{MonMenPer2006} used polarimetry to study the relative disc alignment in T Tauri wide binary and multiple systems.  They all found a preference for disc alignment for binaries.  However, \citet{Jensenetal2004} also found that the wide components of triples and quadruples appear to have random orientations.

For the main calculation (either at the end or at $t=1.038 t_{\rm ff}$) we find no significant dependence of the relative orientation of the two sink particle spins on mass ratio, semi-major axis, period, or eccentricity.  The relative orientations appear to be random.  We do not explicitly consider the relative orientation of the sink particle spins and the orbital plane since if the sink particle spins are uncorrelated with each other, then by definition they cannot (both) be closely correlated with the orbital axis.  The mean relative orientation angle for the 146 binaries (including those that are components of triple and quadruple systems) is $88^\circ\pm 3^\circ$ at the end of the main calculation and $79^\circ\pm 7^\circ$ at $t=1.038 t_{\rm ff}$ (37 binaries).  For the re-run calculation, with smaller accretion radii and orbital periods, the mean angle is $73^\circ\pm 7^\circ$ (43 binaries) and there is a hint that short-period binaries (periods less than a few years) may have preferentially aligned spins but it is not statistically significant (see Figure \ref{spinspin}).  For all of the calculations there is also a hint that the most massive binaries have preferentially aligned spins, but only for the re-run calculation is the reduction in the mean relative angle statistically significant.  In this case, the mean angle for most massive quartile of binaries (11 out of 43, having total binary masses greater than $\approx 0.6$ M$_\odot$) is $38^\circ\pm 12^\circ$ which differs from a random value of $90^\circ$ by more than $4\sigma$, while the mean angle for the other three quartiles are each within $0.5\sigma$ of $90^\circ$ (see the left panel of Figure \ref{spinspin}).  Within the competitive accretion paradigm, the reason that the most massive binaries tend to have aligned rotation axes is presumably that they have both accreted a lot of gas from a common reservoir in order to become massive binaries and that any initial variation in their rotation axes has been decreased by the long period of accretion.  The components of less massive binaries, on the other hand, still largely retain their initial (randomly orientated) rotation axes.  Unfortunately, the observational surveys mentioned above are somewhat ambiguous on whether or not there is a dependence of alignment of the stellar rotation axes on the total binary mass.

\section{Conclusions}
\label{conclusions}

We have presented results from the largest hydrodynamical simulation of star cluster formation to date that resolves the opacity limit for fragmentation.  It also resolves protoplanetary discs (radii $\geq 10$ AU) and binaries with separations as small as 1 AU.  The calculation produced 1254 stars and brown dwarfs.  This large number of objects allows detailed comparison of the statistical properties of the stars, brown dwarfs and multiple systems with the results of observational surveys.  We also re-ran part of the simulation with smaller sink particles and no gravitational softening between sink particle allowing discs with radii $\geq 1$ AU to be resolved and binaries as close as 0.02 AU to test the dependence of the results on the sink particle approximation.  Our conclusions are as follows.

\begin{enumerate}
\item The calculations produce an IMF with a similar form to the observed IMF, including a Salpeter-type slope at the high-mass end, but they over-produce brown dwarfs.  The brown dwarf to star ratio is 3:2 from the main calculation, whereas observationally it is estimated to be more like 1:3. This does not appear to be a result of using sink particles.  Rather it is likely due to the absence of radiative feedback and/or magnetic fields in the calculations.
\item As in previous, smaller calculations, the IMF originates from competition between accretion and ejection which terminates the accretion and sets an object's final mass.  Stars and brown dwarfs form the same way, with similar accretion rates from the molecular cloud, but stars accrete for longer than brown dwarfs before undergoing the dynamical interactions that terminate their accretion.
\item We examine the dependence of binarity, velocity dispersion and the IMF on the distance from the centre of the resulting stellar cluster.  We find that the binarity and velocity dispersion is constant throughout the bulk of the cluster, but beyond 3 half-mass radii (the outer 20\% of the stellar mass) the binarity decreases and the velocity dispersion increases because these objects have been ejected.  We find that stars have a slightly higher velocity dispersion than VLM objects, and binaries have a significantly lower velocity dispersion than single objects.  Contrary to the expectations of competitive accretion, we find no evidence of mass segregation.  This may be because the stellar cluster was formed from the merger of 5 sub-clusters shortly before the calculation was stopped.
\item We examine the potential effect of dynamical interactions on protoplanetary disc sizes.  We find that the typical truncation radius decreases with increasing stellar mass (i.e. more massive stars have had closer encounters).  It is difficult to directly associate the closest encounter with the radii of protostellar discs because many stars accrete new discs after suffering a close encounter.  This is particularly true for the more massive stars.  However, for VLM objects, dynamical encounters usually occur soon after their formation and terminate their accretion so their truncation radii may more closely reflect their disc radii.  Under this assumption we find that at least 10\% of VLM objects should have disc radii $>40$ AU. In lower density star-forming environments this fraction may be expected to be larger.  More massive stars that undergo close encounters and do not subsequently accrete new discs may be the progenitors of WTTS with very young ages ($\lsim 1$ Myr).
\item We find that multiplicity strongly increases with primary mass.  The results from the main calculation are in good agreement with the observed multiplicities of G, K, and M dwarfs.  For VLM objects with primary masses $0.03-0.10$ M$_\odot$ the multiplicity fraction is $0.10\pm 0.03$ which is lower than observations by a factor of two.  However, when smaller accretion radii are used the VLM multiplicity is rises to $0.19\pm0.05$, in good agreement with observations.  Therefore, we conclude that hydrodynamical simulations are able to match the observed multiplicities if the resolution is adequate.  We also predict that the multiplicity continues to drop below 30 Jupiter masses.  We expect a multiplicity no more than $\approx 7$\% for objects with masses 10-30 Jupiter masses, and less than 3\% for primaries of less than 10 Jupiter masses.
\item We find very low frequencies of VLM companions to stars and we find that the frequency does not depend strongly on primary mass.  However, the median star-VLM separation strongly increases as primary mass increases from less than 10 AU for $0.1-0.2$ M$_\odot$ primaries to $\sim 50$ AU for masses $\approx 0.4$ M$_\odot$ and $>100$ AU for solar-type stars.
\item We examine the separation distributions of binaries, triples and quadruples.  We find that the median separation decreases with decreasing primary mass with stellar systems having a median separation of $\approx 26$ AU and VLM systems $\approx 10$ AU.  This trend is in agreement with observed systems, but is not as strong.  At small separations the distributions are dependent on the sink particle parameters.  Better agreement is obtained with smaller sink particle accretion radii and gravitational softening.
\item The mass ratio distribution of M-dwarf binaries is roughly flat and consistent with observations.  VLM systems have a strong preference for equal masses, but not as strong as appears to be the case for observed systems.  However, for K and G-type primaries the calculations underproduce unequal mass systems.  We find that closer binaries tend to have a higher proportion of equal mass components in broad agreement with observed trends.  We also find reasonable agreement with observations on the mass ratios of triples and quadruples, but with relatively large uncertainties from both the simulation and observations.
\item We find that the separations and mass ratios of VLM binaries evolve during their formation from wide systems with unequal masses towards close, equal mass systems. 
\item The main calculation produces a strong excess of short-period highly eccentric binaries.  However, when smaller sink particle accretion radii and gravitational softening is used this excess disappears leaving a reasonable eccentricity distribution with a mean eccentricity that is in agreement with observations.  We also find a weak link between mass ratio and eccentricity such that `twins' have lower eccentricities, as is observed.
\item We investigate the relative orientation of the orbital planes of triple systems.  We obtain a mean orientation angle of $\langle\Phi\rangle=65^\circ\pm 6^\circ$ from the main calculation in excellent agreement with the observed value. Thus, triples have a small tendency for orbital alignment.  The distribution of orientation angles is also in agreement with observations.  There is an absence of relative angles greater than $\approx 140^\circ$ in the simulated triples.
\item Finally, we study the relation orientations of sink particle angular momentum vectors in binaries(analogous to the rotation axes of stars and their inner discs).  We find no significant tendency towards alignment.  However, there is weak evidence that the most massive binaries and/or the shorter period systems may have a tendency for alignment.  Observations suggest that shorter period binaries have a tendency towards alignment.
\end{enumerate}

Overall, the hydrodynamical star cluster formation simulations display good agreement with a wide range of the observed statistical properties of stellar systems.  There are only two of areas of poor agreement: the over production of brown dwarfs relative to stars, and the lack of unequal-mass K and G-dwarf binaries.  The former of these is likely due to the absence of radiative feedback and/or magnetic fields in the simulations, but the reason for the latter is unclear.

Finally, we note that from this point forward, numerical simulations of star formation should be capable of producing precise predictions for the statistical properties stars.  The precision of observational surveys will soon become the limiting factor in comparing the results of numerical simulations with observations.  The results of large observational surveys of stellar properties will be needed in the near future.

\section*{Acknowledgments}

MRB is grateful to the anonymous referee, whose careful reading uncovered several errors in the original version of the paper, and to A.\ Tokovinin, N.\ Siegler, and M.\ Meyer for comments which improved the paper.
The computations reported here were performed using the UK Astrophysical Fluids Facility (UKAFF). 
MRB is grateful for the support of a Philip Leverhulme Prize and a EURYI Award. 
This work, conducted as part of the award ``The formation of stars and planets: Radiation hydrodynamical and magnetohydrodynamical simulations" made under the European Heads of Research Councils and European Science Foundation EURYI (European Young Investigator) Awards scheme, was supported by funds from the Participating Organisations of EURYI and the EC Sixth Framework Programme. 
This publication has made use of the VLM Binaries Archive maintained by Nick Siegler at http://www.vlmbinaries.org/.

\bibliography{mbate}

\end{document}